\documentclass[useAMS,usegraphicx,usenatbib]{mn2e}
\usepackage{amsmath}

\newcommand{\Msun}{\ensuremath{\,{\rm M}_\odot}}           
\newcommand{\Rsun}{\ensuremath{\,{\rm R}_\odot}}           
\newcommand{\psun}{\ensuremath{\,\rho_\odot}}              
\newcommand{\Teff}{\ensuremath{T_{\rm eff}}}               
\newcommand{\logg}{\ensuremath{\log g}}                    
\newcommand{\Mjup}{\ensuremath{\,{\rm M}_{\rm Jup}}}       
\newcommand{\Rjup}{\ensuremath{\,{\rm R}_{\rm Jup}}}       
\newcommand{\pjup}{\ensuremath{\,\rho_{\rm Jup}}}          
\newcommand{\kms}{\,km\,s$^{-1}$}                          
\newcommand{\ms}{\,m\,s$^{-1}$}                            
\newcommand{\cms}{\,cm\,s$^{-2}$}                          
\newcommand{\mc}[1]{\multicolumn{2}{c}{#1}}
\newcommand{\er}[3]{\ensuremath{#1^{+#2}_{-#3}}}
\newcommand{\erm}[3]{\mc{\ensuremath{#1^{+#2}_{-#3}}}}

\newcommand{\FeH}{\ensuremath{\left[\frac{\rm Fe}{\rm H}\right]}}       
\newcommand{\Porb}{\ensuremath{P_{\rm orb}}}               
\newcommand{\Mlt}{\ensuremath{\alpha_{\rm MLT}}}            
\newcommand{\reff}[1]{{#1}}                                  
\newcommand{\refff}[1]{{#1}}                                  

\title[Homogeneous studies of transiting extrasolar planets. II.]
      {Homogeneous studies of transiting extrasolar planets. II. Physical properties}

\author[John Southworth]
       {John Southworth\thanks{E-mail: j.k.taylor@warwick.ac.uk} \\
        Department of Physics, University of Warwick, Coventry, CV4 7AL, UK}

\begin{document} \maketitle 

\begin{abstract}
I present an homogeneous determination of the physical properties of fourteen transiting extrasolar planetary systems for which good photometric and spectroscopic data are available. The input quantities for each system are the results of the light curve analyses presented in Paper\,I, and published measurements of the stellar velocity amplitude, effective temperature and metal abundance. The physical properties are determined by interpolating within tabulated predictions from stellar theory to find the optimal match to these input data. Statistical uncertainties are found using a perturbation algorithm, which gives a detailed error budget for every output quantity. Systematic uncertainties are assessed for each quantity by comparing the values found using several independent sets of stellar models. As a theory-free alternative, physical properties are also calculated using an empirical mass--radius relation constructed from high-precision studies of low-mass eclipsing binary stars. \\
I find that the properties of the planets depend mostly on parameters measured from the light and radial velocity curves, and have a relatively minor sensitivity to theoretical predictions. In contrast, the orbital semimajor axes and stellar masses have a strong dependence on theoretical predictions, and their systematic uncertainties can be substantially larger than the statistical ones. Using the empirical mass--radius relation instead, the \reff{semimajor axes and stellar masses} are smaller by up to 15\%. Thus our understanding of extrasolar planets is currently limited by our lack of understanding of low-mass stars. \\
Using the properties of all known transiting extrasolar planets, I find that correlations between their orbital periods, masses and surface gravities are significant at the 2--3\,$\sigma$ level. However, the separation of the known planets into two classes according to their Safronov number is weaker than previously found, and may not be statistically significant. Three systems, HAT-P-2, WASP-14 and XO-3, form their own little group of outliers, with eccentric orbits, massive planets, and stars with masses $\sim$1.3\Msun. \\
The detailed error budgets calculated for each system show where further observations are needed. XO-1 and WASP-1 could do with new transit light curves. TrES-2 and WASP-2 would benefit from more precise stellar temperature and abundance measurements. Velocity measurements of the parent stars are vital for determining the planetary masses: TrES-1, XO-1, WASP-1, WASP-2 and the OGLEs need additional data. The homogeneous analysis presented here is a step towards large-scale statistical studies of transiting extrasolar planetary systems, in preparation for the expected deluge of new detections from {\it CoRoT} and {\it Kepler}.
\end{abstract}



\section{Introduction}                                                                                 \label{sec:intro}

The discovery of extrasolar planets, made possible through high-precision radial velocity observations of dozens of stars \citep{MayorQueloz95nat}, is one of the great scientific achievements of the twentieth century. Radial velocity surveys have have been remarkably successful so far \citep{UdrySantos07araa}, discovering nearly 300 extrasolar planets at the time of writing%
\footnote{See {\tt http://exoplanet.eu/} for a list of known extrasolar planets.}.
The shortcoming of this technique is that it does not allow us to obtain a detailed understanding of individual objects. For each system it is typically possible to obtain only the orbital period and eccentricity, and lower limits on the mass and orbital separation of the planet.

The detection of the first transiting extrasolar planetary system, HD\,209458 \citep{Charbonneau+00apj,Henry+00apj}, has demonstrated the solution to this problem. By modelling the light curve of a transiting extrasolar planetary system (TEP), adding in radial velocity measurements of the star, and adopting one additional constraint from elsewhere, it is possible to determine the masses and radii of both the star and planet. This information allows the study of the chemical compositions of the two components, and thus the formation and evolution of stellar and planetary systems.

Approximately fifty TEPs are currently known, the majority discovered through wide-field photometric variability surveys. Some estimates of the masses and radii of the components are available for each system, but these have been determined in a variety of ways and using a wide range of additional constraints besides photometric and radial velocity measurements. We are now at the threshold of statistical studies of the properties of TEPs, for which {\em homogeneous} analyses are a fundamental requirement. This work is the second instalment of a series of papers intended to provide an homogeneous study of the known TEPs. A recent paper by \reff{\citet*[][hereafter TWH08]{Torres++08apj}} has the same goal but differences in the method of analysis, \reff{particularly concerning the light curve modelling process}.

Each individual TEP is here studied in a two-stage process, the first stage being detailed modelling of all available good light curves of the system, and the second stage being the inclusion of additional observational and theoretical information to derive the physical properties of both star and planet. Whilst the first stage has little or no dependence on theoretical calculations, the second stage, presented here, is reliant on the predictions of stellar evolutionary models.

In Paper\,I \citep{Me08mn} I tackled stage one: a detailed analysis of the light curves of the fourteen TEPs for which good light curves were then available. The modelling process was performed using the {\sc jktebop} code%
\footnote{{\sc jktebop} is available from \\ {\tt http://www.astro.keele.ac.uk/$\sim$jkt/codes.html}}
\citep{Me++04mn,Me++04mn2}, which represents the components of an eclipsing binary system using biaxial spheroids  \citep{NelsonDavis72apj,PopperEtzel81aj}. Random errors were assessed using Monte Carlo simulations \citep{Me+04mn3,Me+05mn} and systematic errors using a residual-permutation algorithm \citep{Jenkins++02apj}. Careful thought was give to the treatment of limb darkening: five different limb darkening laws were tried \citep[see][]{Me++07aa} and the coefficients of the laws were empirically determined where possible. Theoretically predicted limb darkening coefficients were found to be in harmony with those obtained for most TEPs, but were clearly unable to match the results of the highest-quality data ({\it Hubble Space Telescope} observations of HD\,209458).


The results found in Paper\,I were generally in good agreement with published studies, but for both HD\,189733 and HD\,209458 the analysis of several independent light curves gave divergent results. The discrepancy was the worst for the ratio of the stellar and planetary radii, and amounted to 6.7$\sigma$ for HD\,189733 and 5.6$\sigma$ for HD\,209458. As the ratio of the radii is \reff{primarily dependent on} the transit depth, this means either that the available light curves are affected by some systematic error which is not noticeable from the reduced data alone, or that undetected starspots exist on the stellar surface. In either case, this disagreement can only be adequately dealt with if three or more independent light curves are available for a single TEP. It will be possible to determine whether the discrepancy arises from undetectable systematic errors or starspots by obtaining several independent light curves, each covering the same transit event of a TEP.

In this work I describe and perform the second stage of the analysis: derivation of the physical properties of the fourteen TEPs studied in Paper\,I. This uses the results of the light curve analyses, radial velocity measurements, and additional constraints from theoretical stellar model predictions. Several different sets of stellar models are used, allowing the systematic error inherent in this method to be assessed for every output quantity. I also calculate detailed error budgets for each TEP, showing what further observations will be useful for each system. As a theory-free alternative to stellar model calculations, I also consider an empirical mass--radius relation obtained from high-accuracy studies of \reff{0.2--1.6\Msun} eclipsing binary star systems. The constraints are discussed in Section\,\ref{sec:method}, and applied to each TEP in Section\,\ref{sec:results}. This leads to an homogeneous set of physical properties for the fourteen TEPs (Section\,\ref{sec:absdim}). Finally, the properties of all known TEPs are compiled and studied in Section\,\ref{sec:analysis}.


\section{Method of analysis}                                                                          \label{sec:method}

The modelling of a set of light curves of a TEP gives four quantities which are important here%
\footnote{Throughout this work I identify stellar parameters with a subscripted `A'
and planetary parameters with a subscripted 'b' to conform to IAU nomenclature.}
(Paper\,I): the orbital period (\Porb), the inclination of the orbit with respect to the observer ($i$), and the fractional radii of the star and planet, which are defined to be
\begin{equation} r_{\rm A} = \frac{R_{\rm A}}{a} \qquad \qquad r_{\rm b} = \frac{R_{\rm b}}{a} \end{equation}
where $R_{\rm A}$ and $R_{\rm b}$ are the (absolute) stellar and planetary radii and $a$ is the orbital semimajor axis. To first order, there are four quantities that are directly measurable from a transit light curve (separation in time, depth, overall duration and duration of totality) and four derived quantities (\Porb, $i$, $r_{\rm A}$ and $r_{\rm b}$), so these derived quantities are well determined when the available data give the shape of the transit reliably. The light curve alone does not (apart from \Porb) have any direct dependence on the absolute scale of the system. Note that $i$ is well constrained by a single good light curve, contrary to some statements in the literature \reff{(see Paper\,I)}.

As well as \Porb, $i$, $r_{\rm A}$ and $r_{\rm b}$, it is possible to measure the orbital velocity amplitude of the star, $K_{\rm A}$, from radial velocity measurements. However, one additional quantity or constraint is needed to be able to calculate the physical properties of the system. This constraint is normally derived from stellar evolution theory, but in some cases an accurate {\it Hipparcos} parallax or angular diameter is available which allows stellar theory to be circumvented \reff{\citep[e.g.][]{Baines+07apj}}. Alternatively, an empirical mass--radius relation can be used.

\reff{Once this additional constrant has been specified, it is possible to calculate the physical properties of the planet and star: their masses ($M_{\rm A}$ and $M_{\rm b}$), radii ($R_{\rm A}$ and $R_{\rm b}$), surface gravities ($\log g_{\rm A}$ and $g_{\rm b}$), densities ($\rho_{\rm A}$ and $\rho_{\rm b}$), and the semimajor axis of the orbit ($a$). The interaction with the stellar models also allows an age to be assigned to each TEP based on the properties of the parent star.}

\subsection{Constraints from stellar theory}                                                    \label{sec:method:model}

The mass and radius of the stellar component of a TEP can be constrained by comparing its observed effective temperature (\Teff), surface gravity ($\log g_{\rm A}$), and a measure of its metal abundance (here taken to be \FeH), to the predictions of a set of stellar evolutionary models. Values for \Teff, $\log g_{\rm A}$ and \FeH\ are generally obtained from the analysis of high-dispersion spectra of the star, frequently using the same observational material as for the radial velocity measurements. Knowledge of the stellar parameters then allows the properties of the planet to be determined.

This procedure often produces imprecise results so can be an important source of uncertainty in the final physical properties of the system \citep[e.g.][]{Alonso+04apj,Sato+05apj}. An improved method is to use the stellar density instead of $\log g_{\rm A}$, which is precisely calculable from the results of the light curve analysis \citep{SeagerMallen03apj} and is a good indicator of the mass of a main sequence star. An excellent discussion and example of this process is given by \citet{Sozzetti+07apj} and \citet{Holman+07apj} in their study of TrES-2.

The method of analysis used in the present work is to input the measured quantities \reff{(\Porb, $i$, $r_{\rm A}$, $r_{\rm b}$, $K_{\rm A}$, and \Teff\ and \FeH)} into a code for calculating the physical properties of the system ({\sc jktabsdim}; \reff{\citealt{Me++05aa}}). A reasonable value for the velocity amplitude of the {\em planet} ($K_{\rm b}$) is chosen and used to determine a provisional set of physical properties. The code then interpolates within a tabulated set of stellar model predictions to obtain the expected stellar radius and \Teff\ for the provisional stellar mass and \FeH. \reff{It is done this way because mass and \FeH\ are input quantities for stellar model codes whereas radius and \Teff\ are output quantities.} The $K_{\rm b}$ is then adjusted until the best match is found to the model-predicted stellar radius and \Teff. In practise it is difficult to include the age of the star in the optimisation process because of the strongly nonlinear dependence of stellar properties on age. Therefore the above procedure is performed for a series of ages, starting at 0.1\,Gyr and incrementing in 0.1\,Gyr chunks until the star has evolved well beyond the main sequence. The final set of physical properties corresponds to the best-fitting stellar radius, \Teff\ {\em and} age.

Several notes on this procedure are relevant. (1) Eccentric orbits can be treated without problem. (2) The velocity amplitude of the planet, $K_{\rm b}$, is used purely as a fitting parameter: its clear physical meaning is not useful as it is not directly observable with current technology. (3) Linear interpolation is used within the tables of model predictions, as it is the most robust and reliable technique. This in turn requires the model tabulations to have a dense coverage of parameter space. (4) $\log g_{\rm A}$ values from spectral analysis are not used as constraints because there are questions over the reliability of this procedure (for example see the conflicting observational results for XO-3; \citealt{Johnskrull+08apj}; \citealt{Winn+08apj2}). (5) The very slow evolutionary timescales of K dwarfs and M dwarfs means that their ages are essentially unconstrained and in fact have a negligible effect on the results. G and F dwarfs evolve more quickly, meaning their evolutionary state has a significant effect on the results, and can therefore be determined more precisely. (6) The procedure implicitly applies the constraint on the stellar density obtained from the light curve modelling process.

\subsubsection{Which theoretical stellar evolutionary predictions to use?}                \label{sec:method:model:which}

\begin{table*}
\caption{\label{table:models} Physical ingredients and coverage of the stellar models
used in this work. Note that the {\sf Cambridge\,2007} and {\sf Claret} model sets are
extensions to lower masses which were calculated upon request (J.\ Eldridge, 2007, private
communication; A.\ Claret, 2007, private communication). To ensure homogeneity, these
calculations have not been supplemented with previously published models. Four columns
give some physical quantities adopted by the model sets: $Y_{\rm ini}$ is the primordial
helium abundance, $\Delta Y/\Delta Z$ is the helium-to-metals enrichment ratio, $Z_{\sun}$
is the solar metal abundance (fraction by mass) and \Mlt\ is the mixing length parameter.}
\setlength{\tabcolsep}{5pt}
\begin{tabular}{lllllllll} \hline \hline
Model set             & Reference                                             & Range in        & Range in metal    &                  $Y_{\rm ini}$ & \underline{$\Delta Y$} & $Z_{\sun}$ & \Mlt  & Notes                    \\
\                     &                                                       & mass (\Msun)    & abundance ($Z$)   &                                &            $\Delta Z$  &            &       &                          \\
\hline
{\sf Padova}          & \citet{Girardi+00aas}                                 & 0.15 to 7.0     & 0.0004 to 0.03    &     0.23      &           2.25         &   0.019    & 1.68  &                          \\
{\sf Siess}           & \citet{Siess++00aa}                                   & 0.1 to 7.0      & 0.01 to 0.04      &     0.235     &           2.1          &   0.02     & 1.605 & Includes pre-MS phase    \\
{\sf Y$^2$}           & \citet{Demarque+04apjs}                               & 0.4 to 5.2      & 10$^{-5}$ to 0.08 &     0.23      &           2.0          &   0.02     & 1.743 & Scaled-solar abundances  \\
{\sf Cambridge\,2000} & \citet{Pols+98mn}                                     & 0.5 to 50       & 10$^{-4}$ to 0.03 &     0.24      &           2.0          &   0.0188   & 2.0   & Models with overshooting \\
{\sf Cambridge\,2007} & \citet{EldridgeTout04mn2}                             & 0.5 to 2.0      & 0.01 to 0.05      &     0.24      &           2.0          &   0.0188   & 2.0   & Calculated for this work \\
{\sf Claret}          & \citet{Claret04aa,Claret05aa,Claret06aa2,Claret07aa2} & 0.2 to 1.5      & 0.01 to 0.05      &     0.24      &           2.0          &   0.02     & 1.68  & Calculated for this work \\
\hline \hline \end{tabular} \end{table*}

The method outlined above of determining the properties of TEP has a clear dependence on stellar models. It is consequently important to use a reliable set of models, as any errors will propagate into the final results in full strength. This choice is unfortunately not as straightforward as it might seem at first glance, because there is a long-standing discrepancy between the observed and predicted properties of low-mass \reff{(0.2--1.1\Msun)} eclipsing and interferometric binaries. This disagreement is demonstrated and discussed further in the next section.

For this work I have obtained six different sets of stellar model predictions from five independent groups. The intercomparison of results obtained using different sets of models allows any systematic differences to be identified and their effects on our understanding of TEPs quantified. References and basic characteristics of the sets of models are contained in Table\,\ref{table:models}. Three of the model sets ({\sf Padova}, {\sf Siess}, and {\sf Y$^2$}) have been calculated with substantial emphasis on accurate modelling of low-mass stars, whereas the other three sets ({\sf Cambridge\,2000}, {\sf Cambridge\,2007}, and {\sf Claret}) were originally aimed at the study of more massive stars. I am very grateful to Dr.\ J.\ Eldridge and Dr.\ A.\ Claret for calculating model sets at my request. The models of \citet{Baraffe+98aa} were not considered due to their limited coverage in metal abundance.

Whilst the models sets have been calculated by independent research groups, there are many similarities in the way physical effects have been treated so the final results are certainly not independent. In all six cases the opacities used are from the OPAL group \citep{RogersIglesias92apj,IglesiasRogers93apj,IglesiasRogers96apj} at higher temperatures and from \citet{AlexanderFerguson94apj} at lower temperatures. The metal abundances are scaled-solar and use the solar abundances of \citet{AndersGrevesse89}, \citet{GrevesseNoels93} or \citet{Grevesse++96aspc}. The more recent and controversial solar chemical composition measurements presented by \citet{Asplund++06} have not yet been adopted in the stellar models used here. Five of the six model sets incorporate moderate convective core overshooting (the exception being {\sf Siess}) -- the differing implementations means it is not possible to directly compare the strengths of the effect adopted by the different groups. The {\sf Cambridge\,2000} models are also available without overshooting, but this alternative was not used as I found it made a negligible difference. The {\sf Y$^2$} models are available with enhanced abundances of the $\alpha$-elements; to avoid complication this possibility was again not used here.

The procedure in the work has been to derive the physical properties of TEPs using each of the model sets separately, allowing a clear comparison of the results. To save the reader looking ahead, I find that the {\sf Padova}, {\sf Y$^2$} and {\sf Claret} models generally agree very well, but that the {\sf Cambridge\,2007} and {\sf Siess} model sets display clear and diverse discrepancies for several TEPs. The {\sf Cambridge\,2000} models are not available for masses below 0.5\Msun\ or for metal abundances above $Z = 0.03$ so cannot be used for many of the TEPs studied here. \reff{I have therefore used the {\sf Padova}, {\sf Y$^2$} and {\sf Claret} models for the final results presented below, in an attempt to reach consensus and to show how important systematics are even between models which seem to be in happy agreement. I have furthermore adopted the {\sf Claret} models as the baseline set, and used the results from the other two model sets to infer the systematic errors present in the physical properties. This is because the {\sf Padova} models do not stretch to high enough metal abundances to cover all of the TEPs studied in this work, and also to provide some variety compared to most literature studies which consider only the {\sf Y$^2$} models.}

This is the first time that systematic uncertainties have been presented for the full set of physical properties of a sample of TEPs. However, they should be treated with some caution as they are based on only three different sets of stellar models. Work is underway to improve the situation for later papers in this series. TWH08 presented an homogeneous study of 23 TEPs in which both the {\sf Y$^2$} and {\sf Padova} models were used, and found no major differences between the results.

\subsection{Empirical mass--radius--\Teff\ relations from eclipsing binary stars}                  \label{sec:method:eb}

\begin{table*}
\caption{\label{tab:MRT:eb} Properties of the sample of stars used to
determine the empirical mass--radius and mass--\Teff\ relations.}
\begin{tabular}{l r@{\,$\pm$\,}l r@{\,$\pm$\,}l r@{\,$\pm$\,}l l} \hline \hline
Star                      & \mc{Mass (\Msun)} & \mc{Radius (\Rsun)} & \mc{$\log$\Teff\ (K)} & Reference \\
\hline
CM Dra B                  & 0.2135 & 0.0010 & 0.2347 & 0.0019 & 3.498 & 0.014 & \citet{Metcalfe+96apj} \\
CM Dra A                  & 0.2306 & 0.0011 & 0.2516 & 0.0020 & 3.498 & 0.014 & \citet{Metcalfe+96apj} \\
CU Cnc B                  & 0.3980 & 0.0014 & 0.3908 & 0.0094 & 3.495 & 0.021 & \citet{Ribas03aa} \\
CU Cnc A                  & 0.4333 & 0.0017 & 0.4317 & 0.0052 & 3.500 & 0.021 & \citet{Ribas03aa} \\
NSVS 010317 B             & 0.4982 & 0.0025 & 0.5088 & 0.0030 & 3.546 & 0.004 & \citet{Lopez+06xxx} \\
NSVS 010317 A             & 0.5428 & 0.0027 & 0.5260 & 0.0028 & 3.558 & 0.008 & \citet{Lopez+06xxx} \\
YY Gem A                  & 0.5975 & 0.0047 & 0.6196 & 0.0057 & 3.582 & 0.011 & \citet{TorresRibas02apj} \\
YY Gem B                  & 0.6009 & 0.0047 & 0.6036 & 0.0057 & 3.582 & 0.011 & \citet{TorresRibas02apj} \\
GU Boo B                  & 0.599  & 0.006  & 0.620  & 0.020  & 3.581 & 0.015 & \citet{LopezRibas05apj} \\
GU Boo A                  & 0.610  & 0.007  & 0.623  & 0.016  & 3.593 & 0.014 & \citet{LopezRibas05apj} \\
2MASS J05162881+2607387 B & 0.770  & 0.009  & 0.817  & 0.010  & 3.618 & 0.025 & \citet{BaylessOrosz06apj} \\
2MASS J05162881+2607387 A & 0.787  & 0.012  & 0.788  & 0.015  & 3.623 & 0.020 & \citet{BaylessOrosz06apj} \\
RW Lac B                  & 0.870  & 0.004  & 0.964  & 0.004  & 3.745 & 0.012 & \citet{Lacy+05aj} \\
HS Aur B                  & 0.879  & 0.017  & 0.873  & 0.024  & 3.716 & 0.006 & \citet{Popper+86aj} \\
HS Aur A                  & 0.900  & 0.019  & 1.004  & 0.024  & 3.728 & 0.006 & \citet{Popper+86aj} \\
V1061 Cyg B               & 0.9315 & 0.0068 & 0.974  & 0.020  & 3.724 & 0.012 & \citet{Torres+06apj} \\
FL Lyr B                  & 0.960  & 0.012  & 0.962  & 0.028  & 3.724 & 0.008 & \citet{Popper+86aj} \\
Sun                       &   \mc{1.0}      &   \mc{1.0}      & 3.762 & 0.001 & \citet{Smalley07msais} \\
V432 Aur A                & 1.080  & 0.014  & 1.230  & 0.006  & 3.825 & 0.006 & \citet{Siviero+04aa} \\
EW Ori A                  & 1.194  & 0.014  & 1.141  & 0.011  & 3.776 & 0.007 & \citet{Popper+86aj} \\
HS Hya B                  & 1.2186 & 0.0070 & 1.2161 & 0.0071 & 3.806 & 0.003 & \citet{Torres+97aj} \\
AD Boo B                  & 1.237  & 0.013  & 1.211  & 0.018  & 3.775 & 0.007 & \citet{Lacy97aj} \\
HS Hya A                  & 1.2552 & 0.0078 & 1.2747 & 0.0072 & 3.813 & 0.003 & \citet{Torres+97aj} \\
HD 71636 B                & 1.285  & 0.007  & 1.361  & 0.008  & 3.809 & 0.009 & \citet{Henry+06aj} \\
YZ Cas B                  & 1.350  & 0.010  & 1.348  & 0.015  & 3.826 & 0.016 & \citet{Lacy81apj} \\
V442 Cyg B                & 1.410  & 0.023  & 1.662  & 0.033  & 3.833 & 0.006 & \citet{LacyFrueh87aj} \\
FS Mon B                  & 1.462  & 0.010  & 1.629  & 0.012  & 3.816 & 0.007 & \citet{Lacy+00aj} \\
TZ Men B                  & 1.504  & 0.010  & 1.432  & 0.015  & 3.857 & 0.018 & \citet{Andersen++87aa} \\
GV Car B                  & 1.540  & 0.020  & 1.430  & 0.060  & 3.889 & 0.019 & \citet{MeClausen06conf} \\
V1229 Tau B               & 1.586  & 0.042  & 1.565  & 0.015  & 3.861 & 0.022 & \citet{Groenewegen+07aa} \\
\hline \hline \end{tabular} \end{table*}

The main drawback to using theoretical stellar models in the derivation of the physical properties of TEPs is that any shortcomings of the models propagate directly into systematic errors in the resulting TEP parameters. This is particularly worrying because there is a clear discrepancy between the predicted and observed properties of \reff{0.2--1.1\Msun} stars. Our primary source of information on the properties of normal stars is the study of eclipsing binaries, where it is possible to measure the masses and radii of stars empirically and to accuracies of 1\% or better \citep[e.g.][]{Lacy++08aj,MeClausen07aa}.

{\reff The radii of low-mass eclipsing binary stars are observed to be larger than predicted by theoretical models, by up to 15\% \citep{Hoxie73aa,Clausen98conf,TorresRibas02apj,Ribas+08conf}. Their \Teff s are correspondingly smaller, as the predicted luminosities are close to those observed.} There is a gathering consensus that this discrepancy is due to stellar activity \citep{Ribas06apss,Lopez07apj,Chabrier++07aa}, which is stronger in quickly-rotating stars such as young ones \citep{Morales++08aa} and those in eclipsing binaries. A different conclusion was reached by \citet{Berger+06apj}, whose interferometric observations showed that the radius discrepancy existed in a sample of slowly-rotating field M dwarfs and was correlated with metallicity (a surrogate for opacity). The radius discrepancy was also found for the host star of the TEP GJ\,436, despite this being a slowly-rotating and inactive M dwarf \citep{Torres07apj}.

As an alternative to the use of stellar models, in this work I present physical properties of TEPs calculated using a stellar mass--radius relation constructed from well-studied eclipsing binaries. This procedure is actually much simpler than the one using stellar models (outlined in Section\,\ref{sec:method:model}), and requires only that the value of $K_{\rm b}$ (the orbital velocity amplitude of the planet) is chosen \reff{for which the stellar properties satisfy} a given mass--radius relation. Variations in age and metallicity cannot be accommodated because such values are not in general directly observable for low-mass eclipsing binaries. Instead, the stellar components of the TEPs are assumed to represent the same stellar population as the stars in eclipsing binaries.

\begin{figure} \includegraphics[width=0.48\textwidth,angle=0]{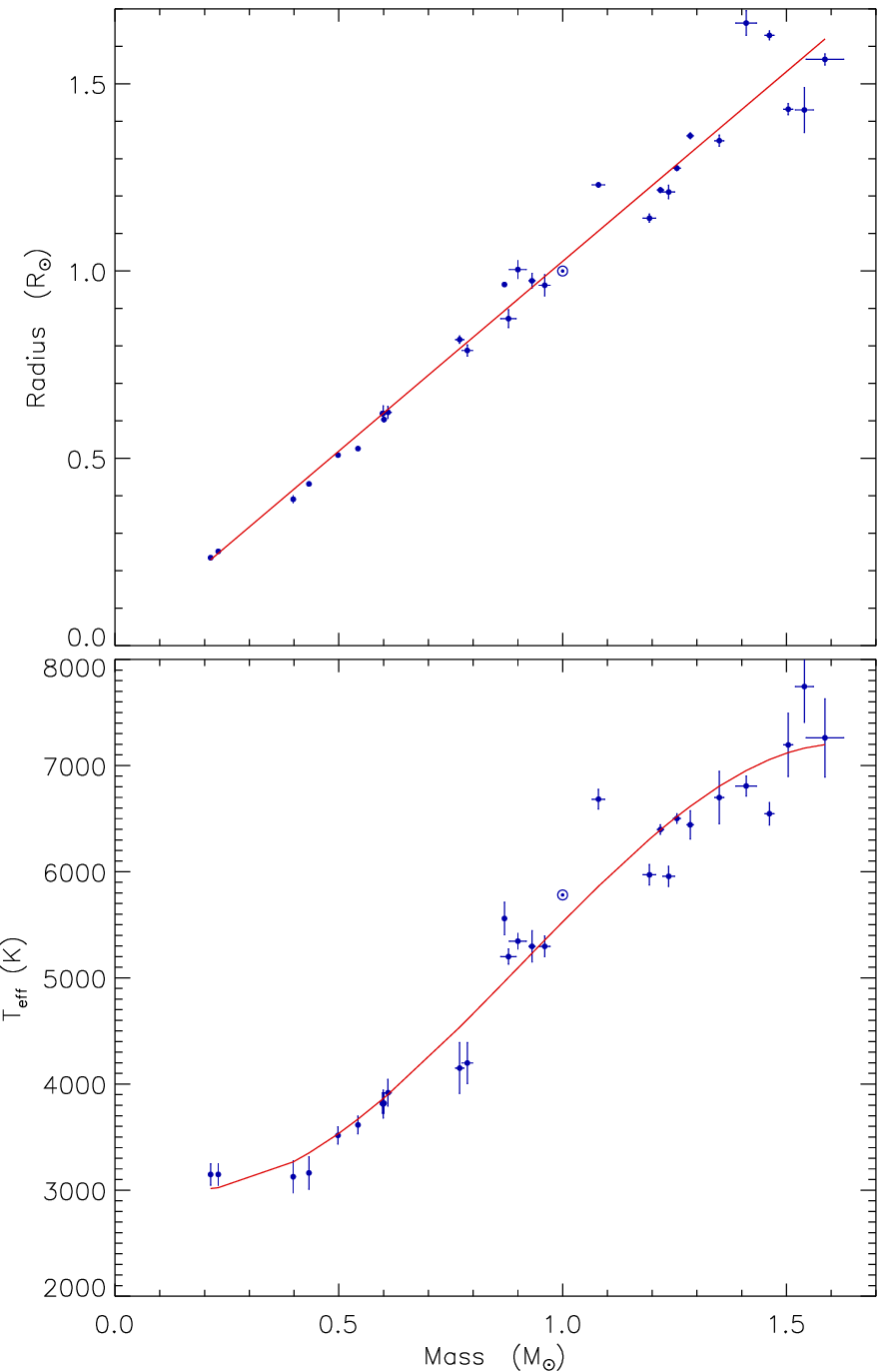}
\caption{\label{fig:MRT:EB} Mass--radius and mass--\Teff\ diagrams showing
the sample of stars given in Table\,\ref{tab:MRT:eb}. The filled circles
show the properties of stars in eclipsing binary systems and the Sun is
represented by a $\odot$. The solid lines represent the mass--radius and
mass--\Teff\ relations obtained from the data (see text for details).}
\end{figure}

\begin{figure} \includegraphics[width=0.48\textwidth,angle=0]{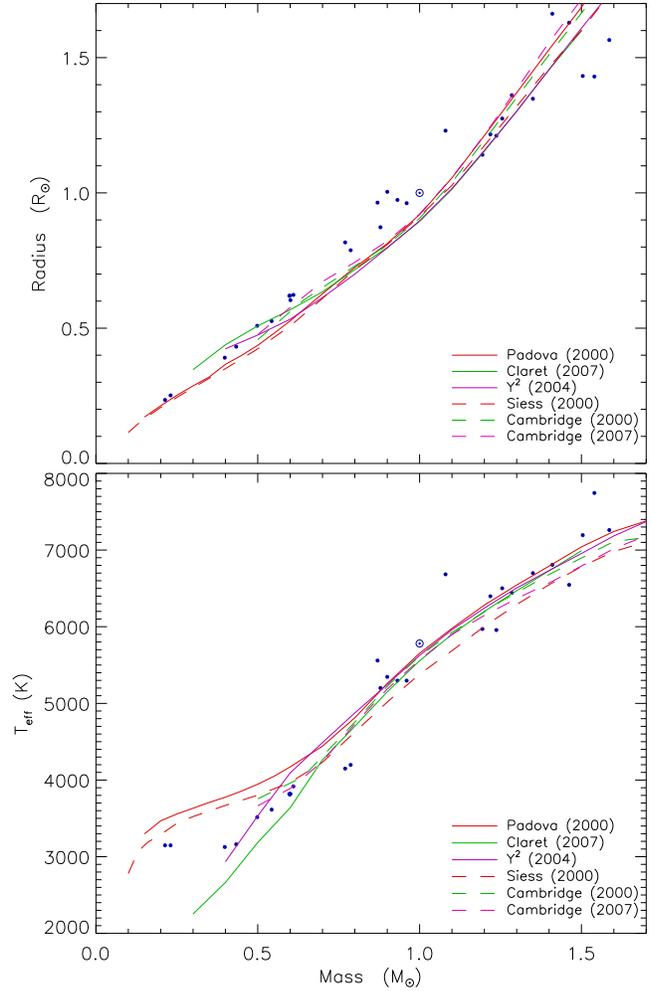}
\caption{\label{fig:MRT:model} Mass--radius and mass--\Teff\ diagrams comparing the observed
properties of low-mass eclipsing binaries and the Sun to theoretical model predictions. The
symbols are as in Fig\,\ref{fig:MRT:EB}. The lines show the predictions of the six sets of
stellar models used in this work, for an age of 1\,Gyr and an approximately solar chemical
composition. Solid lines show predictions from the model sets used for the final results
({\sf Padova}, {\sf Y$^2$} and {\sf Claret}) and dashed lines show predictions from other
the model sets considered in this work ({\sf Cambridge\,2000}, {\sf Cambridge\,2007},
{\sf Siess}). Note that the models do not match the solar properties as they are
calculated for an age much younger than that of the Sun.} \end{figure}

To construct an empirical mass-radius relation for low-mass stars I have compiled the physical properties of all stars in eclipsing binaries%
\footnote{A catalogue of well-studied eclipsing binary systems is available at \\
{\tt http://www.astro.keele.ac.uk/$\sim$jkt/debdata/debs.html}}
which have masses of $M < 1.6$\Msun, and masses and radii determined to accuracies of no worse than 3\%. These data are supplemented by the Sun and are given in Table\,\ref{tab:MRT:eb}. The sample of well-studied eclipsing binaries is biased towards more evolved systems \citep{Andersen91aar} because they are brighter, and their larger radii means that there is a greater probability that they will eclipse. This bias basically affects only those stars with $M \ga 1$\,M$_\odot$, due to age effects, but must be removed from the sample. A simple cut in \logg\ could be used to reject more evolved systems, but the particular choice of cut has a substantial effect on the resulting mass--radius relation. Instead, for eclipsing binaries consisting of two $>$1\,M$_\odot$ stars, I have used only the secondary component and also imposed the requirement that the mass ratio $q = \frac{M_{\rm B}}{M_{\rm A}} < 0.9$. In this way the secondary star is guaranteed to be only part-way through its main sequence lifetime, because if it were not then the primary star would have evolved to or beyond the giant stage. This procedure allows the construction of a sample of stars which are unevolved without making a direct and statistically problematic cut to reject evolved ones. \reff{As an example, TZ\,Men\,A is a 2.5\Msun\ main sequence star so must have an age below about 0.5\,Gyr. This means that TZ\,Men\,B (see Table\,\ref{tab:MRT:eb}) must be no older than 20\% of its total main sequence lifetime of roughly 2.4\,Gyr.}

The above criteria result in the sample of eclipsing binary star components given in Table\,\ref{tab:MRT:eb}. V1229\,Tau\,B was included as its membership of the Pleiades provides independent evidence that it is unevolved \citep{Groenewegen+07aa,Me++05aa}. The resulting sample of 29 stars in eclipsing binaries (plus the Sun) covers the masses 0.214\,M$_\odot$ to 1.586\,M$_\odot$ and -- apart from any effects due to binarity -- is representative of low-mass stars in the Solar neighbourhood.

Low-order polynomials have been fitted to the data in Table\,\ref{tab:MRT:eb} to define mass--radius and mass--\Teff\ relations. The scatter around the relations is much larger than the measurement errors, demonstrating that the stars in the sample have substantial `cosmic scatter' due to differing evolutionary stage, chemical composition, \reff{activity level} and other properties between stars. The measurement errors were therefore not used in calculating the final relations. A first-order polynomial (straight line) is a satisfactory fit to the mass--radius relation, and results in:
\begin{equation} R = (0.00676 \pm 0.03408) + (1.01824 \pm 0.03368) \cdot M \label{eq:mr} \end{equation}
where $M$ and $R$ represent stellar mass and radius in solar units. The quoted uncertainties are 1$\sigma$ errors and the rms scatter about the best fit is 0.073\,R$_\odot$. A large number of significant figures are deliberately included, to avoid problems with rounding off. The mass--\Teff\ relation requires a third-order polynomial to get a good fit, and is:
\begin{eqnarray}
\Teff & = & (3217 \pm 564) - (2427 \pm 2304) \cdot M \nonumber\\
      &   & + (7509 \pm 2802) \cdot M^2 - (2771 \pm 1030) \cdot M^3
\end{eqnarray}
where the rms scatter about the best fit is 328\,K. The mass--\Teff\ relation is less robust than the mass--radius relation due to the large variety of ways in which \Teff s have been observationally measured. I therefore do not consider the mass--\Teff\ relation further.

The mass--radius and mass--\Teff\ relations are compared to the eclipsing binary data in Fig.\,\ref{fig:MRT:EB}. These data are also contrasted in Fig.\,\ref{fig:MRT:model} with the predictions of the six sets of stellar evolutionary models used in this work. It can be seen that the agreement between theory and observation is poor, particularly in the interval 0.7--1.0\Msun. The agreement seems to be better at lower masses, but the radius discrepancy is still visible in new results for the 0.22\,M$_\odot$ eclipsing binary CM\,Dra (Dr. I.\ Ribas, private communication).

\subsection{Physical properties and the pervasive influence of systematic errors}               \label{sec:method:syserr}

\reff{The previous two sections have discussed how to obtain the extra constraint which is needed to transform the directly measured quantities into the physical properties of the star and planet for each system. Following a request from the referee, I now show how important this extra constraint is for each of the physical properties.

The directly observed quantities from the light and velocity curves are \Porb, $K_1$, $e$, $r_{\rm A}$, $r_{\rm b}$ and $i$, where $e$ represents the orbital eccentricity. The extra constraint is derived from several observable and theoretical inputs, which in the current approach are combined to specify the quantity $K_{\rm b}$. Thus the specification of $K_{\rm b}$ contains {\em all} of the indirect constraints which depend on the stellar models (or empirical mass--radius relations) used. Any model-dependent systematic errors will act on $K_{\rm b}$ and so infect most of the quantities below.

From Kepler's third law and the definitions of fractional radius, surface gravity and density, the investment of a small amount of algebra results in the following equations for the semimajor axis and the properties of the star and planet. The equations are in the S.I.\ system so do not include scaling factors to account for the use of astronomical units.}
\begin{equation}  \label{eq:a}
a = a_{\rm A} + a_{\rm b} = \left(\frac{\Porb}{2\pi}\right) \frac{(1-e^2)^\frac{1}{2}}{\sin i} (K_{\rm A} + K_{\rm b})
\end{equation}
\begin{equation}  \label{eq:m1}
M_{\rm A} = \frac{1}{G} \left(\frac{\Porb}{2\pi}\right) \frac{(1-e^2)^\frac{3}{2}}{\sin^3 i} (K_{\rm A} + K_{\rm b})^2 K_{\rm b}
\end{equation}
\begin{equation}  \label{eq:m2}
M_{\rm b} = \frac{1}{G} \left(\frac{\Porb}{2\pi}\right) \frac{(1-e^2)^\frac{3}{2}}{\sin^3 i} (K_{\rm A} + K_{\rm b})^2 K_{\rm A}
\end{equation}
\begin{equation}  \label{eq:r1}
R_{\rm A} = r_{\rm A} a = \left(\frac{\Porb}{2\pi}\right) \frac{(1-e^2)^\frac{1}{2}}{\sin i} r_{\rm A} (K_{\rm A} + K_{\rm b})
\end{equation}
\begin{equation}  \label{eq:r2}
R_{\rm b} = r_{\rm b} a = \left(\frac{\Porb}{2\pi}\right) \frac{(1-e^2)^\frac{1}{2}}{\sin i} r_{\rm b} (K_{\rm A} + K_{\rm b})
\end{equation}
\begin{equation}  \label{eq:g1}
g_{\rm A} = \frac{GM_{\rm A}}{R_{\rm A}^{\ 2}}
          = \left(\frac{2\pi}{\Porb}\right) \frac{(1-e^2)^\frac{1}{2}}{\sin i} \frac{K_{\rm b}}{r_{\rm A}^{\ 2}}
\end{equation}
\begin{equation}  \label{eq:g2}
g_{\rm b} = \frac{GM_{\rm b}}{R_{\rm b}^{\ 2}}
          = \left(\frac{2\pi}{P}\right) \frac{(1-e^2)^\frac{1}{2}}{\sin i} \frac{K_{\rm A}}{r_{\rm b}^{\ 2}}
\end{equation}
\begin{equation}  \label{eq:rho1}
\rho_{\rm A} = \frac{GM_{\rm A}}{R_{\rm A}^{\ 3}}
             = \left(\frac{2\pi}{\Porb}\right)^2 \frac{1}{r_{\rm A}^{\ 3}} \frac{K_{\rm b}}{(K_{\rm A} + K_{\rm b})}
\end{equation}
\begin{equation}  \label{eq:rho2}
\rho_{\rm b} = \frac{GM_{\rm b}}{R_{\rm b}^{\ 3}}
             = \left(\frac{2\pi}{\Porb}\right)^2 \frac{1}{r_{\rm b}^{\ 3}} \frac{K_{\rm A}}{(K_{\rm A} + K_{\rm b})}
\end{equation}
\reff{where Eq.\,\ref{eq:g2} is taken from \citet{Me++07mn}.}

\refff{When interpreting these equations it must be remembered that $K_{\rm A}$ is generally a few hundred m\,s$^{-1}$ and $K_{\rm b}$ is typically 150\kms. Representing the uncertainty in a quantity $x$ as $\sigma(x)$, this means that:
\begin{equation}  \label{eq:k1+k2}
K_{\rm A} + K_{\rm b} \approx K_{\rm b} \qquad \qquad \sigma(K_{\rm A} + K_{\rm b}) \approx \sigma(K_{\rm b})
\end{equation}
and
\begin{equation}  \label{eq:k1/k2}
\frac{K_{\rm b}}{K_{\rm A} + K_{\rm b}} \approx 1 \qquad \qquad \sigma\left(\frac{K_{\rm b}}{K_{\rm A} + K_{\rm b}}\right) \to 0
\end{equation}

Two properties can be picked out which depend mainly on $K_{\rm b}$ and thus are strongly affected by model-dependent systematic errors: $a$ and $M_{\rm A}$. In the usual case where \Porb\ and $\sin i$ have a negligible uncertainty and $e$ is assumed to be zero, the relation in Eq.\,\ref{eq:k1+k2} means that the precisions of $a$ and $M_{\rm A}$ depend only on the precision of $K_{\rm b}$. It is important to remember that both $a$ and $M_{\rm A}$ have quite a strong dependence on the measured $r_{\rm A}$: this is not explicit in Eqs.\ \ref{eq:a} and \ref{eq:m1} but happens because $r_{\rm A}$ has an important role in determining $K_{\rm b}$ (Section\,\ref{sec:method:model}).

The other properties of the star are less affected by systematic errors. $R_{\rm A}$ and $g_{\rm A}$ are proportional to $K_{\rm b}$ but also depend on $r_{\rm A}$. In general, the uncertainty in $r_{\rm A}$ dominates that in $K_{\rm b}$ so the model dependence is less important than the quality of the light curve. $\rho_{\rm A}$ is a special case as its dependence on outside constraints is negligible (Eq.\,\ref{eq:rho1} and Eq.\,\ref{eq:k1/k2}), as is already well known \citep{SeagerMallen03apj}.

The properties of the planet are all in general more strongly affected by the quality of the observations than by the model-dependent systematics. This is the case for $R_{\rm b}$ and $\rho_{\rm b}$ for all the TEPs discussed here. The situation for the planet's mass is more complicated: using Eq.\,\ref{eq:k1/k2} gives $M_{\rm b} \propto K_{\rm A} K_{\rm b}^{\ 2}$. Whilst $K_{\rm b}$ is typically known to a precision of 1.5\%, $K_{\rm A}$ is measured to accuracies of between 1.2\% and 30\% for these TEPs. In most cases the uncertainty in $K_{\rm A}$ dominates that in $K_{\rm b}^{\ 2}$, so the model-dependence of the properties of the planet is relatively unimportant. The surface gravity of the planet, $g_{\rm b}$, is a special case as it is only dependent on observable quantities so can be freed from any influence of systematic errors \citep{Me+04mn3,Me++07mn}.
}

\subsection{Error analysis}                                                                    \label{sec:method:errors}

A robust error analysis method is a vital tool in high-precision studies. The error analysis is performed using the {\sc jktabsdim} code by perturbing \reff{every input parameter by its uncertainty, whilst keeping all other parameters at their input values,} to measure the effect it has on all output quantities. The resulting individual error contributions are summed in quadrature to provide a final uncertainty for each of the physical properties calculated by {\sc jktabsdim}. The uncertainty in the mass--radius relation is taken to be the 1$\sigma$ errors on the two coefficients.

From the full error budget calculated by the above procedure, it is possible to see for each system exactly how much effect each input quantity has on the final uncertainties, and therefore what observations could be taken to improve the measurement of the physical properties for each system. As $K_{\rm b}$ is used as a fitting parameter, its output uncertainty is a measure of the uncertainty in the stellar properties resulting from the observational errors in \Teff\ and \FeH. Example error budgets are given below for the TEPs TrES-1 and HD\,209458.

As a general rule, if the input parameters have symmetric errors, the resulting output parameters also have approximately symmetric uncertainties. In some cases the light curve analyses in Paper\,I resulted in separate upper and lower error estimates for the photometric parameters, which propagate through into asymmetric error estimates for the quantities calculated in this work. The exception is the model-derived stellar age, which has asymmetric errorbars in almost all cases.


\section{Physical properties of the TEPs}                                                            \label{sec:results}

\begin{table*} \caption{\label{tab:abspar:LCresults} Parameters from the
light curve modelling presented in Paper\,I and used in this work.}
\begin{tabular}{l l r@{\,$\pm$\,}l r@{\,$\pm$\,}l r@{\,$\pm$\,}l}
\hline \hline
System & Orbital period (days) & \mc{Orbital inclination (degrees)} & \mc{Fractional stellar radius, $r_{\rm A}$} & \mc{Fractional planetary radius, $r_{\rm b}$} \\
\hline
TrES-1      & 3.030065   & 88.67 & 0.71   & 0.0964 & 0.0018              & 0.01331 & 0.00035            \\
TrES-2      & 2.47063    & 83.71 & 0.42   & 0.1296 & 0.0038              & 0.01643 & 0.00046            \\
XO-1        & 3.941534   & 89.06 & 0.84   & 0.0886 & 0.0019              & 0.01166 & 0.00035            \\
WASP-1      & 2.519961   & 88.0 & 2.0     & \erm{0.1737}{0.0057}{0.0089} & \erm{0.0182}{0.0007}{0.0011} \\
WASP-2      & 2.152226   & 84.83 & 0.53   & 0.1245 & 0.0058              & 0.01635 & 0.00093            \\
HAT-P-1     & 4.46543    & 86.26 & 0.24   & 0.0930 & 0.0028              & 0.01043 & 0.00033            \\
OGLE-TR-10  & 3.101278   & 83.87 & 0.69   & 0.157 & 0.009                & 0.0182 & 0.0011              \\
OGLE-TR-56  & 1.211909   & 79.8 & 2.4     & 0.245 & 0.026                & 0.0241 & 0.0034              \\
OGLE-TR-111 & 4.0144479  & 88.11 & 0.66   & 0.0842 & 0.0038              & 0.01107 & 0.00067            \\
OGLE-TR-132 & 1.689868   & 83.3 & 2.4     & 0.211 & 0.020                & 0.0198 & 0.0024              \\
GJ\,436     & 2.64385    & 86.43 & 0.18   & 0.0731 & 0.0027              & 0.00605 & 0.00023            \\
HD\,149026  & 2.8758882  & 88.0 & 2.0     & \erm{0.140}{0.012}{0.006}    & \erm{0.0068}{0.011}{0.008}   \\
HD\,189733  & 2.2185733  & 85.78 & 0.25   & 0.1113 & 0.0031              & 0.0175 & 0.0005              \\
HD\,209458  & 3.52474859 & 86.590 & 0.046 & 0.11384 & 0.00041            & 0.01389 & 0.00006            \\
\hline \hline \end{tabular} \end{table*}

\begin{table*} \caption{\label{tab:abspar:obsstar} Measured quantities for
the parent stars which were adopted in the analysis presented in this work.}
\setlength{\tabcolsep}{2pt}
\begin{tabular}{l r@{\,$\pm$\,}l l r@{\,$\pm$\,}l l r@{\,$\pm$\,}l l}
\hline \hline
System & \multicolumn{3}{l}{Velocity amplitude (\ms)} & \mc{\Teff\ (K)} & Reference & \mc{\FeH} & Reference \\
\hline
TrES-1 & 115.2 & 6.2 & \citet{Alonso+04apj} & 5226 & 50 & \citet{Santos+06aa} & 0.06 & 0.05 & \citet{Santos+06aa} \\
TrES-2 & 181.3 & 2.6 & \citet{Odonovan+06apj} & 5850 & 50 & \citet{Sozzetti+07apj} & $-$0.15&0.10&\citet{Sozzetti+07apj}\\
XO-1 & 116.0 & 9.0 & \citet{Mccullough+06apj} & 5750 & 50 & \citet{Mccullough+06apj} &0.015&0.05&\citet{Mccullough+06apj}\\
WASP-1 & 114 & 13 & \citet{Cameron+07mn} & 6110 & 50 & \citet{Stempels+07mn} & 0.23 & 0.08 & \citet{Stempels+07mn} \\
WASP-2 & 155 & 11 & \citet{Cameron+07mn} & 5200 & 200 & \citet{Cameron+07mn} & 0.00 & 0.15 & (assumed) \\
HAT-P-1 & 60.3 & 2.1 & \citet{Bakos+07apj} & 5975 & 50 & \citet{Bakos+07apj} & 0.13 & 0.05 & \citet{Bakos+07apj} \\
OGLE-TR-10 & 80 & 17 & \citet{Konacki+05apj} & 6075 & 86 & \citet{Santos+06aa} & 0.28 & 0.10 & \citet{Santos+06aa} \\
OGLE-TR-56 & 212 & 22 & \citet{Bouchy+05aa2} & 6119 & 62 & \citet{Santos+06aa} & 0.25 & 0.08 & \citet{Santos+06aa} \\
OGLE-TR-111 & 78 & 14 & \citet{Pont+04aa} & 5044 & 83 & \citet{Santos+06aa} & 0.19 & 0.07 & \citet{Santos+06aa} \\
OGLE-TR-132 & 141 & 42 & \citet{Bouchy+04aa} & 6210 & 59 & \citet{Gillon+07aa3} & 0.37 & 0.07 & \citet{Gillon+07aa3} \\
GJ\,436 & 18.34 & 0.52 & \citet{Maness+07pasp} & 3350 & 300 & \citet{Maness+07pasp} & $-$0.03 &0.2&\citet{Bonfils+05aa}\\
HD\,149026 & 43.3 & 1.2 & \citet{Sato+05apj} & 6147 & 50 & \citet{Sato+05apj} & 0.36 & 0.05 & \citet{Sato+05apj} \\
HD\,189733 & 205 & 6 & \citet{Bouchy+05aa} & 5050 & 50 & \citet{Bouchy+05aa} & $-$0.03 & 0.05 & \citet{Bouchy+05aa} \\
HD\,209458 & 85.1 & 1.0 & \citet{Naef+04aa} & 6117 & 50 & \citet{Santos++04aa} & 0.02 & 0.05 & \citet{Santos++04aa} \\
\hline \hline \end{tabular} \end{table*}

The analyses presented in this work require measurements of \Porb, $i$, $r_{\rm A}$ and $r_{\rm b}$ from transit light curve modelling. These measurements were obtained in Paper\,I and are given in Table\,\ref{tab:abspar:LCresults}. The analyses also require values for the stellar properties $K_{\rm A}$, \Teff\ and \FeH. These have been taken from the literature, and are listed and referenced in Table\,\ref{tab:abspar:obsstar}. The minimum uncertainties in \Teff\ and \FeH\ have been set at 50\,K and 0.05\,dex, respectively, following the recommendations of TWH08 and Dr.\ B.\ Smalley (private communication). These limits transpire from studies of the reliability of the \Teff\ scale for FGK dwarfs \citep[e.g.][]{RamirezMelendez05apj,RamirezMelendez05apj2}, and the fact that \Teff\ and \FeH\ are significantly correlated in spectral synthesis analyses \citep[e.g.][]{Buzzoni+01pasp,Holman+07apj2}.

The analyses are presented below for the same fourteen TEPs as in Paper\,I, and following strictly the same order as that work. Example full error budgets are presented for the first and last TEPs (TrES-1 and HD\,209458). The planetary surface gravity is specified completely by parameters measured from the light and radial velocity curves \citep{Me+04mn3,Me++07mn} so has no dependence on theoretical models or mass--radius relations%
\footnote{The literature contains statements that calculation of planetary surface gravity is `almost' or `virtually' independent of theoretical calculations. It is actually totally independent -- depending on how limb darkening is accounted for in the light curve analysis -- and is applicable to any opaque and approximately circular body, including stars, planets, moons, and tennis balls.}.
The stellar density is similarly almost totally free of theoretical dependence \citep{SeagerMallen03apj}, the `almost' stemming from the requirement that the mass of the planet is negligible compared to that of the star.

In this sort of study is it is important to compare the results against literature determinations, to find and investigate any discrepancies. The resulting tables of results are bulky and have been exiled to the Appendix (electronic only).


\subsection{TrES-1}                                                                                    \label{sec:tres1}

\begin{table*}
\caption{\label{tab:abspar:tres1} Derived physical properties for the TrES-1 system. $a$ is the
orbital semimajor axis. The stellar mass, radius, gravity and density are denoted by $M_{\rm A}$,
$R_{\rm A}$, $\log g_{\rm A}$ and $\rho_{\rm A}$, respectively. The corresponding planetary
quantities are given by $M_{\rm b}$, $R_{\rm b}$, $g_{\rm b}$ and $\rho_{\rm b}$.}
\begin{tabular}{l r@{\,$\pm$\,}l r@{\,$\pm$\,}l r@{\,$\pm$\,}l r@{\,$\pm$\,}l r@{\,$\pm$\,}l r@{\,$\pm$\,}l}
\hline \hline
                      & \mc{Mass--radius} & \mc{{\sf Padova} models} & \mc{{\sf Siess} models} & \mc{{\sf Y$^2$} models}
                            & \mc{{\sf Cambridge\,2007}} & \mc{{\sf Claret} models} \\
\hline      
$a$         (AU)      & 0.0373& 0.0011& 0.03954& 0.00036& 0.04035\,$^*$& 0.00023 & 0.03946&0.00034& 0.03942& 0.00017& 0.04000& 0.00043\\[2pt]  
$M_{\rm A}$ (\Msun)   & 0.752 & 0.066 & 0.897  & 0.025  & 0.954\,$^*$  &  0.016  & 0.892  & 0.023 & 0.890  & 0.012  & 0.929  & 0.030  \\       
$R_{\rm A}$ (\Rsun)   & 0.772 & 0.031 & 0.819  & 0.015  & 0.836\,$^*$  &  0.020  & 0.818  & 0.015 & 0.817  & 0.015  & 0.829  & 0.015  \\       
$\log g_{\rm A}$ (cgs)& 4.539 & 0.017 & 4.564  & 0.018  & 4.573\,$^*$  &  0.014  & 4.563  & 0.018 & 4.563  & 0.016  & 4.569  & 0.018  \\       
$\rho_{\rm A}$ (\psun)& 1.632 & 0.093 & 1.632  & 0.092  & 1.632\,$^*$  &  0.092  & 1.632  & 0.092 & 1.632  & 0.092  & 1.632  & 0.092  \\[2pt]  
$M_{\rm b}$ (\Mjup)   & 0.678 & 0.054 & 0.763  & 0.043  & 0.795\,$^*$  &  0.044  & 0.760  & 0.043 & 0.759  & 0.041  & 0.781  & 0.045  \\       
$R_{\rm b}$ (\Rjup)   & 1.038 & 0.041 & 1.101  & 0.031  & 1.124\,$^*$  &  0.030  & 1.099  & 0.030 & 1.098  & 0.029  & 1.114  & 0.032  \\       
$g_{\rm b}$ (\ms)     & 15.6  & 1.2   & 15.6   & 1.2    & 15.6         &  1.2    & 15.6   & 1.2   & 15.6   & 1.2    & 15.6   & 1.2    \\       
$\rho_{\rm b}$ (\pjup)& 0.606 & 0.060 & 0.572  & 0.055  & 0.560\,$^*$  &  0.054  & 0.573  & 0.055 & 0.573  & 0.055  & 0.565  & 0.054  \\[2pt]       
Age (Gyr)             & \mc{ } & \erm{1.0}{3.9}{1.6} & \erm{0.1\,{^*}\,}{0.0}{0.0} & \erm{3.4}{2.1}{2.1} & \erm{0.1}{1.8}{0.0} & \erm{1.2}{4.1}{1.8} \\ 
\hline \hline \end{tabular} \\
\begin{flushleft} $^*$\,These quantities have been calculated using stellar models for stars with small ages. Their values and uncertainties may therefore be unreliable due to edge effects within the grid of model tabulations. \end{flushleft}
\end{table*}

TrES-1 was discovered to be a TEP by the Trans-Atlantic Exoplanet Survey \citep{Alonso+04apj}. An excellent $z$-band light curve of TrES-1, obtained by \citet{Winn++07apj} and chosen for analysis in Paper\,I, gives the fractional radii of the star and planet to 2\% and 3\%, respectively. The physical properties of the two components and of the system have been calculated using five sets of stellar models ({\sf Padova}, {\sf Siess}, {\sf Y$^2$}, {\sf Cambridge\,2007}, and {\sf Claret}) and are listed in Table\,\ref{tab:abspar:tres1}.

The agreement between three sets of models ({\sf Padova}, {\sf Y$^2$} and {\sf Claret}) is good, but the other two models are slightly discrepant and predict very young ages which are towards the edge of the range of possible ages. \reff{The ages found for the three good sets of models agree to within their (rather large) errors; note that} it is possible to have a larger errorbar than value for age because individual error contributions are added in quadrature. The age \reff{of TrES-1} is therefore constrained only to be less than about 5.5\,Gyr. A comparison with published results is available in the Appendix (Table\,\ref{apptab:abspar:tres1}) and shows a reasonable agreement.

The physical properties found using the eclipsing binary mass--radius relation (Eq.\,\ref{eq:mr}) are quite different to those found using stellar models, as TrES-1\,A is in the mass regime where models and observational data match poorly (Fig.\,\ref{fig:MRT:EB}). The discrepancy amounts to 15\% in $M_{\rm A}$, 6\% in $R_{\rm A}$, 13\% in $M_{\rm b}$ and 6\% in $R_{\rm b}$. The stellar and planetary masses are thus much more strongly affected than their respective radii. The planet's surface gravity, $g_{\rm b}$, is not affected because this depends only on observed properties \citep{Me+04mn3,Me++07mn}, but the planet density is affected by 6\%. It is important to note that this disagreement is a manifestation of the radius discrepancy seen in low-mass eclipsing and field stars (Section\,\ref{sec:method:eb}), so is a fundamental limitation due to the current state of knowledge in stellar astrophysics. Until this discrepancy can be cleared up, we cannot claim to measure the parameters of TEPs to better than the percentages given above (depending on the system). Thus {\em our understanding of planets is limited by our understanding of low-mass stars}.


\subsubsection{Error budget}                                                                       \label{sec:tres1:err}

\begin{table}
\caption{\label{tab:abspar:tres1:errmod} Detailed error budget for the calculation
of the system properties of TrES-1 from the light curve parameters, stellar velocity
amplitude, and the predictions of the {\sf Padova} stellar models. Each number in the table
is the fractional contribution to the final uncertainty of an output parameter from
the errorbar of an input parameter. The final uncertainty for each output parameter
(not given) is the quadrature sum of the individual contributions from each input
parameter. Whilst the orbital period is an input parameter, it is not included here
as its uncertainty too small to register.}
\begin{tabular}{l|rrrrrrr}
\hline \hline
Output            & \multicolumn{6}{c}{Input parameter} \\
parameter         & $K_{\rm A}$ & $i$   & $r_{\rm A}$ & $r_{\rm b}$ & \Teff & \FeH  \\
\hline
Age               &             &       &    0.737    &             & 0.468 & 0.468 \\
$a$               &    0.002    &       &    0.336    &             & 0.715 & 0.542 \\[2pt]
$M_{\rm A}$       &             &       &    0.335    &             & 0.715 & 0.542 \\
$R_{\rm A}$       &    0.001    &       &    0.886    &             & 0.367 & 0.278 \\
$\log g_{\rm A}$  &    0.001    &       &    0.980    &             & 0.159 & 0.121 \\
$\rho_{\rm A}$    &    0.001    &       &    1.000    &             &       &       \\[2pt]
$M_{\rm b}$       &    0.945    & 0.005 &    0.107    &             & 0.229 & 0.173 \\
$R_{\rm b}$       &    0.001    &       &    0.110    &    0.943    & 0.234 & 0.177 \\
$g_{\rm b}$       &    0.715    & 0.004 &             &    0.699    &       &       \\
$\rho_{\rm b}$    &    0.560    & 0.003 &    0.032    &    0.823    & 0.068 & 0.052 \\
\hline \hline
\end{tabular} \end{table}

\begin{table}
\caption{\label{tab:abspar:tres1:errMR} Detailed error budget for the calculation
of the system properties of TrES-1 from the light curve parameters, stellar velocity
amplitude, and mass--radius relation. The layout of the table is the same as that
for Table\,\ref{tab:abspar:tres1:errmod}. The quantities MR0 and MR1 refer to the
constant and linear coefficients of the mass--radius relation.}
\begin{tabular}{l|rrrrrrr}
\hline \hline
Output            & \multicolumn{6}{c}{Input parameter} \\
parameter         & $K_{\rm A}$ & $i$   & $r_{\rm A}$ & $r_{\rm b}$ & MR0   & MR1   \\
\hline
$a$               &    0.001    &       &    0.321    &             & 0.760 & 0.564 \\[2pt]
$M_{\rm A}$       &             &       &    0.321    &             & 0.760 & 0.565 \\
$R_{\rm A}$       &    0.001    &       &    0.710    &             & 0.565 & 0.419 \\
$\log g_{\rm A}$  &    0.001    &       &    0.706    &             & 0.568 & 0.421 \\
$\rho_{\rm A}$    &    0.001    &       &    1.000    &             &       &       \\[2pt]
$M_{\rm b}$       &    0.675    & 0.004 &    0.237    &             & 0.561 & 0.417 \\
$R_{\rm b}$       &    0.001    &       &    0.239    &    0.666    & 0.567 & 0.421 \\
$g_{\rm b}$       &    0.715    & 0.004 &             &    0.699    &       &       \\
$\rho_{\rm b}$    &    0.537    & 0.003 &    0.094    &    0.790    & 0.224 & 0.166 \\
\hline \hline
\end{tabular} \end{table}

The analysis procedure used here (see Section\,\ref{sec:method:model}) yields a full error budget detailing the effect on each input parameter on each output quantity. For TrES-1 I include these error budgets to give an example of what information can be grasped from such numbers. The error budget calculated using the {\sf Padova} stellar models is given in Table\,\ref{tab:abspar:tres1:errmod} and using the mass--radius relation in Table\,\ref{tab:abspar:tres1:errMR}.

From Table\,\ref{tab:abspar:tres1:errmod} it can be seen that the uncertainty in the orbital inclination ($i$) is totally unimportant to the calculated quantities. \reff{The uncertainty in $K_{\rm A}$ dominates the uncertainty in $M_{\rm b}$, and is important also for $g_{\rm b}$ and $\rho_{\rm b}$.} Unsurprisingly, $r_{\rm A}$ is a vital input in calculating the physical properties of the star, in particular its age and density, and $r_{\rm b}$ is similarly critical to our understanding of the planet's properties. Of most interest is the effect of uncertainty in \Teff\ and \FeH\ -- the quantities which provide the all-important final constraint for calculation of the full physical properties of the system. These are relevant for most of the output quantities, but are by a long way the most important input for calculation of the stellar mass and the orbital separation ($a$). It is reassuring that the effect of \Teff\ and \FeH\ is relatively limited for our understanding of the properties of the planet.

The error budget for the calculations involving the mass--radius relation (Table\,\ref{tab:abspar:tres1:errMR}) is similar to that using stellar models. As a general rule, the final results are less accurate as the uncertainty in the coefficients of the relation is more important than the uncertainty in \Teff\ and \FeH. These coefficients therefore take on a more important role in calculation of the physical properties, and dominate the uncertainties in $a$ and $M_{\rm A}$. The effect of uncertainties on the observed input quantities ($K_{\rm A}$, $i$, $r_{\rm A}$, $r_{\rm b}$) are consequently less important. This effect would be even stronger if the `cosmic scatter' in the calculation were included in full strength (i.e.\ if the uncertainty in the mass--radius calibration did not decrease with the inclusion of more data).

The final conclusions of this section are that the best way to improve our knowledge of the parameters of the planet is to get a better measurement for $r_{\rm b}$. Similarly, a more precise $r_{\rm A}$ is needed to better understand the star. As these quantities are already accurately measured, it will need a very good light curve to improve this situation. However, a more accurate measurement of $K_{\rm A}$ is well within reach and would much improve the measurement of the mass of the planet.


\subsection{TrES-2}                                                                                    \label{sec:tres2}

\begin{table*}
\caption{\label{tab:abspar:tres2} Derived physical properties for TrES-2.
The symbols are as in Table\,\ref{tab:abspar:tres1}. In every case the
planetary surface gravity is $g_{\rm b} = 19.9 \pm 1.2$\ms\ and the
stellar density is $\rho_{\rm A} = 1.008 \pm 0.092$\psun.}
\begin{tabular}{l r@{\,$\pm$\,}l r@{\,$\pm$\,}l r@{\,$\pm$\,}l r@{\,$\pm$\,}l r@{\,$\pm$\,}l r@{\,$\pm$\,}l r@{\,$\pm$\,}l}
\hline \hline
                      & \mc{Mass--radius} & \mc{{\sf Padova} models} & \mc{{\sf Siess} models} & \mc{{\sf Y$^2$} models}
                            & \mc{{\sf Cambridge\,2007}}  & \mc{{\sf Claret} models} \\
\hline                                                                                                                                    
$a$         (AU)      & 0.0352&0.0010 & 0.03537&0.00061 & 0.03666&0.00063 & 0.03559&0.00051 & 0.03708&0.00024 & 0.03568&0.00056 \\[2pt]  
$M_{\rm A}$ (\Msun)   & 0.958 & 0.081 & 0.966  & 0.050  & 1.075  & 0.056  & 0.984  & 0.043  & 1.113  & 0.022  & 0.991  & 0.047  \\       
$R_{\rm A}$ (\Rsun)   & 0.983 & 0.049 & 0.985  & 0.031  & 1.021  & 0.032  & 0.991  & 0.031  & 1.033  & 0.033  & 0.994  & 0.031  \\       
$\log g_{\rm A}$ (cgs)& 4.435 & 0.022 & 4.436  & 0.028  & 4.451  & 0.028  & 4.439  & 0.027  & 4.456  & 0.025  & 4.440  & 0.027  \\[2pt]       
$M_{\rm b}$ (\Mjup)   & 1.180 & 0.069 & 1.186  & 0.045  & 1.274  & 0.048  & 1.201  & 0.039  & 1.303  & 0.026  & 1.206  & 0.042  \\       
$R_{\rm b}$ (\Rjup)   & 1.213 & 0.048 & 1.216  & 0.040  & 1.260  & 0.041  & 1.224  & 0.039  & 1.275  & 0.037  & 1.226  & 0.040  \\       
$\rho_{\rm b}$ (\pjup)& 0.662 & 0.060 & 0.660  & 0.057  & 0.636  & 0.055  & 0.655  & 0.057  & 0.629  & 0.054  & 0.654  & 0.057  \\[2pt]       
Age (Gyr)         & \mc{ } & 4.5 & 2.1 & \erm{1.5}{1.9}{2.0} & \erm{4.5}{1.3}{1.7} & \erm{0.1 }{1.2}{0.1} & \erm{5.3}{1.6}{2.6} \\
\hline \hline \end{tabular}
\end{table*}

The second TEP discovered by the TrES survey \citep{Odonovan+06apj} is of interest because its relatively low orbital inclination (high impact parameter) means that more accurate results can be obtained from a light curve of a given quality. TrES-2 is also unusual in that the star has a subsolar metal abundance ($\FeH = -0.15 \pm 0.10$; \citealt{Sozzetti+07apj}). In Paper\,I the analysis of the $z$-band transit light curve from \citet{Holman+07apj} resulted in measurements of $r_{\rm A}$ and $r_{\rm b}$ to 3\%.

The physical properties of TrES-2 are presented in Table\,\ref{tab:abspar:tres2}, where good agreement is again found between the {\sf Padova}, {\sf Y$^2$} and {\sf Claret} models for ages \reff{around 5\,Gyr}. As before, the {\sf Cambridge\,2007} and {\sf Siess} models predict a smaller radius for a given mass in this mass regime, resulting in overall larger values for the physical properties. The mass--radius relation result differs in the opposite sense, but this time provide a reasonable agreement with the calculations involving theoretical predictions. The agreement between the calculations involving the {\sf Padova}, {\sf Y$^2$} and {\sf Claret} models, and literature results is good (Table\,\ref{apptab:abspar:tres2}).

The error budgets calculated by {\sc jktabsdim} tell a similar story to that of TrES-1. The exceptions are that $K_{\rm A}$ is measured more accurately for TrES-2, and a more precise value for \FeH\ would be the best way to improve knowledge of the properties of the system.

%
%


\subsection{XO-1}

\begin{table*}
\caption{\label{tab:abspar:xo1} Derived physical properties for XO-1.
In every case $g_{\rm b} = 15.8 \pm 1.5$\ms\ and $\rho_{\rm A} = 1.242 \pm 0.080$\psun.}
\begin{tabular}{l r@{\,$\pm$\,}l r@{\,$\pm$\,}l r@{\,$\pm$\,}l r@{\,$\pm$\,}l r@{\,$\pm$\,}l r@{\,$\pm$\,}l}
\hline \hline
                      & \mc{Mass--radius} & \mc{{\sf Padova} models} & \mc{{\sf Siess} models} & \mc{{\sf Y$^2$} models}
                            & \mc{{\sf Cambridge\,2007}}  & \mc{{\sf Claret} models} \\
\hline                                                                                                                                    
$a$         (AU)      & 0.0465& 0.0013 & 0.04909&0.00026 & 0.05002\,$^*$&0.00035 & 0.04929&0.00023 & 0.04909&0.00031 & 0.04990&0.00029\\[2pt]   
$M_{\rm A}$ (\Msun)   & 0.863 & 0.072  & 1.015  & 0.016  & 1.074\,$^*$  & 0.023  & 1.028  & 0.015  & 1.015  & 0.019  & 1.066  & 0.018 \\        
$R_{\rm A}$ (\Rsun)   & 0.886 & 0.037  & 0.935  & 0.025  & 0.953\,$^*$  & 0.027  & 0.939  & 0.024  & 0.935  & 0.025  & 0.950  & 0.022 \\        
$\log g_{\rm A}$ (cgs)& 4.480 & 0.018  & 4.503  & 0.016  & 4.511\,$^*$  & 0.016  & 4.505  & 0.017  & 4.503  & 0.016  & 4.510  & 0.018 \\[2pt]        
$M_{\rm b}$ (\Mjup)   & 0.818 & 0.078  & 0.911  & 0.071  & 0.946\,$^*$  & 0.075  & 0.918  & 0.072  & 0.911  & 0.072  & 0.941  & 0.074 \\        
$R_{\rm b}$ (\Rjup)   & 1.135 & 0.046  & 1.198  & 0.037  & 1.220\,$^*$  & 0.038  & 1.203  & 0.037  & 1.198  & 0.037  & 1.218  & 0.038 \\        
$\rho_{\rm b}$ (\pjup)& 0.559 & 0.068  & 0.530  & 0.063  & 0.520\,$^*$  & 0.062  & 0.528  & 0.063  & 0.530  & 0.063  & 0.521  & 0.062 \\[2pt]        
Age (Gyr)             & \mc{ } & \erm{0.6}{0.1}{0.0} & \erm{0.1\,{^*}\,}{0.2}{0.0} & \erm{1.4}{0.7}{1.3} & \erm{0.3}{0.2}{0.1} & \erm{0.1}{1.3}{0.0} \\
\hline \hline \end{tabular}
\begin{flushleft} $^*$\,These quantities have been calculated using stellar models for stars with small ages. Their values and uncertainties may therefore be unreliable due to edge effects within the grid of model tabulations. \end{flushleft}
\end{table*}

XO-1 was the first TEP discovered by the XO survey \citep{Mccullough+06apj} and contains a star with properties very similar to those of the Sun. The photometric analysis in Paper\,I was performed using the $R$- and $Z$-band light curves presented by \citet{Holman+06apj}. The very low uncertainties in the \Teff\ and \FeH\ quoted by \citet{Mccullough+06apj} have been increased to 50\,K and $\pm$0.05\,dex here.

The properties of the star in XO-1 are very close to those of the Sun, meaning that all the stellar models (which were calibrated on the Sun) agree well. The physical properties of the system are given in Table\,\ref{tab:abspar:xo1}, \reff{and the system seems to be quite young}. A comparison to literature results is given in Table\,\ref{apptab:abspar:xo1} and shows a good agreement.

The error budget indicates that a more precise measurement of $K_{\rm A}$, which is observationally straightforward, would be useful. A improved $r_{\rm A}$ determination would also help.



\subsection{WASP-1}

\begin{table*}
\caption{\label{tab:abspar:wasp1} Derived physical properties for WASP-1.
In each case $g_{\rm b} = \er{10.0}{1.6}{1.2}$\ms\ and $\rho_{\rm A} = \er{0.403}{0.069}{0.037}$\psun.}
\begin{tabular}{l r@{\,$\pm$\,}l r@{\,$\pm$\,}l r@{\,$\pm$\,}l r@{\,$\pm$\,}l r@{\,$\pm$\,}l r@{\,$\pm$\,}l}
\hline \hline
                      & \mc{Mass--radius} & \mc{{\sf Siess} models} & \mc{{\sf Y$^2$} models}
                           & \mc{{\sf Cambridge\,2007}} & \mc{{\sf Claret} models} \\
\hline
$a$         (AU)      & \erm{0.0417}{0.0011}{0.0014}& \erm{0.03955}{0.00040}{0.00044}& \erm{0.03944}{0.00037}{0.00044}& \erm{0.03815}{0.00035}{0.00047}& \erm{0.03933}{0.00031}{0.00032}\\[2pt]
$M_{\rm A}$ (\Msun)   & \erm{1.52}{0.11}{0.12}      & \erm{1.299}{0.040}{0.043}      & \erm{1.288}{0.037}{0.043}      & \erm{1.166}{0.033}{0.043}      & \erm{1.278}{0.031}{0.031}      \\
$R_{\rm A}$ (\Rsun)   & \erm{1.56}{0.08}{0.12}      & \erm{1.477}{0.060}{0.082}      & \erm{1.473}{0.056}{0.086}      & \erm{1.424}{0.054}{0.083}      & \erm{1.469}{0.055}{0.083}      \\
$\log g_{\rm A}$ (cgs)& \erm{4.236}{0.035}{0.023}   & \erm{4.213}{0.044}{0.026}      & \erm{4.212}{0.043}{0.027}      & \erm{4.197}{0.043}{0.027}      & \erm{4.211}{0.044}{0.026}      \\[2pt]
$M_{\rm b}$ (\Mjup)   & \erm{1.02}{0.11}{0.12}      & \erm{0.917}{0.090}{0.090}      & \erm{0.912}{0.089}{0.090}      & \erm{0.853}{0.083}{0.084}      & \erm{0.907}{0.088}{0.088}      \\
$R_{\rm b}$ (\Rjup)   & \erm{1.59}{0.07}{0.11}      & \erm{1.506}{0.060}{0.093}      & \erm{1.502}{0.060}{0.092}      & \erm{1.453}{0.057}{0.089}      & \erm{1.498}{0.059}{0.091}      \\
$\rho_{\rm b}$ (\pjup)& \erm{0.255}{0.058}{0.037}   & \erm{0.268}{0.061}{0.039}      & \erm{0.269}{0.061}{0.039}      & \erm{0.278}{0.063}{0.040}      & \erm{0.270}{0.061}{0.039}      \\[2pt]
Age (Gyr)             & \mc{ } & \erm{2.7}{0.4}{0.6} & \ \ \ \ \ \ 2.8 & 0.5 & \ \ \ \ \ \ 2.9 & 0.4 & \erm{3.1}{0.4}{0.5} \\
\hline \hline \end{tabular} \end{table*}

WASP-1 was discovered by the SuperWASP consortium \citep{Cameron+07mn} and decent light curves have been obtained by \citet{Shporer+07mn} and \citet{Charbonneau+07apj}. This TEP has a high inclination (low impact factor), which means that solutions of its light curve are quite degenerate. In Paper\,I the Shporer and Charbonneau data were analysed but yielded measurements of $r_{\rm A}$ and $r_{\rm b}$ to only 5\% accuracy. The asymmetric errorbars from this analysis have been explicitly carried through the analysis presented here, and the measured physical properties of the system are given in Table\,\ref{tab:abspar:wasp1}.

The {\sf Padova} and {\sf Cambridge\,2000} models do not go to a high enough metal abundance ($Z = 0.03$) so have not been used. Calculations using the {\sf Siess}, {\sf Y$^2$} and {\sf Claret} models agree well, and find ages \reff{in the region of 3\,Gyr}. Physical properties calculated using the mass--radius relation are in poor agreement. Literature results, however, are in accord with the results presented here (Table\,\ref{apptab:abspar:wasp1}).

The error budget indicates that better light {\em and} radial velocity curves are needed to improve measurements of the physical properties of WASP-1. Analyses involving more sets of models with high metal abundances will also be useful.

%
%
%


\subsection{WASP-2}

\begin{table*}
\caption{\label{tab:abspar:wasp2} Derived physical properties for WASP-2.
In every case $g_{\rm b} = 19.7 \pm 2.7$\ms\ and $\rho_{\rm A} = 1.50 \pm 0.21$\psun.}
\begin{tabular}{l r@{\,$\pm$\,}l r@{\,$\pm$\,}l r@{\,$\pm$\,}l r@{\,$\pm$\,}l r@{\,$\pm$\,}l r@{\,$\pm$\,}l r@{\,$\pm$\,}l}
\hline \hline
                      & \mc{Mass--radius} & \mc{{\sf Padova} models} & \mc{{\sf Siess} models} & \mc{{\sf Y$^2$} models}
                            & \mc{{\sf Cambridge\,2007}}  & \mc{{\sf Claret} models} \\
\hline                                                                                                                            
$a$         (AU)      & 0.0301&0.0011 & 0.0309&0.0013 & 0.0321&0.0010 & 0.0310&0.0011 & 0.0312&0.0009 & 0.0312&0.0013 \\[2pt]     
$M_{\rm A}$ (\Msun)   & 0.784 & 0.085 & 0.846 & 0.110 & 0.951 & 0.089 & 0.858 & 0.090 & 0.877 & 0.078 & 0.875 & 0.109 \\          
$R_{\rm A}$ (\Rsun)   & 0.805 & 0.061 & 0.852 & 0.044 & 0.859 & 0.046 & 0.830 & 0.044 & 0.836 & 0.042 & 0.835 & 0.045 \\          
$\log g_{\rm A}$ (cgs)& 4.520 & 0.032 & 4.531 & 0.050 & 4.548 & 0.044 & 4.534 & 0.046 & 4.537 & 0.045 & 4.536 & 0.049 \\[2pt]          
$M_{\rm b}$ (\Mjup)   & 0.841 & 0.085 & 0.885 & 0.100 & 0.956 & 0.094 & 0.893 & 0.089 & 0.906 & 0.084 & 0.905 & 0.099 \\          
$R_{\rm b}$ (\Rjup)   & 1.030 & 0.069 & 1.056 & 0.076 & 1.098 & 0.073 & 1.061 & 0.071 & 1.069 & 0.069 & 1.068 & 0.075 \\          
$\rho_{\rm b}$ (\pjup)& 0.77  & 0.15  & 0.75  & 0.14  & 0.72  & 0.14  & 0.75  & 0.14  & 0.74  & 0.14  & 0.74  & 0.14  \\[2pt]          
\hline \hline \end{tabular} \end{table*}

WASP-2 was discovered by \citet{Cameron+07mn}, and a good $z$-band light curve was obtained by \citet{Charbonneau+07apj} and modelled in Paper\,I. In contrast to WASP-1, the relatively low inclination allows more precise measurements of $r_{\rm A}$ and $r_{\rm b}$. Unfortunately, the stellar \Teff\ measurement is accurate to only $\pm$200\,K, and no metal abundance measurement is available. As ``the abundances do not appear to be substantially different from solar'' \citep{Cameron+07mn}, I have here adopted $\FeH = 0.00 \pm 0.15$ (a slightly different value to TWH08). A detailed spectroscopic analysis is urgently needed.

The physical properties listed in Table\,\ref{tab:abspar:wasp2} (see also Table\,\ref{apptab:abspar:wasp2}) show a generally good agreement both internally (except for the {\sf Siess} or mass--radius approaches) and with literature studies. The values are comparatively imprecise, and improved spectroscopic parameters ($K_{\rm A}$, \Teff, \FeH) and photometry are required to improve this. New data have been obtained but not yet published \citep{Hrudkova+08xxx}. The age of the system is unconstrained by the present data, so WASP-2 could be anywhere between 0 and 12\,Gyr old.

%
%
%
%


\subsection{HAT-P-1}

\begin{table*}
\caption{\label{tab:abspar:hat1} Derived physical properties of the HAT-P-1 system.
In each case $g_{\rm b} = 9.05 \pm 0.66$\ms\ and $\rho_{\rm A} = 0.837 \pm 0.076$\psun.}
\begin{tabular}{l r@{\,$\pm$\,}l r@{\,$\pm$\,}l r@{\,$\pm$\,}l r@{\,$\pm$\,}l r@{\,$\pm$\,}l r@{\,$\pm$\,}l}
\hline \hline
                      & \mc{Mass--radius} & \mc{{\sf Padova} models} & \mc{{\sf Siess} models} & \mc{{\sf Y$^2$} models}
                            & \mc{{\sf Cambridge\,2007}}  & \mc{{\sf Claret} models} \\
\hline                                                                                                                                                   
$a$         (AU)      & 0.05401&0.00149 & 0.05540&0.00038 & 0.05669&0.00033 & 0.05529&0.00043 & 0.05464&0.00050 & 0.05570&0.00049 \\[2pt]    
$M_{\rm A}$ (\Msun)   & 1.054  & 0.087  & 1.137  & 0.023  & 1.218  & 0.021  & 1.130  & 0.026  & 1.091  & 0.030  & 1.156  & 0.030  \\         
$R_{\rm A}$ (\Rsun)   & 1.080  & 0.055  & 1.107  & 0.034  & 1.133  & 0.040  & 1.105  & 0.034  & 1.092  & 0.032  & 1.113  & 0.032  \\         
$\log g_{\rm A}$ (cgs)& 4.394  & 0.022  & 4.405  & 0.026  & 4.415  & 0.024  & 4.404  & 0.026  & 4.399  & 0.027  & 4.408  & 0.027  \\[2pt]         
$M_{\rm b}$ (\Mjup)   & 0.507  & 0.033  & 0.533  & 0.020  & 0.558  & 0.021  & 0.531  & 0.020  & 0.519  & 0.020  & 0.539  & 0.021  \\         
$R_{\rm b}$ (\Rjup)   & 1.179  & 0.049  & 1.209  & 0.039  & 1.237  & 0.040  & 1.207  & 0.039  & 1.192  & 0.039  & 1.216  & 0.040  \\         
$\rho_{\rm b}$ (\pjup)& 0.309  & 0.033  & 0.302  & 0.031  & 0.295  & 0.030  & 0.302  & 0.031  & 0.306  & 0.031  & 0.300  & 0.031  \\[2pt]         
Age (Gyr)             & \mc{ } & \erm{0.7}{1.4}{1.0} & \erm{0.1}{0.4}{0.0} & \erm{2.1}{0.8}{1.0} & \erm{1.1}{1.5}{1.1} & \erm{1.6}{1.1}{1.3} \\
\hline \hline \end{tabular} \end{table*}

This system was the first TEP discovered by the HAT survey \citep{Bakos+07apj} and excellent light curves have been put forward by \citet{Winn+07aj2}. The analysis of these data (Paper\,I) has cemented the position of HAT-P-1 as one of the best-understood TEPs, with fractional component radii measured to 3\%. The stellar properties are also well-known \citep{Bakos+07apj}.

The physical properties of HAT-P-1 in Table\,\ref{tab:abspar:hat1} are all in reasonable agreement with each other; those descending from the {\sf Padova}, {\sf Y$^2$} and {\sf Claret} models in particular are highly compatible. There is also good agreement with literature values (Table\,\ref{apptab:abspar:hat1}). The {\sf Claret} models propose a young age of \er{1.6}{1.1}{1.3}\,Gyr and other models are in agreement. It is noticeable that I find an uncertainty in $M_{\rm A}$ which is substantially smaller than for all literature determinations; in the case of TWH08 this arises from their adoption of even larger uncertainties in \Teff\ and \FeH.

The error budget shows that this is a well-understood TEP. Whilst improvements could be made to the light and velocity curves, other systems would benefit more from a contribution of telescope time.

%
%
%
%
%
%


\subsection{OGLE-TR-10}                                                                     \label{sec:ogle10}

\begin{table*}
\caption{\label{tab:abspar:ogle10} Physical properties of the OGLE-TR-10 system.
In every case $g_{\rm b} = 10.2 \pm 2.7$\ms\ and $\rho_{\rm A} = 0.59 \pm 0.11$\psun.}
\begin{tabular}{l r@{\,$\pm$\,}l r@{\,$\pm$\,}l r@{\,$\pm$\,}l r@{\,$\pm$\,}l r@{\,$\pm$\,}l r@{\,$\pm$\,}l}
\hline \hline
                      & \mc{Mass--radius} & \mc{{\sf Siess} models} & \mc{{\sf Y$^2$} models}
                            & \mc{{\sf Cambridge\,2007}} & \mc{{\sf Claret} models} \\
\hline
$a$         (AU)      & 0.0449 &0.0017 & 0.04464&0.00051 & 0.04476&0.00064 & 0.04278&0.00071 & 0.04471&0.00059 \\[2pt]
$M_{\rm A}$ (\Msun)   & 1.256  & 0.141 & 1.233  & 0.042  & 1.243  & 0.053  & 1.085  & 0.054  & 1.239  & 0.049  \\
$R_{\rm A}$ (\Rsun)   & 1.286  & 0.121 & 1.278  & 0.079  & 1.281  & 0.086  & 1.225  & 0.082  & 1.280  & 0.082  \\
$\log g_{\rm A}$ (cgs)& 4.319  & 0.041 & 4.316  & 0.053  & 4.317  & 0.051  & 4.297  & 0.052  & 4.317  & 0.053  \\[2pt]
$M_{\rm b}$ (\Mjup)   & 0.67   & 0.15  & 0.66   & 0.14   & 0.67   & 0.14   & 0.61   & 0.13   & 0.66   & 0.14   \\
$R_{\rm b}$ (\Rjup)   & 1.28   & 0.11  & 1.27   & 0.10   & 1.27   & 0.10   & 1.22   & 0.10   & 1.27   & 0.10   \\
$\rho_{\rm b}$ (\pjup)& 0.32   & 0.11  & 0.32   & 0.11   & 0.32   & 0.11   & 0.34   & 0.11   & 0.32   & 0.11   \\[2pt]
Age (Gyr)             & \mc{ } & \erm{2.0}{1.2}{1.5} & \ \ \ \ \ \ 2.0 & 1.1 & \erm{2.6}{1.2}{1.1} & \erm{2.0}{1.5}{1.4} \\
\hline \hline \end{tabular} \end{table*}

We now leave the realm of bright and easy-to-study TEPs. The OGLE systems have many interesting features but are much more challenging to observe, due mainly to their faintness and locations in crowded fields. OGLE-TR-10 \citep{Udalski+02aca,Konacki+03apj} is good example of this: the planet is of much interest due to its very low density, but it has not yet been possible to obtain definitive light curves or spectroscopy of the system. In Paper\,I I studied two sets of photometry: Magellan observations from \citet{Holman+07apj2} and VLT data from \citet{Pont+07aa}. The Magellan light curves are marred by a systematic underestimation of the eclipse depth arising from the use of image-subtraction photometry \citep{Pont+07aa}. The VLT data are reliable but unfortunately sparse and cover only half of one transit. Here I adopt the results from Paper\,I for the VLT $V$-band and $R$-band photometry obtained by \citet{Pont+07aa}.

The high metal abundance of OGLE-TR-10 \citep{Santos+06aa} means that the {\sf Cambridge\,2000} and {\sf Padova} models could not be used. Results via the {\sf Siess}, {\sf Y$^2$} and {\sf Claret} models are in good agreement with each other (Table\,\ref{tab:abspar:ogle10}) and with the literature (Table\,\ref{apptab:abspar:ogle10}), but all results have large uncertainties due to the limited observational data available. OGLE-TR-10\,b is one of the lowest-density planets known. The error budget suggests that good light curve are urgently needed, as are more extensive velocity observations and a more precise spectral synthesis study. In light of this, OGLE-TR-10 cannot be allowed onto the list of well-understood planetary systems.



\subsection{OGLE-TR-56}

\begin{table*}
\caption{\label{tab:abspar:ogle56} Physical properties for OGLE-TR-56.
In each case $g_{\rm b} = 22.3 \pm 7.0$\ms\ and $\rho_{\rm A} = 0.62 \pm 0.21$\psun.}
\begin{tabular}{l r@{\,$\pm$\,}l r@{\,$\pm$\,}l r@{\,$\pm$\,}l r@{\,$\pm$\,}l r@{\,$\pm$\,}l r@{\,$\pm$\,}l}
\hline \hline
                       & \mc{Mass--radius} & \mc{{\sf Siess} models} & \mc{{\sf Y$^2$} models}
                             & \mc{{\sf Cambridge\,2007}} & \mc{{\sf Claret} models} \\
\hline
$a$         (AU)       & 0.02381&0.00137 & 0.02407&0.00030 & 0.02390&0.00034 & 0.02303&0.00031 & 0.02395&0.00029\\[2pt]
$M_{\rm A}$ (\Msun)    & 1.225  & 0.212  & 1.266  & 0.047  & 1.238  & 0.053  & 1.108  & 0.045  & 1.247  & 0.045 \\
$R_{\rm A}$ (\Rsun)    & 1.294  & 0.201  & 1.268  & 0.144  & 1.258  & 0.148  & 1.213  & 0.139  & 1.261  & 0.140 \\
$\log g_{\rm A}$ (cgs) & 4.330  & 0.070  & 4.334  & 0.089  & 4.331  & 0.088  & 4.315  & 0.089  & 4.332  & 0.089 \\[2pt]
$M_{\rm b}$ (\Mjup)    & 1.29   & 0.18   & 1.32   & 0.14   & 1.30   & 0.14   & 1.21   & 0.13   & 1.31   & 0.14  \\
$R_{\rm b}$ (\Rjup)    & 1.20   & 0.18   & 1.21   & 0.17   & 1.21   & 0.17   & 1.16   & 0.17   & 1.21   & 0.17  \\
$\rho_{\rm b}$ (\pjup) & 0.75   & 0.35   & 0.74   & 0.34   & 0.74   & 0.35   & 0.77   & 0.36   & 0.74   & 0.35  \\[2pt]
Age (Gyr)              & \mc{ } & \erm{1.2}{1.4}{1.2} & \erm{1.8}{0.9}{1.4} & \erm{2.0}{0.8}{2.0} & \erm{1.6}{1.3}{1.5} \\
\hline \hline \end{tabular} \end{table*}

Similarly to OGLE-TR-10, OGLE-TR-56 \citep{Udalski+02aca3,Konacki+03nat} has only sparse $V$- and $R$-band photometry from the VLT \citep{Pont+07aa}, but in this case the data cover a full transit and therefore yield more useful results. Using the photometric solution from Paper\,I, the physical properties from the mass--radius relation, {\sf Siess}, {\sf Y$^2$} and {\sf Claret} models (Table\,\ref{tab:abspar:ogle56}) agree well with each other and with the independent analysis of TWH08. Published results are collected in Table\,\ref{apptab:abspar:ogle56}.

The error budget shows that OGLE-TR-56 would benefit from further observations of all types, particularly more extensive photometry. However, the faintness of OGLE-TR-56 and the extreme field crowding it suffers from argue in favour of using such telescope time on brighter and less complicated TEPs.

%
%
%
%


\subsection{OGLE-TR-111}

\begin{table*}
\caption{\label{tab:abspar:ogle111} Derived properties of the OGLE-TR-111 system.
In every case $g_{\rm b} = 11.5 \pm 2.5$\ms\ and $\rho_{\rm A} = 1.40 \pm 0.19$\psun.}
\begin{tabular}{l r@{\,$\pm$\,}l r@{\,$\pm$\,}l r@{\,$\pm$\,}l r@{\,$\pm$\,}l r@{\,$\pm$\,}l r@{\,$\pm$\,}l r@{\,$\pm$\,}l}
\hline \hline
                    & \mc{Mass--radius} & \mc{{\sf Padova} models} & \mc{{\sf Siess} models} & \mc{{\sf Y$^2$} models}
                          & \mc{{\sf Cambridge\,2007}} & \mc{{\sf Claret} models} \\
\hline
$a$         (AU)      & 0.04616&0.00162 & 0.04650&0.00112 & 0.04745&0.00088 & 0.04676&0.00067 & 0.04511&0.00077 & 0.04702&0.00073 \\[2pt]
$M_{\rm A}$ (\Msun)   & 0.814  & 0.085  & 0.832  & 0.060  & 0.884  & 0.049  & 0.846  & 0.036  & 0.760  & 0.039  & 0.860  & 0.040  \\
$R_{\rm A}$ (\Rsun)   & 0.835  & 0.061  & 0.842  & 0.035  & 0.859  & 0.038  & 0.846  & 0.037  & 0.816  & 0.043  & 0.851  & 0.032  \\
$\log g_{\rm A}$ (cgs)& 4.505  & 0.031  & 4.508  & 0.045  & 4.517  & 0.042  & 4.510  & 0.041  & 4.495  & 0.037  & 4.513  & 0.044  \\[2pt]
$M_{\rm b}$ (\Mjup)   & 0.532  & 0.103  & 0.540  & 0.100  & 0.562  & 0.103  & 0.546  & 0.099  & 0.508  & 0.093  & 0.552  & 0.101  \\
$R_{\rm b}$ (\Rjup)   & 1.069  & 0.075  & 1.077  & 0.070  & 1.099  & 0.070  & 1.083  & 0.067  & 1.045  & 0.066  & 1.089  & 0.068  \\
$\rho_{\rm b}$ (\pjup)& 0.44   & 0.11   & 0.44   & 0.11   & 0.52   & 0.14   & 0.43   & 0.11   & 0.45   & 0.11   & 0.43   & 0.11   \\[2pt]
Age (Gyr)             & \mc{ } & \erm{10.8}{9.9}{7.9} & \erm{10.0}{3.4}{8.8} & \erm{9.9}{4.2}{4.4} & \erm{14.3}{2.0}{6.2} & \erm{11.5}{5.4}{7.0} \\
\hline \hline \end{tabular} \end{table*}

OGLE-TR-111 was detected in a survey towards the Carina constellation \citep{Udalski+02aca2,Pont+04aa} rather than the Galactic centre, so suffers from less field crowding than OGLEs TR-10 and TR-56. It consequently has better photometry (from \citealt{Winn++07aj}), although its faintness means that $K_{\rm A}$ is not known precisely. In Paper\,I I presented a solution of the \citet{Winn++07aj} light curve, which is used here. The spectral synthesis study of \citet{Santos+06aa} found $\Teff = 5044 \pm 83$\,K and $\FeH = 0.19 \pm 0.07$, in disagreement (3.1$\sigma$) with the $\Teff = 4650 \pm 95$\,K given by \citet{Gallardo+05aa}. I have adopted the former results as these include a determination of \FeH. A new spectral analysis study is needed to investigate this discrepancy and show which results are reliable.

The physical properties of OGLE-TR-111 (Table\,\ref{tab:abspar:ogle111}) from using the {\sf Padova}, {\sf Siess}, {\sf Y$^2$} and {\sf Claret} models agree internally and with literature values (Table\,\ref{apptab:abspar:ogle111}). The {\sf Claret} models yield an age of \er{11.5}{5.4}{7.0}\,Gyr -- the poor precision of this value comes from the very long evolutionary timescales of 0.8\Msun\ stars. If the lower \Teff\ from \citet{Gallardo+05aa} is adopted, the masses and radii of both star and planet decrease by approximately 5\% and the age becomes even larger. Either way, OGLE-TR-111\,b is one of the lowest-density planets known.

Aside from the need for a third spectral synthesis analysis, a more precise value for $K_{\rm A}$ is needed for OGLE-TR-111. Additional photometry would be useful but of a lower priority than new velocity measurements.

%
%


\subsection{OGLE-TR-132}

\begin{table*}
\caption{\label{tab:abspar:ogle132} Derived properties of the OGLE-TR-132 system.
In all cases $g_{\rm b} = 15.6 \pm 6.1$\ms\ and $\rho_{\rm A} = 0.50 \pm 0.15$\psun.}
\begin{tabular}{l r@{\,$\pm$\,}l r@{\,$\pm$\,}l r@{\,$\pm$\,}l r@{\,$\pm$\,}l r@{\,$\pm$\,}l r@{\,$\pm$\,}l r@{\,$\pm$\,}l}
\hline \hline
              & \mc{Mass--radius} & \mc{{\sf Y$^2$} models} & \mc{{\sf Cambridge\,2007} models} & \mc{{\sf Claret} models} \\
\hline
$a$         (AU)      & 0.03081&0.00160 & 0.03050&0.00034 & 0.02879&0.00044 & 0.03040&0.00033 \\[2pt]
$M_{\rm A}$ (\Msun)   & 1.365  & 0.213  & 1.325  & 0.044  & 1.114  & 0.051  & 1.311  & 0.043  \\
$R_{\rm A}$ (\Rsun)   & 1.397  & 0.201  & 1.38   & 0.14   & 1.31   & 0.14   & 1.38   & 0.14   \\
$\log g_{\rm A}$ (cgs)& 4.283  & 0.062  & 4.278  & 0.079  & 4.253  & 0.079  & 4.277  & 0.079  \\[2pt]
$M_{\rm b}$ (\Mjup)   & 1.02   & 0.32   & 1.00   & 0.30   & 0.89   & 0.27   & 1.00   & 0.30   \\
$R_{\rm b}$ (\Rjup)   & 1.28   & 0.17   & 1.26   & 0.15   & 1.19   & 0.15   & 1.26   & 0.15   \\
$\rho_{\rm b}$ (\pjup)& 0.49   & 0.24   & 0.50   & 0.24   & 0.53   & 0.26   & 0.50   & 0.24   \\[2pt]
Age (Gyr)             & \mc{ } & \erm{1.4}{0.6}{1.1} & \erm{1.9}{0.5}{1.3} & \erm{1.5}{0.6}{1.4} \\
\hline \hline \end{tabular} \end{table*}

OGLE-TR-132 was discovered by \cite{Udalski+03aca} and its planetary nature was confirmed by \citet{Bouchy+04aa}. Its high metal abundance ($\FeH = 0.37 \pm 0.07$; \citealt{Gillon+07aa3}) means that only the {\sf Y$^2$}, {\sf Cambridge\,2007} and {\sf Claret} models could be used here. The photometric solution from Paper\,I was based on the VLT light curve of \citet{Gillon+07aa3}.

The physical properties calculated using the {\sf Y$^2$} and {\sf Claret} models (Table\,\ref{tab:abspar:ogle132}) are in good accord and indicate a slightly low stellar surface gravity ($\logg_{\rm A} = 4.277 \pm 0.080$). This low gravity is a bit different to those found from spectral synthesis analyses ($\logg_{\rm A} = 4.86 \pm 0.50$, \citealt{Bouchy+04aa}; $4.51 \pm 0.27$, \citealt{Gillon+07aa3}), indicating that the \Teff s and \FeH s measured by these analyses may be biased. Literature results (Table\,\ref{apptab:abspar:ogle132}) are in agreement with the physical properties presented here.

Aside from the need for a new spectral synthesis analysis, OGLE-TR-132 would also benefit from additional photometry (covering all of the transit) and particularly velocity measurements.

%


\subsection{GJ 436}

\begin{table*}
\caption{\label{tab:abspar:gj436} Derived properties of the GJ\,436 TEP system.
In each case $g_{\rm b} = 13.7 \pm 1.1$\ms\ and $\rho_{\rm A} = 4.92 \pm 0.55$\psun.}
\setlength{\tabcolsep}{2pt}
\begin{tabular}{l r@{\,$\pm$\,}l r@{\,$\pm$\,}l r@{\,$\pm$\,}l r@{\,$\pm$\,}l r@{\,$\pm$\,}l r@{\,$\pm$\,}l}
\hline \hline
                      & \mc{Mass--radius} & \mc{{\sf Padova} models} & \mc{{\sf Siess} models}
                            & \mc{{\sf Y$^2$} models} & \mc{{\sf Claret} models} \\
\hline
$a$         (AU)      & 0.0282 & 0.0013 & 0.02928&0.00070 & 0.02995&0.00042 & 0.02882&0.00066 & 0.02972& 0.0010 \\[2pt]
$M_{\rm A}$ (\Msun)   & 0.429  & 0.060  & 0.479  & 0.035  & 0.513  & 0.022  & 0.457  & 0.031  & 0.501  & 0.053  \\
$R_{\rm A}$ (\Rsun)   & 0.443  & 0.031  & 0.460  & 0.026  & 0.471  & 0.023  & 0.453  & 0.021  & 0.467  & 0.025  \\
$\log g_{\rm A}$ (cgs)& 4.777  & 0.031  & 4.793  & 0.025  & 4.803  & 0.027  & 4.786  & 0.032  & 4.799  & 0.034  \\[2pt]
$M_{\rm b}$ (\Mjup)   & 0.0704 & 0.0069 & 0.0757 & 0.0042 & 0.0793 & 0.0032 & 0.0734 & 0.0039 & 0.0781 & 0.0059 \\
$R_{\rm b}$ (\Rjup)   & 0.357  & 0.022  & 0.371  & 0.017  & 0.379  & 0.015  & 0.365  & 0.016  & 0.376  & 0.019  \\
$\rho_{\rm b}$ (\pjup)& 1.54   & 0.20   & 1.49   & 0.18   & 1.45   & 0.17   & 1.51   & 0.18   & 1.47   & 0.18   \\
\hline \hline \end{tabular} \end{table*}

GJ\,436 is the most important of the known TEPs as it contains the smallest and least massive planet {\em and} star. The eccentric orbit of the system is also a surprise as tidal effects are expected to have circularised a binary system with such a short orbital period. This and additional phenomena mean that GJ\,436 is a candidate for a multiple-planet system \citep{Ribas++08apj}, but the possible properties of a putative third planet are the subject of intense discussion \citep{BeanSeifahrt08aa,Alonso+08aa,Ribas+08conf2}. In the current analysis I followed Paper\,I in adopting an orbital eccentricity of $e = 0.14 \pm 0.01$ \citep{Demory+07aa}.

The main limitation in our understanding of GJ\,436 is our knowledge of the stellar \Teff\ and \FeH, quantities which are notoriously difficult to measure for M dwarfs \citep[e.g.][]{Bonfils+05aa}. Adopting the photometric solution from Paper\,I, I find good agreement between the {\sf Padova}, {\sf Siess} and {\sf Claret} models for the physical properties of the TEP (Table\,\ref{tab:abspar:gj436}). Using the {\sf Y$^2$} models gives a lower $M_{\rm A}$ and thus $R_{\rm A}$, $M_{\rm b}$ and $R_{\rm b}$, due to the slightly larger stellar radius for a 0.5\Msun\ star in these models. The {\sf Y$^2$} results are in decent agreement with literature studies (Table\,\ref{apptab:abspar:gj436}), most of which used these models in their analysis. \citet{Torres07apj} also found a more massive star in one of his solutions, but discounted it due to a poorer agreement with the {\it Hipparcos} parallax of GJ\,436. Given this, the physical properties found using the empirical mass--radius relation may be closest to the truth, as they return a low mass of $M_{\rm A} = 0.429$\Msun. However, I retain the results using the {\sf Claret} models for consistency with the rest of the analysis in this work. The age is unconstrained because 0.5\Msun\ stars exhibit negligible evolutionary effects in a Hubble time.

Further photometry of GJ\,436 has become available since publication of Paper\,I, so revised properties of GJ\,436 will be presented in the future. There seems no need to obtain further light or velocity curves, but improvements are required to our understanding of the spectral characteristics of M dwarfs and their dependence on \Teff\ and \FeH.

%


\subsection{HD 149026}

\begin{table*}
\caption{\label{tab:abspar:149026} Physical properties of the HD\,149026 system.
In all cases $g_{\rm b} = \er{23.7}{6.8}{6.2}$\ms\ and $\rho_{\rm A} = \er{0.59}{0.08}{0.13} $\psun.}
\begin{tabular}{l r@{\,$\pm$\,}l r@{\,$\pm$\,}l r@{\,$\pm$\,}l r@{\,$\pm$\,}l r@{\,$\pm$\,}l}
\hline \hline
  & \mc{Mass--radius} & \mc{{\sf Y$^2$} models} & \mc{{\sf Cambridge\,2007} models} & \mc{{\sf Claret} models} \\
\hline
$a$         (AU)      & \erm{0.0427}{0.0020}{0.0013} & \erm{0.04303}{0.00034}{0.00026} & \erm{0.04060}{0.0050}{0.0032} & \erm{0.04294}{0.00037}{0.00021} \\[2pt]
$M_{\rm A}$ (\Msun)   & \erm{1.26}{0.19}{0.11}       & \erm{1.285}{0.031}{0.023}       & \erm{1.079}{0.040}{0.026}     & \erm{1.277}{0.033}{0.019}       \\
$R_{\rm A}$ (\Rsun)   & \erm{1.28}{0.17}{0.09}       & \erm{1.295}{0.121}{0.058}       & \erm{1.221}{0.116}{0.057}     & \erm{1.292}{0.12}{0.054}        \\
$\log g_{\rm A}$ (cgs)& \erm{4.319}{0.030}{0.054}    & \erm{4.322}{0.037}{0.068}       & \erm{4.297}{0.037}{0.068}     & \erm{4.321}{0.039}{0.069}       \\[2pt]
$M_{\rm b}$ (\Mjup)   & \erm{0.353}{0.035}{0.024}    & \erm{0.358}{0.011}{0.011}       & \erm{0.319}{0.012}{0.010}     & \erm{0.357}{0.012}{0.011}       \\
$R_{\rm b}$ (\Rjup)   & \erm{0.608}{0.110}{0.047}    & \erm{0.612}{0.099}{0.072}       & \erm{0.578}{0.094}{0.068}     & \erm{0.611}{0.099}{0.072}       \\
$\rho_{\rm b}$ (\pjup)& \erm{1.57}{0.71}{0.58}       & \erm{1.56}{0.71}{0.57}          & \erm{1.65}{0.76}{0.60}        & \erm{1.56}{0.71}{0.57}          \\[2pt]
Age (Gyr)             & \mc{ } & \erm{1.2}{0.8}{0.6} & \ \ \ \ \ \ \ \ \ \ \ \ 1.2 & 1.0 & \erm{1.9}{0.7}{0.4} \\
\hline \hline \end{tabular} \end{table*}

HD\,149026\,b \citep{Sato+05apj} is an anomalous TEP because the density of the planet is much larger than expected for its mass, suggesting that much of its matter consists of a rock/ice core. Its brightness temperature, measured through occultation observations at 8\,$\mu$m, is also much greater than we would expect it to be \citep{Harrington+07nat}. The very shallow transit exhibited by HD\,149026 means its photometric parameters are not well known -- in Paper\,I I found an inclination of $i \approx 90^\circ$ which is quite different to other determinations. This solution, which has since been supported by \citet{Winn+08apj}, leads to a smaller star and planet compared to previous measurements. I adopt this solution here.

The physical properties of HD\,149026 from the {\sf Y$^2$} and {\sf Claret} models (Table\,\ref{tab:abspar:149026}) are quite different to all previous studies expect that of TWH08, who also found a high-inclination light curve solution. A comparison with the results of \citet{Sato+05apj,Charbonneau+06apj,Wolf+07apj,Winn+08apj} and TWH08 is made in Table\,\ref{apptab:abspar:149026}. I consequently find a planetary density of $\rho_{\rm b} = \er{1.56}{0.71}{0.57}$\,$\rho_{\rm Jup}$ which is the highest of all published determinations and exacerbates the anomaly mentioned above. However, this result should be treated with caution until improved light curves are included in the analysis. The other types of data for this system are all of definitive quality.

%
%


\subsection{HD 189733}

\begin{table*}
\caption{\label{tab:abspar:189733} Derived physical properties of HD\,189733.
In every case $g_{\rm b} = 22.0 \pm 1.4$\ms\ and $\rho_{\rm A} = 1.98 \pm 0.16$\psun.}
\begin{tabular}{l r@{\,$\pm$\,}l r@{\,$\pm$\,}l r@{\,$\pm$\,}l r@{\,$\pm$\,}l r@{\,$\pm$\,}l r@{\,$\pm$\,}l}
\hline \hline
                    & \mc{Mass--radius} & \mc{{\sf Padova} models} & \mc{{\sf Siess} models} & \mc{{\sf Y$^2$} models}
                          & \mc{{\sf Cambridge\,2007}} & \mc{{\sf Claret} models} \\
\hline
$a$         (AU)      & 0.02932&0.00097 & 0.03129&0.00019 & 0.03189$^*$&0.00024 & 0.03122&0.00036 & 0.03144&0.00021 & 0.03175$^*$&0.00019 \\[2pt]     
$M_{\rm A}$ (\Msun)   & 0.682 & 0.067   & 0.830 & 0.016   & 0.878$^*$  & 0.020  & 0.824  & 0.029  & 0.841  & 0.017  & 0.866$^*$  & 0.015  \\          
$R_{\rm A}$ (\Rsun)   & 0.701 & 0.036   & 0.749 & 0.024   & 0.763$^*$  & 0.027  & 0.747  & 0.022  & 0.752  & 0.025  & 0.760$^*$  & 0.022  \\          
$\log g_{\rm A}$ (cgs)& 4.580 & 0.022   & 4.608 & 0.023   & 4.616$^*$  & 0.021  & 4.607  & 0.025  & 4.610  & 0.022  & 4.615$^*$  & 0.024  \\[2pt]          
$M_{\rm b}$ (\Mjup)   & 1.023 & 0.074   & 1.165 & 0.037   & 1.210$^*$  & 0.040  & 1.159  & 0.043  & 1.176  & 0.034  & 1.199$^*$  & 0.038  \\          
$R_{\rm b}$ (\Rjup)   & 1.074 & 0.047   & 1.146 & 0.034   & 1.168$^*$  & 0.035  & 1.143  & 0.035  & 1.151  & 0.034  & 1.163$^*$  & 0.034  \\          
$\rho_{\rm b}$ (\pjup)& 0.827 & 0.080   & 0.774 & 0.071   & 0.760$^*$  & 0.069  & 0.776  & 0.071  & 0.771  & 0.070  & 0.763$^*$  & 0.069  \\[2pt]          
Age (Gyr)             & \mc{ } & \erm{0.1}{2.9}{0.0} & \erm{0.1\,{^*}\,}{0.0}{0.0} & \erm{3.7}{4.2}{4.1} & \erm{0.1\,{^*}\,}{0.0}{0.0} & \erm{0.1}{3.3}{0.0} \\
\hline \hline \end{tabular}
\begin{flushleft} $^*$\,These quantities have been calculated using stellar models for stars with small ages. Their values and uncertainties may therefore be unreliable due to edge effects within the grid of model tabulations. \end{flushleft}
\end{table*}

As the joint-brightest of the TEPs with HD\,209458, HD\,189733 \citep{Bouchy+05aa} has been very well studied. The analysis in Paper\,I was based on definitive light curves from \citet{Bakos+06apj2} and \citet{Winn+06apj}, and gives fractional radii to uncertainties of only 3\%. A note of caution is needed here: the solutions of the different passbands did not agree very well, with the scatter being worst for the ratio of the component radii (6.7\hspace{1pt}$\sigma$). This is either due to systematic errors in the light curves which are not directly detectable, or to the presence of starspots. This discrepancy is taken into account in the errorbars quoted in Paper\,I.

HD\,189733 appears to be a very young system \reff{(0--3\,Gyr)}, and the physical properties presented in Table\,\ref{tab:abspar:189733} are in some cases affected by the lack of availability of theoretical stellar models for negative ages. Only the {\sf Y$^2$} models predict an age greater than the minimum allowed value, so in this respect they are slightly discrepant. If the \Teff\ is lowered by 1$\sigma$ this difficulty disappears, indicating that it has not caused a major problem in the resulting properties. The uncertainties on individual sets of physical properties are definitely affected by these edge effects, but this does not wrap into the final values (see Section\,\ref{sec:analysis}) as the largest uncertainties are always adopted (in this case from the {\sf Y$^2$} model solutions). Given these problems, it is reassuring that all studies have found similar values for the physical properties of the HD\,189733 system (Table\,\ref{apptab:abspar:189733}; \citealt{Bouchy+05aa,Bakos+06apj2,Winn+06apj,Winn++07apj,Torres++08apj}. As an additional check, the interferometrically-measured stellar radius of $R_{\rm A} = 0.779 \pm 0.052$\Rsun\ given by \citet{Baines+07apj} is in good agreement with the results presented here, with the additional bonus that it is almost independent of stellar theory. It may be worthwhile to revisit the \Teff\ determination for HD\,189733, as a slightly lower temperature would obliterate the difficulties found above.

%
%


\subsection{HD 209458}

\begin{table*}
\caption{\label{tab:abspar:209458} Derived properties of the HD\,209458 system.
In all cases $g_{\rm b} = 9.08 \pm 0.17$\ms\ and $\rho_{\rm A} = 0.727 \pm 0.005$\psun.}
\begin{tabular}{l r@{\,$\pm$\,}l r@{\,$\pm$\,}l r@{\,$\pm$\,}l r@{\,$\pm$\,}l r@{\,$\pm$\,}l r@{\,$\pm$\,}l r@{\,$\pm$\,}l r@{\,$\pm$\,}l r@{\,$\pm$\,}l r@{\,$\pm$\,}l r@{\,$\pm$\,}l r@{\,$\pm$\,}l r@{\,$\pm$\,}l}
\hline \hline
                    & \mc{Mass--radius} & \mc{{\sf Padova} models} & \mc{{\sf Siess} models} & \mc{{\sf Y$^2$} models}
                          & \mc{{\sf Cambridge\,2007}} & \mc{{\sf Claret} models} \\
\hline                                                                                                                                       
$a$         (AU)       & 0.04724&0.00105 & 0.04742&0.00045 & 0.04888&0.00032 & 0.04757&0.00034 & 0.04799&0.00029 & 0.04770&0.00041 \\[2pt]   
$M_{\rm A}$ (\Msun)    & 1.132  & 0.076  & 1.144  & 0.033  & 1.253  & 0.025  & 1.155  & 0.025  & 1.186  & 0.021  & 1.165  & 0.030  \\        
$R_{\rm A}$ (\Rsun)    & 1.159  & 0.026  & 1.163  & 0.011  & 1.199  & 0.008  & 1.167  & 0.009  & 1.177  & 0.008  & 1.170  & 0.010  \\        
$\log g_{\rm A}$ (cgs) & 4.364  & 0.010  & 4.365  & 0.005  & 4.378  & 0.003  & 4.367  & 0.004  & 4.370  & 0.003  & 4.368  & 0.004  \\[2pt]        
$M_{\rm b}$ (\Mjup)    & 0.693  & 0.032  & 0.698  & 0.016  & 0.742  & 0.013  & 0.703  & 0.013  & 0.715  & 0.012  & 0.707  & 0.015  \\        
$R_{\rm b}$ (\Rjup)    & 1.376  & 0.032  & 1.381  & 0.016  & 1.424  & 0.014  & 1.386  & 0.014  & 1.398  & 0.013  & 1.389  & 0.015  \\        
$\rho_{\rm b}$ (\pjup) & 0.266  & 0.009  & 0.265  & 0.007  & 0.257  & 0.007  & 0.264  & 0.007  & 0.262  & 0.007  & 0.263  & 0.007  \\[2pt]        
Age (Gyr)              & \mc{ } & \erm{1.7}{0.9}{0.7} & \erm{0.2}{0.6}{0.1} & \ \ \ \ \ \ 2.3 & 0.5 & \erm{0.5}{0.6}{0.3} & \erm{2.3}{0.7}{0.6} \\
\hline \hline \end{tabular} \end{table*}

In contrast to its position as the last object studied in this work, HD\,209458 was the first TEP to be discovered \citep{Charbonneau+00apj,Henry+00apj}. This, combined with its bright apparent magnitude, means that the available light curves of HD\,209458 form a catalogue of observations of remarkable quality. In Paper\,I I analysed the original HST observations from \citet{Brown+01apj}, the stunning ten-band HST data from \citet{Knutson+07apj}, and the high-duty-cycle MOST light curve from \citet{Rowe+06apj}. These datasets allowed determination of the fractional radii to accuracies of better than 1\%, a level which has not been achieved for any other TEP. As with HD\,189733, the high quality of these data allowed the detection of differences between the individual light curves (at the 5.6\hspace{1pt}$\sigma$ level here) which are due either to systematic errors or starspots. These discrepancies are accounted for in the errorbars in Paper\,I.

The results from Paper\,I have been combined with the $K_{\rm A}$ measured by \citet{Naef+04aa}. There are several spectral synthesis studies in the literature \citep{AllendeLambert99aa,Mazeh+00apj,Santos++04aa} and these are in good agreement with each other and with semiempirical results based on infrared photometry \citep{Ribas+03aa}. I adopt $\Teff = 6117 \pm 50$\,K and $\FeH = 0.02 \pm 0.05$ from \citet{Santos++04aa}; this study also finds $\logg_{\rm A} = 4.48 \pm 0.08$. As independent checks, $uvby\beta$ photometry and the calibration grids of \citet{MoonDworetsky85mn} give $\Teff = 6080 \pm 30$\,K and $\logg_{\rm A} = 4.26 \pm 0.06$, and a preliminary analysis using the Infrared Flux Method \citep{BlackwellShallis77mn,Blackwell++80aa} yields $\Teff = 6180 \pm 220$\,K (Dr.\ B.\ Smalley, private communication).

The resulting physical properties (Table\,\ref{tab:abspar:209458}) are very well constrained, and in good agreement with most literature studies (Table\,\ref{apptab:abspar:209458}).

A detailed error budget for the calculations using the {\sf Claret} models is given in Table\,\ref{tab:abspar:209458:errmod}, and shows that most of the uncertainty in the physical parameters comes from the uncertainties in the \Teff\ and \FeH\ measurements. As these are realistically limited to minimum values of $\pm$50\,K and $\pm$0.05\,dex, respectively, due to our understanding of the temperature scale of low-mass stars, there is no immediate hope of improvement. The planetary properties are still mainly dependent on the input $K_{\rm A}$ and $r_{\rm b}$ -- both of these values are already known to very high precision and would require much work to improve. The analysis here therefore results in quantities which are highly reliable and unlikely to change in the near future. HD\,209458 is the best-understood extrasolar planetary system. The error budget for calculations using the empirical mass--radius relation (Table\,\ref{tab:abspar:209458:errMR}) confirms these conclusions.

%
%
%

\begin{table}
\caption{\label{tab:abspar:209458:errmod} Detailed error budget for the calculation
of the system properties of HD\,209458 using the {\sf Claret} (2007) stellar models.
The layout of the table is the same as that for Table\,\ref{tab:abspar:tres1:errmod}.}
\begin{tabular}{l|rrrrrrr}
\hline \hline
Output            & \multicolumn{6}{c}{Input parameter} \\
parameter         & $K_{\rm A}$ & $i$   & $r_{\rm A}$ & $r_{\rm b}$ & \Teff & \FeH  \\
\hline
Age               &             &       &    0.074    &             & 0.738 & 0.664 \\
$a$               &             &       &    0.038    &             & 0.685 & 0.720 \\[2pt]
$M_{\rm A}$       &             &       &    0.038    &             & 0.686 & 0.720 \\
$R_{\rm A}$       &             &       &    0.228    &             & 0.666 & 0.701 \\
$\log g_{\rm A}$  &             &       &    0.452    &             & 0.610 & 0.642 \\
$\rho_{\rm A}$    &    0.001    &       &    1.000    &             & 0.001 & 0.001 \\
$M_{\rm b}$       &    0.568    & 0.004 &    0.031    &             & 0.564 & 0.592 \\[2pt]
$R_{\rm b}$       &             &       &    0.029    &    0.645    & 0.523 & 0.550 \\
$g_{\rm b}$       &    0.633    & 0.004 &             &    0.774    &       &       \\
$\rho_{\rm b}$    &    0.452    & 0.003 &    0.012    &    0.829    & 0.224 & 0.236 \\
\hline \hline
\end{tabular} \end{table}

\begin{table}
\caption{\label{tab:abspar:209458:errMR} Detailed error budget for the calculation
of the system properties of HD\,209458 using the empirical mass--radius relation.
The layout of the table is the same as for Table\,\ref{tab:abspar:tres1:errMR}.}
\begin{tabular}{l|rrrrrrr}
\hline \hline
Output            & \multicolumn{6}{c}{Input parameter} \\
parameter         & $K_{\rm A}$ & $i$   & $r_{\rm A}$ & $r_{\rm b}$ & MR0   & MR1   \\
\hline
$a$               &             &       &    0.048    &             & 0.666 & 0.744 \\[2pt]
$M_{\rm A}$       &             &       &    0.048    &             & 0.665 & 0.745 \\
$R_{\rm A}$       &             &       &    0.141    &             & 0.660 & 0.738 \\
$\log g_{\rm A}$  &             &       &    0.140    &             & 0.660 & 0.738 \\
$\rho_{\rm A}$    &    0.001    &       &    1.000    &             & 0.001 & 0.002 \\
$M_{\rm b}$       &    0.255    & 0.002 &    0.046    &             & 0.643 & 0.720 \\[2pt]
$R_{\rm b}$       &             &       &    0.045    &    0.307    & 0.634 & 0.709 \\
$g_{\rm b}$       &    0.633    & 0.004 &             &    0.774    &       &       \\
$\rho_{\rm b}$    &    0.354    & 0.002 &    0.032    &    0.650    & 0.448 & 0.500 \\
\hline \hline
\end{tabular} \end{table}


\section{Physical properties of the transiting extrasolar planetary systems}                         \label{sec:absdim}

\begin{table*}
\caption{\label{tab:abspar:stars} Physical properties of the stellar components of the TEPs
studied in this work. For each quantity the first uncertainty is derived from a propagation
of all observational errors and the second uncertainty is an estimate of the systematic
errors arising from the variation between results using different sets of stellar models.}
\begin{tabular}{l l@{\,$\pm$\,}l@{\,$\pm$\,}l l@{\,$\pm$\,}l@{\,$\pm$\,}l l@{\,$\pm$\,}l@{\,$\pm$\,}l l@{\,$\pm$\,}l l}
\hline \hline
System & \multicolumn{3}{c}{Mass (\Msun)} & \multicolumn{3}{c}{Radius (\Rsun)} & \multicolumn{3}{c}{$\log g_{\rm A}$ [cm/s]} & \multicolumn{2}{c}{Density (\psun)} & Age (Gyr) \\
\hline
TrES-1      & 0.929 & 0.030 & 0.037 & 0.829 & 0.015 & 0.011 & 4.569 & 0.018 & 0.006 & 1.632 & 0.092 & \er{1.2}{4.1}{1.8} $\pm$ 2.2 \\
TrES-2      & 0.991 & 0.050 & 0.025 & 0.994 & 0.031 & 0.009 & 4.440 & 0.028 & 0.004 & 1.010 & 0.092 & \er{5.3}{1.6}{2.6} $\pm$ 0.8 \\
XO-1        & 1.066 & 0.018 & 0.051 & 0.950 & 0.025 & 0.015 & 4.510 & 0.018 & 0.007 & 1.242 & 0.080 & \er{1.0}{1.3}{0.9} $\pm$ 1.5 \\
WASP-1      & \multicolumn{3}{c}{1.278 $^{+0.040}_{-0.043}$ $^{+0.000}_{-0.021}$} &
              \multicolumn{3}{c}{1.469 $^{+0.060}_{-0.086}$ $^{+0.000}_{-0.008}$} &
              \multicolumn{3}{c}{4.211 $^{+0.044}_{-0.027}$ $^{+0.000}_{-0.002}$} &
              \multicolumn{2}{c}{0.403 $^{+0.069}_{-0.037}$} & \er{3.1}{0.4}{0.5} $\pm$ 0.3 \\
WASP-2      & 0.88  & 0.11  & 0.03  & 0.835 & 0.045 & 0.017 & 4.536 & 0.050 & 0.005 & 1.50  & 0.21  & unconstrained \\
HAT-P-1     & 1.156 & 0.030 & 0.026 & 1.113 & 0.034 & 0.008 & 4.408 & 0.027 & 0.004 & 0.837 & 0.076 & \er{1.6}{1.1}{1.3} $\pm$ 0.9 \\
OGLE-TR-10  & 1.239 & 0.053 & 0.006 & 1.280 & 0.086 & 0.002 & 4.317 & 0.053 & 0.001 & 0.590 & 0.110 & \er{2.0}{1.5}{1.4} $\pm$ 0.0 \\
OGLE-TR-56  & 1.247 & 0.053 & 0.019 & 1.26  & 0.15  & 0.01  & 4.332 & 0.089 & 0.002 & 0.620 & 0.210 & \er{1.6}{1.3}{1.5} $\pm$ 0.2 \\
OGLE-TR-111 & 0.860 & 0.060 & 0.028 & 0.851 & 0.037 & 0.009 & 4.513 & 0.045 & 0.005 & 1.400 & 0.190 & \er{11.5}{5.4}{7.0} $\pm$ 1.6 \\
OGLE-TR-132 & 1.311 & 0.044 & 0.014 & 1.38  & 0.14  & 0.00  & 4.277 & 0.079 & 0.001 & 0.50  & 0.15  & \er{1.5}{0.6}{1.4} $\pm$ 0.1 \\
GJ 436      & 0.501 & 0.053 & 0.044 & 0.467 & 0.026 & 0.014 & 4.799 & 0.034 & 0.013 & 4.92  & 0.55  & unconstrained \\
HD 149026   & \multicolumn{3}{c}{1.277 $^{+0.033}_{-0.023}$ $^{+0.000}_{-0.008}$} &
              \multicolumn{3}{c}{1.292 $^{+0.121}_{-0.058}$ $^{+0.000}_{-0.003}$} &
              \multicolumn{3}{c}{4.321 $^{+0.039}_{-0.069}$ $^{+0.000}_{-0.001}$} &
              \multicolumn{2}{c}{0.592 $^{+0.083}_{-0.129}$} & 1.2 $\pm$ 1.0 $\pm$ 0.0 \\
HD 189733   & 0.866 & 0.029 & 0.042 & 0.760 & 0.024 & 0.013 & 4.615 & 0.025 & 0.008 & 1.980 & 0.170 & 0.0--3.4 $\pm$ 3.6 \\
HD 209458   & 1.165 & 0.033 & 0.021 & 1.170 & 0.011 & 0.007 & 4.368 & 0.005 & 0.003 & 0.727 & 0.005 & \er{2.3}{0.7}{0.6} $\pm$ 0.6 \\
\hline \hline \end{tabular} \end{table*}

\begin{table*}
\caption{\label{tab:abspar:planets} Physical properties of the planetary components of the TEPs
studied in this work. For each quantity the first uncertainty is derived from a propagation
of all observational errors and the second uncertainty is an estimate of the systematic
errors arising from the variation between results using different sets of stellar models.}
\begin{tabular}{l l@{\,$\pm$\,}l@{\,$\pm$\,}l l@{\,$\pm$\,}l@{\,$\pm$\,}l l@{\,$\pm$\,}l@{\,$\pm$\,}l l@{\,$\pm$\,}l l@{\,$\pm$\,}l@{\,$\pm$\,}l}
\hline \hline
System & \multicolumn{3}{c}{Semimajor axis (AU)} & \multicolumn{3}{c}{Mass (\Mjup)} & \multicolumn{3}{c}{Radius (\Rjup)} & \multicolumn{2}{c}{Gravity (\ms)} & \multicolumn{3}{c}{Density (\pjup)} \\
\hline
TrES-1\,b      & 0.04000 & 0.00043 & 0.00054 & 0.781 & 0.045 & 0.021 &  1.114 & 0.032 & 0.015 & \ \ \ \ 15.6  & 1.2  & 0.565 & 0.055 & 0.008 \\
TrES-2\,b      & 0.03568 & 0.00061 & 0.00031 & 1.206 & 0.045 & 0.020 &  1.226 & 0.040 & 0.010 & \ \ \ \ 19.9  & 1.2  & 0.654 & 0.057 & 0.006 \\
XO-1\,b        & 0.04990 & 0.00029 & 0.00081 & 0.941 & 0.074 & 0.030 &  1.218 & 0.038 & 0.020 & \ \ \ \ 15.8  & 1.5  & 0.521 & 0.063 & 0.009 \\
WASP-1\,b      & \multicolumn{3}{c}{0.03933 $^{+0.00040}_{-0.00044}$ $^{+0.00000}_{-0.00022}$} &
                 \multicolumn{3}{c}{0.907 $^{+0.090}_{-0.090}$ $^{+0.000}_{-0.010}$} &
                 \multicolumn{3}{c}{1.498 $^{+0.060}_{-0.093}$ $^{+0.000}_{-0.008}$} &
                 \multicolumn{2}{c}{\ \ \ \ 10.0 $^{+1.6}_{-1.2}$} &
                 \multicolumn{3}{c}{0.270 $^{+0.061}_{-0.039}$ $^{+0.002}_{-0.000}$} \\
WASP-2\,b      & 0.03120 & 0.00130 & 0.00030 & 0.91  & 0.10  & 0.02  &  1.068 & 0.076 & 0.012 & \ \ \ \ 19.7  & 2.7  & 0.74  & 0.14  & 0.01  \\
HAT-P-1\,b     & 0.05570 & 0.00049 & 0.00041 & 0.539 & 0.021 & 0.008 &  1.216 & 0.040 & 0.009 & \ \ \ \  9.05 & 0.66 & 0.300 & 0.031 & 0.002 \\
OGLE-TR-10\,b  & 0.04471 & 0.00064 & 0.00007 & 0.66  & 0.14  & 0.01  &  1.27  & 0.10  & 0.00  & \ \ \ \ 10.1  & 2.7  & 0.32  & 0.11  & 0.00  \\
OGLE-TR-56\,b  & 0.02395 & 0.00034 & 0.00012 & 1.31  & 0.14  & 0.01  &  1.21  & 0.17  & 0.00  & \ \ \ \ 22.3  & 7.0  & 0.74  & 0.35  & 0.00  \\
OGLE-TR-111\,b & 0.0470  & 0.0011  & 0.0005  & 0.55  & 0.10  & 0.01  &  1.089 & 0.070 & 0.012 & \ \ \ \ 11.5  & 2.5  & 0.43  & 0.11  & 0.01  \\
OGLE-TR-132\,b & 0.03040 & 0.00034 & 0.00010 & 1.00  & 0.30  & 0.00  &  1.26  & 0.15  & 0.00  & \ \ \ \ 15.6  & 6.1  & 0.50  & 0.24  & 0.00  \\
GJ\,436\,b     & 0.0297  & 0.0010  & 0.0009  & 0.078 & 0.005 & 0.004 &  0.376 & 0.019 & 0.011 & \ \ \ \ 13.7  & 1.1  & 1.47  & 0.18  & 0.04  \\
HD\,149026\,b  & \multicolumn{3}{c}{0.04294 $^{+0.00037}_{-0.00026}$ $^{+0.00000}_{-0.00009}$} &
                 \multicolumn{3}{c}{0.357 $^{+0.012}_{-0.011}$ $^{+0.000}_{-0.001}$} &
                 \multicolumn{3}{c}{0.611 $^{+0.099}_{-0.072}$ $^{+0.000}_{-0.001}$} &
                 \multicolumn{2}{c}{23.7 $^{+6.8}_{-6.2}$} &
                 \multicolumn{3}{c}{1.56 $^{+0.71}_{-0.57}$ $^{+0.00}_{-0.00}$} \\
HD\,189733\,b  & 0.03175 & 0.00036 & 0.00053 & 1.199 & 0.043 & 0.040 &  1.163 & 0.035 & 0.020 & \ \ \ \ 22.0  & 1.4  & 0.763 & 0.071 & 0.013 \\
HD\,209458\,b  & 0.04770 & 0.00045 & 0.00028 & 0.707 & 0.016 & 0.009 &  1.389 & 0.016 & 0.008 & \ \ \ \  9.08 & 0.17 & 0.263 & 0.007 & 0.002 \\
\hline \hline \end{tabular} \end{table*}

\subsection{Using stellar models}

The final model-dependent physical properties of the TEPs studied in this work are collected in Tables \ref{tab:abspar:stars} and \ref{tab:abspar:planets}. The values of the properties are those from the calculations involving the {\sf Claret} theoretical models. Each quantity has a quoted statistical uncertainty, which arises from the propagation of errors in all input parameters ($\Porb$, $i$, $r_{\rm A}$, $r_{\rm b}$, $K_{\rm A}$, \Teff\ and \FeH). For the statistical uncertainty of each quantity I quote (slightly conservatively) the largest of the uncertainties found from calculations involving the three model sets which show good interagreement ({\sf Claret}, {\sf Padova} and {\sf Y$^2$}).

Every calculated physical property also has a quoted systematic uncertainty, which comes from the variation in physical properties between calculations using these three sets of theoretical models. These systematic uncertainties are lower limits on the true systematics, because the different model sets have many similarities in their calculation methods (see Section\,\ref{sec:method:model:which}. The systematic uncertainties depend on calculations involving only three sets of stellar models; additional model sets will be added in the future to improve the precision. For four systems (OGLE-TR-10, OGLE-TR-56, OGLE-TR-132 and HD\,149026) the {\sf Padova} models did not extend to a high enough \FeH\ so the systematic uncertainty estimates are based on only the {\sf Y$^2$} and {\sf Claret} models. Systematic uncertainties are not quoted for $g_{\rm b}$ or $\rho_{\rm A}$ as these quantities have, respectively, zero and negligible dependence on theoretical calculations.

\reff{In most cases, the lower limits on the systematic errors are quite a bit smaller than the statistical ones.} This is the situation for all the planetary properties ($M_{\rm b}$, $R_{\rm b}$, $g_{\rm b}$, $\rho_{\rm b}$), which means that they should be reliably determinable observationally. The two quantities for which model-dependent systematics are important are the orbital semimajor axis and stellar mass: for these the two types of errors are generally of a similar size. Thus $a$ and $M_{\rm A}$ are the two quantities which are least reliably determined.

\begin{figure} \includegraphics[width=0.48\textwidth,angle=0]{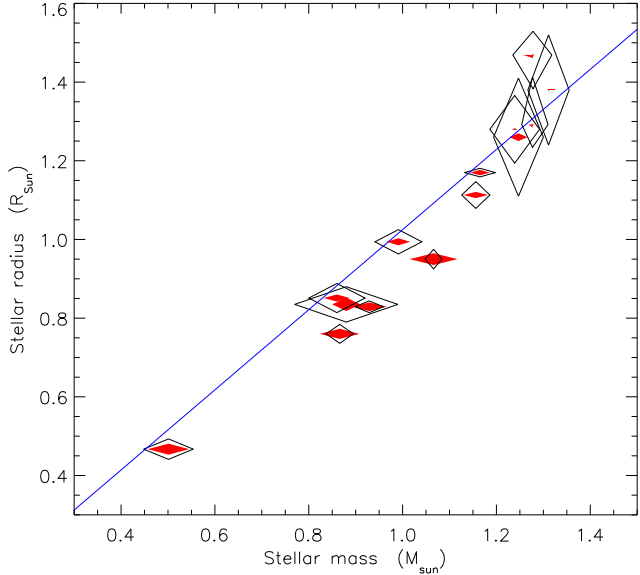}
\caption{\label{fig:absdim:M1R1} Plot of the masses versus the radii of the
stars in the fourteen TEPs studied in this work. The statistical uncertainties
are shown by black open diamonds and the systematic uncertainties by red filled
diamonds. The empirical mass--radius relation derived in Section\,\ref{sec:method:eb})
is shown with a blue unbroken line.} \end{figure}

\begin{figure} \includegraphics[width=0.48\textwidth,angle=0]{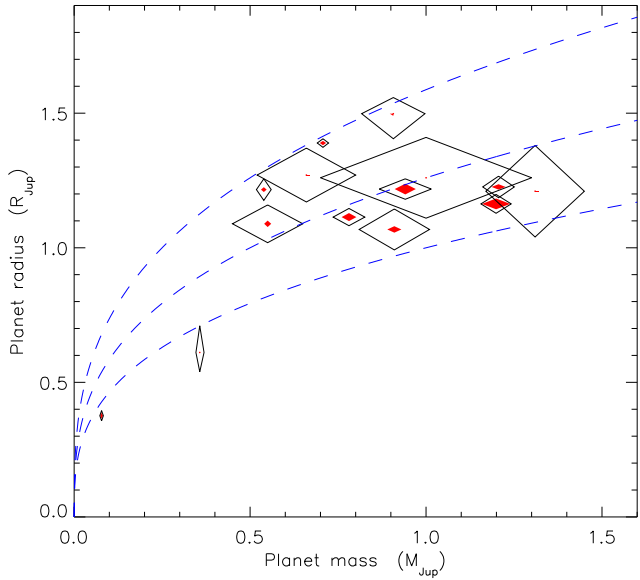}
\caption{\label{fig:absdim:M2R2} Plot of the masses versus the radii of the
planets in the fourteen TEPs studied in this work. The statistical uncertainties
are shown by black open diamonds and the systematic uncertainties by red filled
diamonds. Blue unbroken lines show the loci of constant densities of 0.25, 0.5
and 0.5 \pjup\ (from bottom to top).} \end{figure}

The importance of systematic versus statistical uncertainties is shown in Figures \ref{fig:absdim:M1R1} and \ref{fig:absdim:M2R2}, where statistical uncertainties are shown with black open diamonds and systematic uncertainties with red filled diamonds.

In Fig.\,\ref{fig:absdim:M1R1} it can be seen that the systematic uncertainties are important for only a minority of TEPs, but that in these cases can be substantially larger than the statistical ones. It is noticeable that the stars do not follow the empirical mass--radius relation (Eq.\,\ref{eq:mr}), which is expected because their properties have been determined using stellar models which also do not reproduce this relation.

At larger masses the stars have disproportionately larger radii, suggesting that they are slightly evolved. There are two reasons why the known population of TEPs may be biased in favour of slightly evolved stars. Firstly, they are larger and so a given planet is more likely to be transiting. Secondly, they are brighter so will be over-represented in the magnitude-limited samples which are studied to find TEPs. For {\em more} evolved stars, planetary transits will be shallower and more difficult to detect, so these will be under-represented in observational populations of TEPs.

Fig.\,\ref{fig:absdim:M2R2} shows that systematic uncertainties are generally unimportant compared to statistical uncertainties for the planetary mass and radius measurements, but that they are significant in some cases. The systems which are most affected are GJ\,436, whose low-mass star is not well-understood by current theory, and HD\,189733, which is a young system.

\subsection{Using the empirical mass--radius relation}

Use of the empirical mass--radius relation instead of theoretical predictions yields physical properties which are, generally, smaller. This is most pronounced for the two quantities which are most dependent on outside constraints: $M_{\rm A}$ and $a$. The values of these can diminish by 15\% or more compared to calculations involving stellar models (e.g.\ TrES-1 and XO-1). This effect is less significant for the other quantities. The difference is largest for systems with stellar masses of 0.7--1.1\Msun, where the radius discrepancy between models and observations is most severe.

For completeness the results for each TEP using the mass--radius relation are collected in Tables \ref{tab:abspar:MRstars} and \ref{tab:abspar:MRplanets}. In the rest of this work I adopt the physical properties calculated using stellar models. Whether this step is correct is not yet clear, and will not be until we properly understand why theoretical predictions do not match the measured radii of stars in low-mass eclipsing binaries.

\subsection{Comparing different methods}

\begin{figure} \includegraphics[width=0.48\textwidth,angle=0]{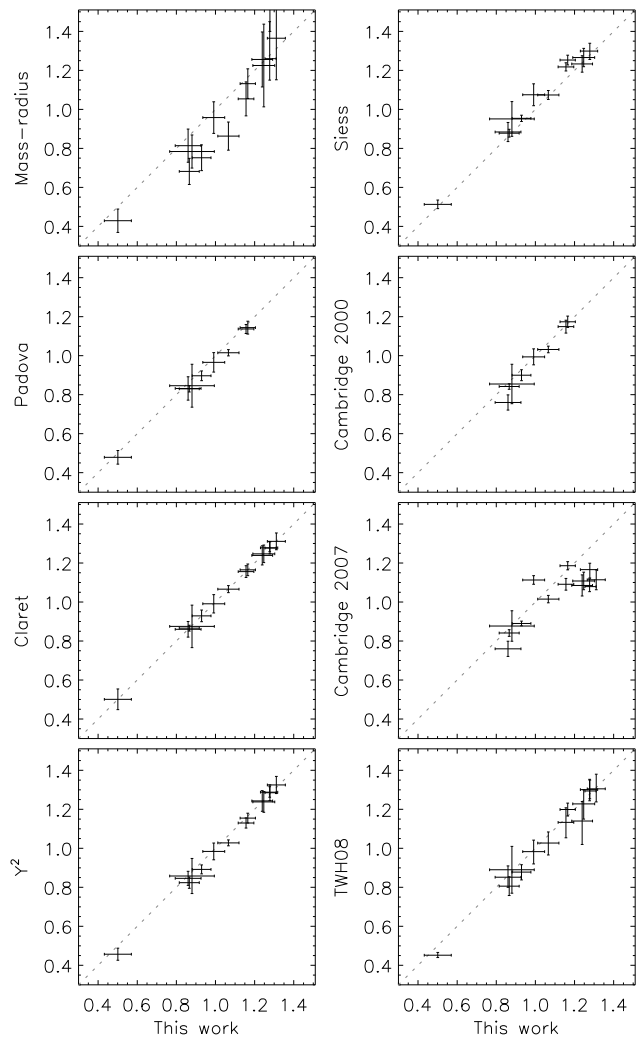}
\caption{\label{fig:absdim:m1} \reff{Comparison between the stellar masses
determined in different ways in this work and with TWH08. `This work' refers
to the masses in Table\,\ref{tab:abspar:stars}, where the random uncertainties
and systematic errors have been added in quadrature. In each case parity is
shown with a dotted line. As the final masses were derived using the Claret
models as a baseline, a comparison with these masses shows perfect agreement.
The corresponding panel has been retained in the plot because the uncertainties
in the final masses include the systematic contribution whereas those derived
using only the Claret models do not.}} \end{figure}

\reff{Fig.\,\ref{fig:absdim:m1} contains a comparison between the stellar masses determined in different ways in this work and in TWH08. The comparison has been undertaken using $M_{\rm A}$ because this is the physical property which is most dependent on stellar models or mass--radius relations, so shows differences most clearly. The various sets of physical properties for each TEP come from the same measured quantities; they differ only in the value of $K_{\rm b}$ adopted to satisfy whichever additional constraint is imposed. 

The left panels in Fig.\,\ref{fig:absdim:m1} compare the final masses against those obtained using the mass--radius relation or one of the three sets of adopted stellar models. \refff{The agreement between the masses found in different ways has been quantified by calculating the mean of the deviations from parity for each comparison. The results from the three model sets agree very well: the mean deviation from the final ({\sf Claret}) masses is 0.020\Msun\ for the {\sf Y$^2$} and 0.030\Msun\ for the {\sf Padova} models. Using the mass--radius relation produces masses which are generally smaller and scattered; the mean deviation here is 0.093\Msun.}

The right panels in Fig.\,\ref{fig:absdim:m1} show comparisons with results from the three sets of models which were not used to determine the final masses. The {\sf Siess} models produce $M_{\rm A}$s which are a bit larger on average than the baseline results: this is because they tend to predict denser stars than the other model sets \refff{(mean deviation 0.036\Msun)}. The {\sf Cambridge 2000} models can be applied to the fewest systems studied here due to their lesser coverage in mass and metallicity, but agree quite well for the eight stars within their remit \refff{(mean deviation 0.029\Msun)}. The {\sf Cambridge 2007} models do not agree well \refff{(mean deviation 0.094\Msun)} and also show quite a lot of scatter. This effect seems to depend on metallicity, with the outliers above and below the line of parity being respectively metal-poor (TrES-2) and metal-rich (HD\,149026 and the OGLEs). It should be remembered that the Cambridge models are targeted towards more massive stars, and should not be criticized too strongly for possible discrepancies in the mass range of interest here.

Finally, a comparison with the stellar masses found by TWH08 (Fig.\,\ref{fig:absdim:m1}, bottom-right panel) shows decent agreement: all $M_{\rm A}$s agree to within their errors: \refff{the mean deviation between the stellar mass values is 0.032\Msun}. The masses from TWH08 are on average slightly lower than those found in this work: this systematic effect is not large enough to worry about at present, but may need attention in the future during statistical analysis of a larger sample of TEPs.
}


\section{Properties of the known transiting extrasolar planetary systems}                           \label{sec:analysis}

\begin{figure} \includegraphics[width=0.48\textwidth,angle=0]{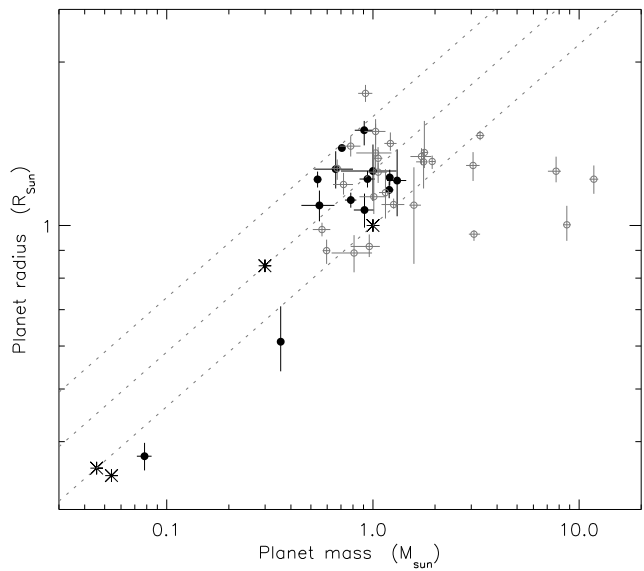}
\caption{\label{fig:analysis:m2r2} Mass--radius plot for the known transiting
extrasolar planets. Both axes are logarithmic. \reff{The fourteen objects
studied in this work are shown with black filled circles and the remaining
ones by grey open circles. The four gas giant planets in our Solar System
are denoted with asterisks.} Dotted lines show loci of constant density
(1.0, 0.5 and 0.25 \pjup, from bottom to top).} \end{figure}

The physical properties of the fourteen TEPs studied in this work have been augmented with results from the refereed literature for the other known TEPs. The overall sample now contains 41 systems, \reff{including only those which are the subject of a study in a refereed journal (including pre-publication papers on preprint servers). The properties of the full sample are reproduced in Tables \ref{tab:abspar:stars} and \ref{tab:abspar:planets}.}

The mass--radius diagram for all known extrasolar planets is reproduced in Fig.\,\ref{fig:analysis:m2r2}, and shows a strong clustering of objects in the region of 0.5--1.1\Mjup\ and 0.9--1.5\Rjup. The presence of outliers to larger mass but comparable radius is consistent with expectations that massive planets, brown dwarfs and low-mass stars have similar radii \citep{ChabrierBaraffe00ara,Pont+05aa2}. The planet with the largest radius is TrES-4 \citep{Mandushev+07apj}; it is the only one to have a density less than a quarter that of Jupiter. The reason why TrES-4 is an outlier remains unclear.

\subsection{Correlations with orbital period}

\begin{figure*} \includegraphics[width=\textwidth,angle=0]{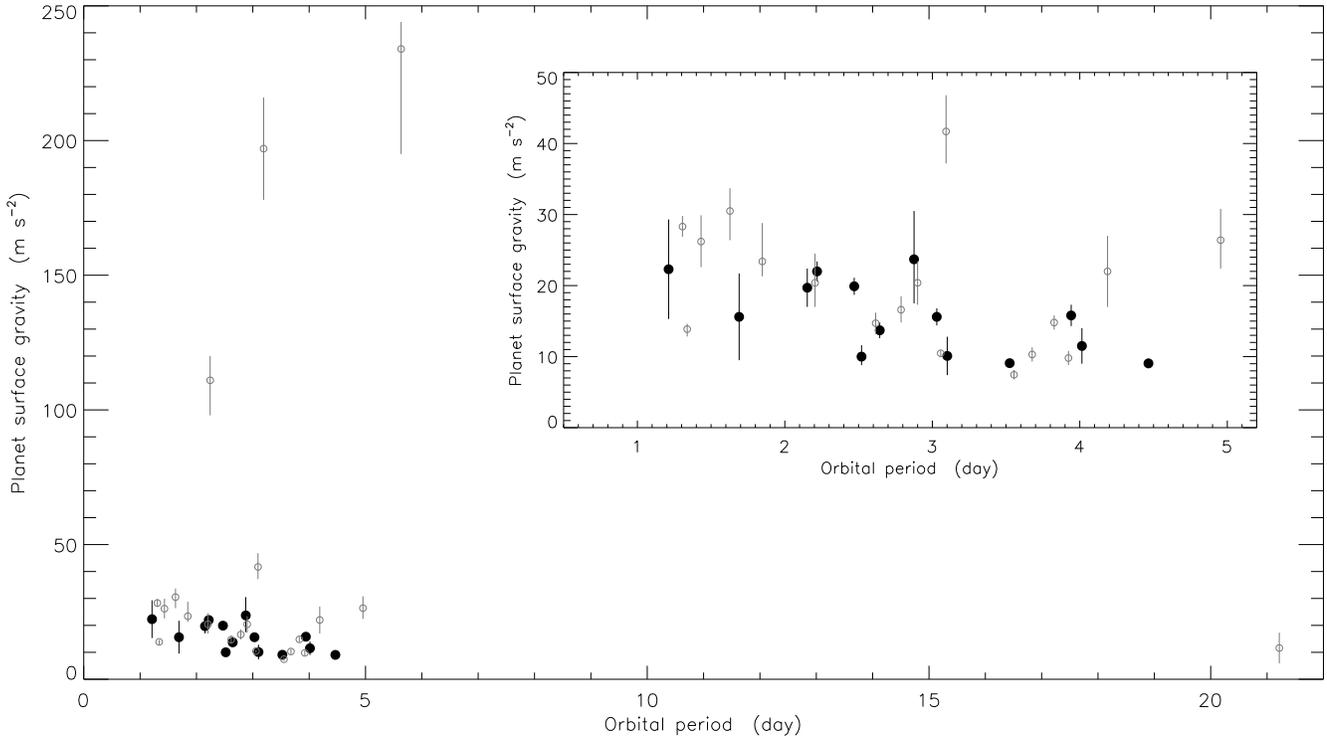}
\caption{\label{fig:analysis:pg2} Plot of the orbital periods versus the
surface gravities of all TEPs for which this information is available.
\reff{Black filled circles denote the fourteen systems studied in this
work and grey open circles represent numbers taken from the literature}.
The inset panel is an enlarged view of the region occupied by the bulk
of the TEPs.} \end{figure*}

A correlation between the orbital periods and surface gravities of exoplanets was found by \citet{Me++07mn} and confirmed in Paper\,I. The impressive influx of new discoveries in early 2008 means this possibility can now be investigated on 41 TEPs instead of the 30 involved in Paper\,I. The corresponding diagram is plotted in Fig.\,\ref{fig:analysis:pg2}. There are three planets which have much higher masses and thus surface gravities: HAT-P-2\,b ($M_{\rm b} = 11.8$\Mjup), WASP--14\,b (7.73\Mjup) and XO-3\,b (11.8\Mjup). These unusual properties mean that they are clear outliers in Fig.\,\ref{fig:analysis:pg2}, so I have discounted them when assessing the statistical significance of the period--gravity correlation. For this I have used two rank coefficient correlation tests: Spearman's $\rho$ returns a probability of 98.3\% and Kendall's $\tau$ a probability of 98.9\% that this correlation is significant.

\begin{figure*} \includegraphics[width=\textwidth,angle=0]{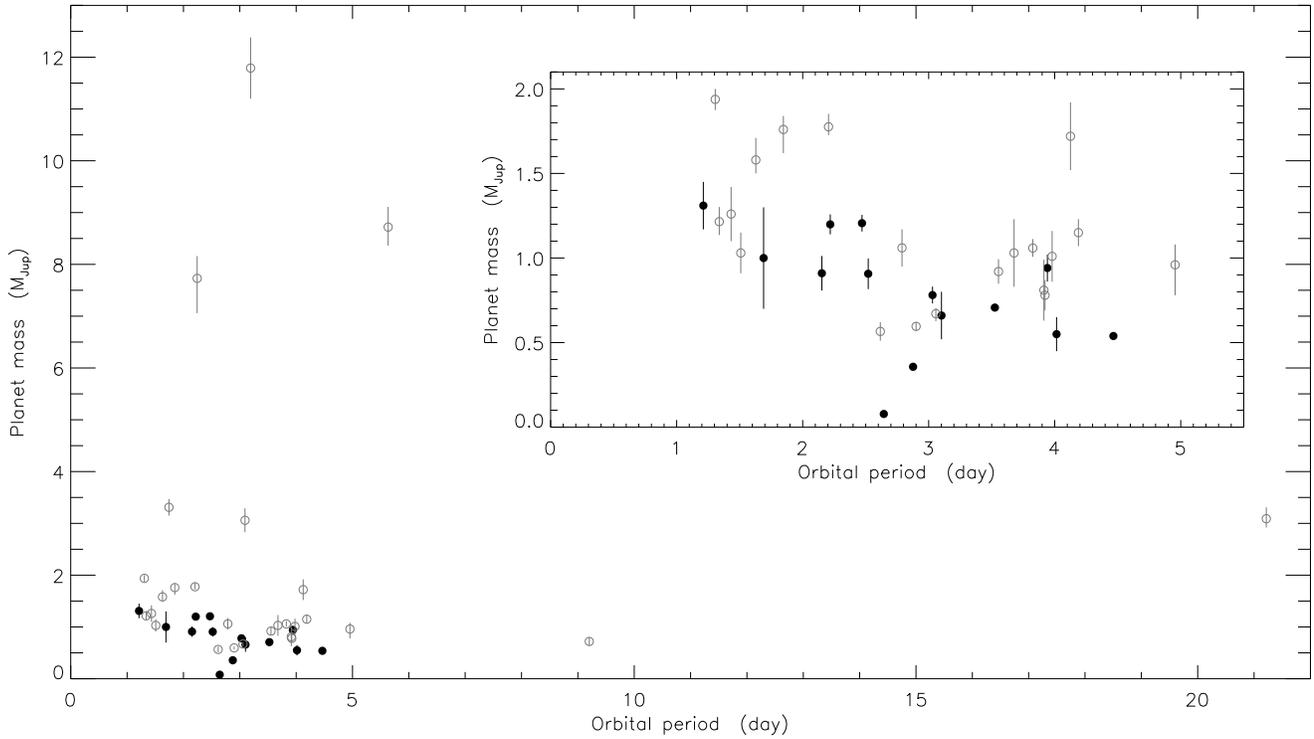}
\caption{\label{fig:analysis:pm2} \reff{Plot of the orbital periods versus
the planetary masses of all known TEPs. The symbols and inset panel are as
in Fig.\,\ref{fig:analysis:pm2}.}} \end{figure*}

A correlation of planet mass with orbital period was demonstrated by \citet{Mazeh++05mn}. The corresponding diagram is plotted in Fig.\,\ref{fig:analysis:pm2}, in which the correlation remains discernable. Using the above sample of 41 TEPs with the three high-mass outliers removed, Spearman's $\rho$ and Kendall's $\tau$ return probabilities of 96.2\% and 94.7\%, respectively, that this correlation not circumstantial.

It has previously been noticed \citep{Yoshi+08xxx} that the three planets which do not fit into the period--mass and period--gravity relations all have a high mass (9.72, 7.73 and 11.8\Mjup) and an eccentric orbit ($e = 0.52$, 0.095 and 0.26). It is also the case that their parent stars are very similar, with \Teff s of 6290, 6475 and 6429\,K, and masses of 1.31, 1.32 and 1.21 \Msun, for HAT-P-2, WASP-14 and XO-3 respectively. This may imply something interesting about the formation of planetary systems, but additional objects are needed before any conclusions are arrived at.

\subsection{Safronov numbers}

\begin{figure} \includegraphics[width=0.48\textwidth,angle=0]{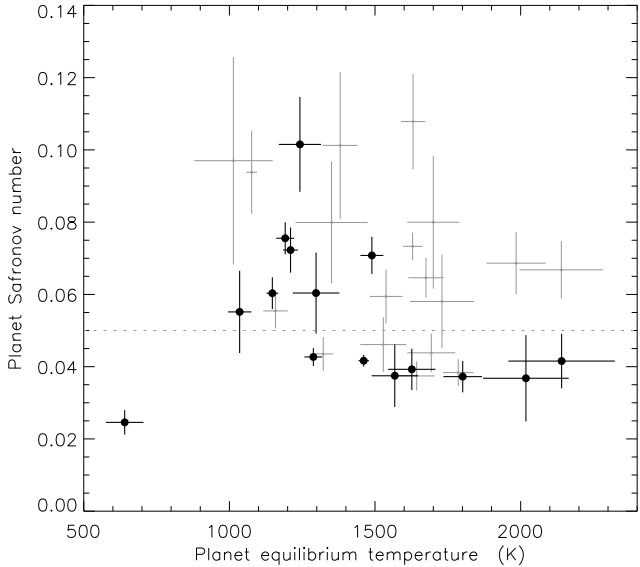}
\caption{\label{fig:analysis:teqsaf} Plot of equilibrium temperature versus
Safronov number for the full sample of planets. \reff{Those studied in this
work are shown with black filled circles and those whose properties are taken
from the the literature are shown with grey errorbars.} The dotted line
represents the divide envisaged by \citet{HansenBarman07apj}. Five planets
lie outside this plot towards large Safronov numbers: HAT-P-2\,b, XO-3\,b,
WASP-10, WASP-14\,b, OGLE-TR-113\,b and HD\,17156\,b. The outlying point at
the lower left is GJ\,436.} \end{figure}

\citet{HansenBarman07apj} presented evidence that there are two classes of extrasolar planet which occupy different regions of a plot of equilibrium temperature ($T_{\rm eq}$) versus \citet{Safronov72} number, the square of the ratio of escape velocity to orbital velocity:
\begin{equation} \Theta = \frac{1}{2} \left(\frac{V_{\rm esc}}{V_{\rm orb}}\right)^2
                        = \frac{a}{R_{\rm b}} \frac{M_{\rm b}}{M_{\rm A}}     \end{equation}
\citet{HansenBarman07apj} found that a sample of 19 TEPs divided naturally into two classes based on their Safronov number, with the split at $\Theta = 0.05$. Those in Class\,II ($\Theta \approx 0.04 \pm 0.01$) were found to have in general smaller planetary mass and larger stellar masses; Class\,II also contains all the planets with anomalously large radii.

The equations given by \citet{HansenBarman07apj} for $T_{\rm eq}$ and $\Theta$ have been used to construct a plot of these quantities, Fig.\,\ref{fig:analysis:teqsaf}, which contains 34 TEPs (five are outside the plot). \reff{The division into Class\,I and Class\,II in Fig.\,\ref{fig:analysis:teqsaf} is not strong. Whilst Class\,II planets (lower $\Theta$) still tend to group together, the distribution of Class\,I planets is much more nebulous than has previously been found, and several TEPs now appear to be intermediate between the two classes. The objects which now fill in the previously clear gap are all recent discoveries: HAT-P-5, HAT-P-6, HAT-P-7, OGLE-TR-211 and WASP-3. Whilst improved observations for these may lower the errorbars and reinstate the divide, on current evidence the separation between Class\,I and Class\,II seems to have been a statistical fluke.} It should also be stated that the finding that Class\,II TEPs have generally larger $R_{\rm b}$ and $M_{\rm A}$, and smaller $M_{\rm b}$, is encouraged by the definition of the Safronov number ($\Theta \propto \frac{M_{\rm b}}{M_{\rm A}R_{\rm b}}$).

TWH08 found that the division between Class\,I and Class\,II was clear in their sample of TEPs, that Class\,II planets are less massive than Class\,I ones for a given $T_{\rm eq}$, and that there is a correlation with the stellar \FeH. Similarly to my finding that the division is less obvious, the data in hand also do not lend strong support to the correlation with \FeH. However, the divide in a plot of $M_{\rm b}$ versus $T_{\rm eq}$ is clear. Further investigation of these results needs a larger sample of well-studied TEPs.


\section{Summary and future directions}

The discovery of the first transiting extrasolar planetary system, HD\,209458 \citep{Charbonneau+00apj,Henry+00apj}, opened up a new area of research in stellar and planetary astrophysics. Whilst we had to wait several years for the second TEP to be identified \citep{Konacki+03nat}, the ensuing flow of new discoveries has gradually increased to the point where roughly fifty of these objects are known. In the near future we can expect this steady stream to build up to a flood as  ground-based searches refine their techniques, and new planets are discovered by the {\it CoRoT} and {\it Kepler} space missions \citep{Baglin+06conf,Basri++05newar}.

At present, TEPs are studied using diverse methods, causing inhomogeneity in the results. This series of papers establishes a single set of methods for the analysis of photometric and spectroscopic observations of TEPs. It is aimed at providing homogeneous and robust physical properties for these objects, which can then be studied statistically in order to elicit the maximum information from existing observational data. Paper\,I presented analyses of the light curves of the fourteen TEPS for which good data were then available. In this work the analysis is extended, using the predictions of theoretical stellar evolutionary models, to produce the physical properties of the stellar and planetary components of each TEP. Emphasis is placed on understanding the statistical and systematic errors in these quantities.

Modelling the light curve of a TEP allows the quantities \Porb, $i$, $r_{\rm A}$ and $r_{\rm b}$ to be measured. Radial velocity measurements contribute $K_{\rm A}$ to this mix, leaving us one piece of information short of being able to derive the masses and radii of the two components. In most cases stellar theory is called upon to provide this missing datum by inferring the mass of the star from its measured \Teff, \FeH, and light-curve-derived density. In this work I determine the stellar mass which gives the optimal fit to all of the above measurements, allowing derivation of the physical properties and age of the system which best satisfy these constraints.

This process has a clear dependence on stellar theory, which may cause systematic errors in the derived properties. To estimate these systematics I have obtained separate solutions, for each of the fourteen TEPs studied in Paper\,I, using tabulated predictions from six different sets of stellar models. The results from using three of these model sets, {\sf Y$^2$}, {\sf Padova} and {\sf Claret}, are always in good agreement. The dispersion between the different results for the three model sets neatly allows a systematic error to be estimated for every output quantity.

The statistical errors have been propagated through the analysis using a perturbation algorithm, which allows a complete error budget to be obtained \reff{for every output quantity in each calculation}. It turns out that the uncertainties in the planetary and stellar radii are dominated by the statistical uncertainties from the light curve modelling, and that the systematic errors are mostly much smaller. The mass of each planet is mainly sensitive to the input $K_{\rm A}$, which is a directly observed quantity. Thus the physical properties of the transiting planets are reliable as they have only a minor dependence on theoretical calculations. In contrast, the semimajor axis and stellar mass are quite sensitive to stellar model predictions, and in some cases the systematic errors are up to three times larger than the statistical ones. Therefore several sets of stellar models should always be used for this kind of analysis, to check for the presence of systematic errors.

A major caveat applies to the previous paragraph. Different sets of stellar models use much of the same input physics (for example opacities, treatment of convective mixing and calibration using our Sun), so cannot be considered totally independent. \reff{Also, only those models which showed a good mutual agreement were considered.} The systematic errors obtained in this work are therefore only lower limits on the potential systematics. Furthermore, stellar theory is notoriously unable to match the accurately-measured properties for 0.7--1.1\Msun\ stars in eclipsing binary systems. The predicted radii are too small by up to 15\%, and the \Teff s are correspondingly too high. (Fig.\,\ref{fig:MRT:model}). To illustrate this, I have also calculated the physical properties of the fourteen TEPs using, instead of stellar theory, an empirical mass--radius relation obtained from well-studied low-mass eclipsing binaries. This different approach can result in stellar masses which are smaller by up to 15\%, and all quantities (except the planet surface gravity) are affected to some extent. This probably sets an upper limit on the systematic errors present in measured properties of TEPs. Until we understand why models fail to match the properties of eclipsing binary stars, we do not really know how big the systematic uncertainties are in the measured properties of the known TEPs. Thus our understanding of planets is limited by our understanding of low-mass stars.

Temporarily putting aside the problems noted in the last paragraph, the physical properties of the fourteen TEPs have been specified using the {\sf Claret} stellar models, statistical uncertainties from the perturbation analysis, and systematic uncertainties from the interagreement between the {\sf Y$^2$}, {\sf Padova} and {\sf Claret} models (Tables \ref{tab:abspar:stars} and \ref{tab:abspar:planets}). \reff{The agreement with literature results is mostly good, with the expection of the tricky system HD\,149026. In particular, my results are in excellent agreement with the homogeneous study of TEPs published by TWH08, despite substantial methodological variations between the two studies.}

The resulting properties of the fourteen TEPs have been augmented with literature results for all others for which a published study is available. The mass--radius plot for the planets shows a heavy clustering in the region 0.9--1.1\Mjup\ and 0.9--1.5\Rjup, with two outliers to lower masses (HD\,149026\,b and GJ\,436\,b) and a number of objects with higher masses but similar radii. Observational selection effects mean that these systems represent only the tip of the iceberg, so not many conclusions can be drawn from such a diagram at present.

The correlation between orbital period and planetary surface gravity \citep{Me++07mn} was revisited and found to be significant at the 98--99\% level. The related correlation between period and mass for TEPs \citep{Mazeh++05mn} is weaker but still significant at the 94--95\% (2$\sigma$) level. Both of these significance assessments were calculated after rejecting three systems (HAT-P-2, WASP-14 and XO-3) with eccentric orbits, unusually high planetary masses, and very similar stellar \Teff s. These three systems are obvious outliers on most plots of the properties of TEPs and may represent a group of planetary systems which formed in a different way to most TEPs. \citet{HansenBarman07apj} found that the known TEPs have a bimodal distribution of Safronov numbers. This division into two classes has weakened considerably with the addition of newly-discovered TEPs, and may not be statistically significant.

The detailed error budgets calculated for each TEP studied in this work have allowed an assessment of which object would most benefit from what type of follow-up observations. TrES-2, WASP-2 and GJ\,436 are good targets for new spectral synthesis studies. Several systems have only a limited number of radial velocity measurements, which are sufficient to confirm their planetary nature but are rather inaccurate for determining their physical properties. These include TrES-1, XO-1, WASP-1 and WASP-2; they are bright enough that more extensive velocity measurements will not require too much effort. Obtaining additional light curves would be a good idea for XO-1 and WASP-1. The four OGLE planets studied here (OGLE-TR-10, 56, 111 and 132) would all benefit from further observations of all types, but are inherently more difficult to study as they are fainter and in more crowded fields than the other TEPs. It therefore may be better to use the finite reservoir of available telescope time on other systems which are equally deserving but also more straightforward to observe.


On the methodological side, no changes are envisaged to the photometric analyses presented in Paper\,I. The method in the current work, however, could be improved in several ways. Firstly, an increase in the number of stellar model sets used will allow more robust systematic errors to be put forward. It may also be necessary to investigate the effects of using different helium abundances \citep{Claret95aas,Bertelli+08aa} or $\alpha$-enhancement \citep{Demarque+04apjs,Pietrinferni+06apj} in the stellar models. Secondly, the empirical mass--radius relation (Eq.\,\ref{eq:mr}) would be helped by the inclusion of new results \citep{Ibanoglu+08mn,Clausen+08aa}. Finally, extension of the analysis to other well-observed TEPs will increase the statistical weight of the resulting sample of objects with homogeneously determined properties. Slightly further into the future, the expected deluge of new transiting extrasolar planets from {\it CoRoT} and {\it Kepler} is eagerly awaited.


\section{Acknowledgements}

I am indebted to Peter Wheatley for extensive discussions, and to Tom Marsh, Boris G\"ansicke, Antonio Claret and Josh Winn for useful comments. I acknowledge the anonymous referee for a timely and robust report. I am grateful to John Eldridge and Antonio Claret for calculating large sets of stellar models for me. I acknowledge financial support from STFC in the form of a postdoctoral research assistant position. The following internet-based resources were heavily used in research for this paper: the NASA Astrophysics Data System; the SIMBAD database operated at CDS, Strasbourg, France; and the ar$\chi$iv scientific paper preprint service operated by Cornell University.



\bibliographystyle{mn_new}

\label{lastpage}


\appendix

\section{Comparison between the system properties derived in this work and those presented in the literature.}

This Appendix contains versions of the tables of physical properties of the TEPs in the main section, which have been extended to include comparisons with all previously published values of these properties.

\begin{table*} \begin{flushleft}
\caption{\label{apptab:abspar:tres1} Derived physical properties of the TrES-1 system compared to literature determinations.}
\begin{tabular}{l r@{\,$\pm$\,}l r@{\,$\pm$\,}l r@{\,$\pm$\,}l r@{\,$\pm$\,}l r@{\,$\pm$\,}l r@{\,$\pm$\,}l}
\hline \hline
                    & \mc{This work} & \mc{This work} & \mc{This work} & \mc{This work} & \mc{This work} & \mc{This work}  \\
                      & \mc{(mass--radius)} & \mc{(Padova models)} & \mc{(Siess models)} & \mc{(Y$^2$ models)}
                            & \mc{(Cambridge models)} & \mc{(Claret models)} \\
\hline      
$a$         (AU)      & 0.0373& 0.0001& 0.03954& 0.00036& 0.04035\,$^*$& 0.00023 & 0.03946&0.00034& 0.03942& 0.00017& 0.04000& 0.00043\\[2pt]  
$M_{\rm A}$ (\Msun)   & 0.752 & 0.066 & 0.897  & 0.025  & 0.954\,$^*$  &  0.016  & 0.892  & 0.023 & 0.890  & 0.012  & 0.929  & 0.030  \\       
$R_{\rm A}$ (\Rsun)   & 0.772 & 0.031 & 0.819  & 0.015  & 0.836\,$^*$  &  0.020  & 0.818  & 0.015 & 0.817  & 0.015  & 0.829  & 0.015  \\       
$\log g_{\rm A}$ (cgs)& 4.539 & 0.017 & 4.564  & 0.018  & 4.573\,$^*$  &  0.014  & 4.563  & 0.018 & 4.563  & 0.016  & 4.569  & 0.018  \\       
$\rho_{\rm A}$ (\psun)& 1.632 & 0.093 & 1.632  & 0.092  & 1.632\,$^*$  &  0.092  & 1.632  & 0.092 & 1.632  & 0.092  & 1.632  & 0.092  \\[2pt]  
$M_{\rm b}$ (\Mjup)   & 0.678 & 0.054 & 0.763  & 0.043  & 0.795\,$^*$  &  0.044  & 0.760  & 0.043 & 0.759  & 0.041  & 0.781  & 0.045  \\
\hline
\end{tabular}
\begin{tabular}{l r@{\,$\pm$\,}l r@{\,$\pm$\,}l r@{\,$\pm$\,}l r@{\,$\pm$\,}l r@{\,$\pm$\,}l r@{\,$\pm$\,}l}
                      & \mc{\citet{Alonso+04apj}} & \mc{\citet{Winn++07apj}} & \mc{TWH08} \\
\hline
Orbital semimajor axis $a$ (AU)                 & \mc{0.0393 $\pm$ 0.0011} & \mc{ }                 & \erm{0.03925}{0.00056}{0.00060} \\[2pt]
Stellar mass $M_{\rm A}$ (\Msun)                & \mc{0.88 $\pm$ 0.07}     & \mc{0.81 (fixed)}      & \erm{0.878}{0.038}{0.040}       \\
Stellar radius $R_{\rm A}$ (\Rsun)              & \mc{0.80 to 0.95}        & \mc{0.811 $\pm$ 0.020} & \erm{0.807}{0.017}{0.016}       \\
Stellar surface gravity $\log g_{\rm A}$ (cgs)  & \mc{ }                   & \mc{ }                 & \erm{4.567}{0.012}{0.015}       \\
Stellar density $\rho_{\rm A}$ (\psun)          & \mc{ }                   & \mc{ }                 & \erm{1.704}{0.001}{0.085}       \\[2pt]
Planetary mass $M_{\rm b}$ (\Mjup)              & \mc{0.75 $\pm$ 0.07}     & \mc{ }                 & \erm{0.752}{0.047}{0.046}       \\
Planetary radius $R_{\rm b}$ (\Rjup)            & \erm{1.08}{0.18}{0.04}   & \mc{1.081 $\pm$ 0.029} & \erm{1.067}{0.022}{0.021}       \\
Planetary surface gravity  $g_{\rm b}$ (\ms)    & \mc{ }                   & \mc{ }                 & \erm{16.6}{0.9}{1.0}            \\
Planetary density $\rho_{\rm b}$ (\pjup)        & \mc{ }                   & \mc{ }                 & \erm{0.620}{0.056}{0.052}       \\
\hline \hline \end{tabular} \\
$^*$\,These quantities have been calculated using stellar models for stars with small ages. Their values and uncertainties may therefore be unreliable due to edge effects within the grid of model tabulations. \end{flushleft}
\end{table*}

\begin{table*} \begin{flushleft}
\caption{\label{apptab:abspar:tres2} Derived physical properties of the TrES-2 system compared to literature determinations.}
\begin{tabular}{l r@{\,$\pm$\,}l r@{\,$\pm$\,}l r@{\,$\pm$\,}l r@{\,$\pm$\,}l r@{\,$\pm$\,}l r@{\,$\pm$\,}l r@{\,$\pm$\,}l}
\hline \hline
                    & \mc{This work} & \mc{This work} & \mc{This work} & \mc{This work} & \mc{This work} & \mc{This work} \\
                      & \mc{(mass--radius)} & \mc{(Padova models)} & \mc{(Siess models)} & \mc{(Y$^2$ models)}
                            & \mc{(Cambridge models)}  & \mc{(Claret models)} \\
\hline                                                                                                                                    
$a$         (AU)      & 0.0352&0.0010 & 0.03537&0.00061 & 0.03666&0.00063 & 0.03559&0.00051 & 0.03708&0.00024 & 0.03568&0.00056 \\[2pt]  
$M_{\rm A}$ (\Msun)   & 0.958 & 0.081 & 0.966  & 0.050  & 1.075  & 0.056  & 0.984  & 0.043  & 1.113  & 0.022  & 0.991  & 0.047  \\       
$R_{\rm A}$ (\Rsun)   & 0.983 & 0.049 & 0.985  & 0.031  & 1.021  & 0.032  & 0.991  & 0.031  & 1.033  & 0.033  & 0.994  & 0.031  \\       
$\log g_{\rm A}$ (cgs)& 4.435 & 0.022 & 4.436  & 0.028  & 4.451  & 0.028  & 4.439  & 0.027  & 4.456  & 0.025  & 4.440  & 0.027  \\       
$\rho_{\rm A}$ (\psun)& 1.008 & 0.092 & 1.008  & 0.092  & 1.010  & 0.089  & 1.010  & 0.089  & 1.010  & 0.089  & 1.010  & 0.089  \\[2pt]  
$M_{\rm b}$ (\Mjup)   & 1.180 & 0.069 & 1.186  & 0.045  & 1.274  & 0.048  & 1.201  & 0.039  & 1.303  & 0.026  & 1.206  & 0.042  \\
\hline
\end{tabular}
\begin{tabular}{l r@{\,$\pm$\,}l r@{\,$\pm$\,}l r@{\,$\pm$\,}l r@{\,$\pm$\,}l r@{\,$\pm$\,}l r@{\,$\pm$\,}l}
                      & \mc{\citet{Odonovan+06apj}} & \mc{\citet{Sozzetti+07apj}}
                              & \mc{\citet{Holman+07apj}} & \mc{TWH08} \\
\hline
$a$         (AU)      & \erm{0.0367}{0.0012}{0.0005} &    \mc{ }                 & \mc{ }        & \erm{0.03558}{0.00070}{0.00077} \\[2pt]
$M_{\rm A}$ (\Msun)   & \erm{1.08}{0.11}{0.05}       & \mc{0.980 $\pm$ 0.062}    & \mc{ }        & \erm{0.983}{0.059}{0.063}       \\
$R_{\rm A}$ (\Rsun)   & \erm{1.00}{0.06}{0.04}       & \erm{1.000}{0.036}{0.033} & 1.003 & 0.027 & \erm{1.003}{0.033}{0.033}       \\
$\log g_{\rm A}$ (cgs)& \mc{ }                       & \erm{4.429}{0.021}{0.023} & \mc{ }        & \erm{4.427}{0.019}{0.021}       \\
$\rho_{\rm A}$ (\psun)& \mc{ }                       & \mc{09.76 $\pm$ 0.046}    & \mc{ }        & \erm{0.974}{0.043}{0.042}       \\[2pt]
$M_{\rm b}$ (\Mjup)   & \erm{1.28}{0.09}{0.04}       &    \mc{ }                 & 1.198 & 0.053 & \erm{1.200}{0.051}{0.053}       \\
$R_{\rm b}$ (\Rjup)   & \erm{1.24}{0.09}{0.06}       &    \mc{ }                 & 1.222 & 0.038 & \erm{1.224}{0.041}{0.041}       \\
$g_{\rm b}$ (\ms)     & \mc{ }                       &    \mc{ }                 & 19.76 & 0.91  & \erm{19.9}{0.75}{0.72}          \\
$\rho_{\rm b}$ (\pjup)& \erm{0.67}{0.10}{0.07}       &    \mc{ }                 & \mc{ }        & \erm{0.655}{0.0.77}{0.067}      \\
\hline \hline \end{tabular} \end{flushleft} \end{table*}

\begin{table*} \begin{flushleft}
\caption{\label{apptab:abspar:xo1} Derived physical properties of the XO-1 system compared to literature determinations.}
\begin{tabular}{l r@{\,$\pm$\,}l r@{\,$\pm$\,}l r@{\,$\pm$\,}l r@{\,$\pm$\,}l r@{\,$\pm$\,}l r@{\,$\pm$\,}l}
\hline \hline
                    & \mc{This work} & \mc{This work} & \mc{This work} & \mc{This work} & \mc{This work} & \mc{This work} \\
                      & \mc{(mass--radius)} & \mc{(Padova models)} & \mc{(Siess models)} & \mc{(Y$^2$ models)}
                            & \mc{(Cambridge models)}  & \mc{(Claret models)} \\
\hline                                                                                                                                    
$a$         (AU)      & 0.0465& 0.0013 & 0.04909&0.00026 & 0.05002$^*$&0.00035 & 0.04929&0.00023 & 0.04909&0.00031 & 0.04990&0.00029\\[2pt]   
$M_{\rm A}$ (\Msun)   & 0.863 & 0.072  & 1.015  & 0.016  & 1.074$^*$  & 0.023  & 1.028  & 0.015  & 1.015  & 0.019  & 1.066  & 0.018 \\        
$R_{\rm A}$ (\Rsun)   & 0.886 & 0.037  & 0.935  & 0.025  & 0.953$^*$  & 0.027  & 0.939  & 0.024  & 0.935  & 0.025  & 0.950  & 0.022 \\        
$\log g_{\rm A}$ (cgs)& 4.480 & 0.018  & 4.503  & 0.016  & 4.511$^*$  & 0.016  & 4.505  & 0.017  & 4.503  & 0.016  & 4.510  & 0.018 \\        
$\rho_{\rm A}$ (\psun)& 1.242 & 0.078  & 1.242  & 0.080  & 1.242$^*$  & 0.080  & 1.242  & 0.080  & 1.242  & 0.080  & 1.242  & 0.080 \\[2pt]   
$M_{\rm b}$ (\Mjup)   & 0.818 & 0.078  & 0.911  & 0.071  & 0.946$^*$  & 0.075  & 0.918  & 0.072  & 0.911  & 0.072  & 0.941  & 0.074 \\
\hline
\end{tabular}
\begin{tabular}{l r@{\,$\pm$\,}l r@{\,$\pm$\,}l r@{\,$\pm$\,}l r@{\,$\pm$\,}l r@{\,$\pm$\,}l r@{\,$\pm$\,}l}
                      & \mc{\citet{Mccullough+06apj}} & \mc{\citet{Wilson+06pasp}}
                            & \mc{\citet{Holman+06apj}} & \mc{TWH08} \\
\hline
$a$         (AU)      & 0.0488 & 0.0005 & \mc{ }     & \mc{ }                     & \erm{0.04928}{0.00089}{0.00099} \\[2pt]
$M_{\rm A}$ (\Msun)   &  1.0 & 0.03     & \mc{ }     & \mc{1.00 $\pm$ 0.03 fixed} & \erm{1.027}{0.057}{0.061} \\
$R_{\rm A}$ (\Rsun)   &  1.0 & 0.08     & 1.0 & 0.08 & \erm{0.928}{0.018}{0.013}  & \erm{0.934}{0.037}{0.032} \\
$\log g_{\rm A}$ (cgs)&  4.53 & 0.065   & \mc{ }     & \mc{ }                     & \erm{4.509}{0.018}{0.027} \\
$\rho_{\rm A}$ (\psun)&  \mc{ }         & \mc{ }     & \mc{ }                     & \erm{1.333}{0.010}{0.149} \\[2pt]
$M_{\rm b}$ (\Mjup)   &  0.90 & 0.07    & \mc{ }     & \mc{0.9 fixed}             & \erm{0.918}{0.081}{0.078} \\
$R_{\rm b}$ (\Rjup)   &  1.30 & 0.11    & 1.34 & 0.12& \erm{1.184}{0.028}{0.018}  & \erm{1.206}{0.047}{0.042} \\
$g_{\rm b}$ (\ms)     &  \mc{ }         & \mc{ }     & \mc{ }                     & \erm{16.3}{1.4}{1.5}      \\
$\rho_{\rm b}$ (\pjup)&  0.41 & 0.10    & \mc{ }     & \mc{0.54 $\pm$ 0.06}       & \erm{0.524}{0.077}{0.069} \\
\hline \hline \end{tabular} \\
$^*$\,These quantities have been calculated using stellar models for stars with small ages. Their values and uncertainties may therefore be unreliable due to edge effects within the grid of model tabulations. \end{flushleft}
\end{table*}

\begin{table*} \begin{flushleft}
\caption{\label{apptab:abspar:wasp1} Derived physical properties of the WASP-1 system compared to literature determinations.}
\begin{tabular}{l r@{\,$\pm$\,}l r@{\,$\pm$\,}l r@{\,$\pm$\,}l r@{\,$\pm$\,}l r@{\,$\pm$\,}l r@{\,$\pm$\,}l}
\hline \hline
                      & \mc{This work}              & \mc{This work}                 & \mc{This work}                 & \mc{This work}                 & \mc{This work}                 \\
                      & \mc{(mass--radius)}                   & \mc{(Siess models)}            & \mc{(Y$^2$ models)}            & \mc{(Cambridge models)}        & \mc{(Claret models)}           \\
\hline
$a$         (AU)      & \erm{0.0417}{0.0011}{0.0014}& \erm{0.03955}{0.00040}{0.00044}& \erm{0.03944}{0.00037}{0.00044}& \erm{0.03815}{0.00035}{0.00047}& \erm{0.03933}{0.00031}{0.00032}\\[2pt]
$M_{\rm A}$ (\Msun)   & \erm{1.52}{0.11}{0.12}      & \erm{1.299}{0.040}{0.043}      & \erm{1.288}{0.037}{0.043}      & \erm{1.166}{0.033}{0.043}      & \erm{1.278}{0.031}{0.031}      \\
$R_{\rm A}$ (\Rsun)   & \erm{1.56}{0.08}{0.12}      & \erm{1.477}{0.060}{0.082}      & \erm{1.473}{0.056}{0.086}      & \erm{1.424}{0.054}{0.083}      & \erm{1.469}{0.055}{0.083}      \\
$\log g_{\rm A}$ (cgs)& \erm{4.236}{0.035}{0.023}   & \erm{4.213}{0.044}{0.026}      & \erm{4.212}{0.043}{0.027}      & \erm{4.197}{0.043}{0.027}      & \erm{4.211}{0.044}{0.026}      \\
$\rho_{\rm A}$ (\psun)& \erm{0.403}{0.069}{0.037}   & \erm{0.403}{0.069}{0.037}      & \erm{0.403}{0.069}{0.037}      & \erm{0.403}{0.069}{0.037}      & \erm{0.403}{0.069}{0.037}      \\[2pt]
$M_{\rm b}$ (\Mjup)   & \erm{1.02}{0.11}{0.12}      & \erm{0.917}{0.090}{0.090}      & \erm{0.912}{0.089}{0.090}      & \erm{0.853}{0.083}{0.084}      & \erm{0.907}{0.088}{0.088}      \\
\end{tabular}
\begin{tabular}{l r@{\,$\pm$\,}l r@{\,$\pm$\,}l r@{\,$\pm$\,}l r@{\,$\pm$\,}l r@{\,$\pm$\,}l r@{\,$\pm$\,}l}
\hline
                      & \mc{Collier Cameron} & \mc{\citet{Shporer+07mn}} & \mc{Charbonneau}
                            & \mc{\citet{Stempels+07mn}} & \mc{TWH08} \\
                      & \mc{et al.\ \citeyearpar{Cameron+07mn}} & \mc{ }
                        & \mc{et al.\ \citeyearpar{Charbonneau+07apj}} & \mc{ } & \mc{ } \\
\hline
$a$         (AU)      & \mc{0.0369 to 0.0395}  & \mc{ }                 & \mc{ }        & \mc{ }               & \erm{0.03946}{0.00049}{0.00048} \\[2pt]
$M_{\rm A}$ (\Msun)   & \erm{1.24}{0.68}{0.20} & \mc{1.15 fixed}        & 1.453 & 0.032 & \mc{ 1.25 to 1.35}   & \erm{1.301}{0.049}{0.047}       \\
$R_{\rm A}$ (\Rsun)   & \erm{1.15}{0.24}{0.09} & \mc{1.415 $\pm$ 0.074} & 1.45 & 0.08   & \mc{ }               & \erm{1.517}{0.052}{0.045}       \\
$\log g_{\rm A}$ (cgs)& \mc{ }                 & \mc{ }                 & \mc{ }        & \mc{4.28 $\pm$ 0.15} & \erm{4.190}{0.020}{0.022}       \\
$\rho_{\rm A}$ (\psun)& \mc{ }                 & \mc{ }                 & \mc{ }        & \mc{ }               & \erm{0.390}{0.006}{0.042}       \\
\hline \hline \end{tabular} \end{flushleft} \end{table*}

\begin{table*} \begin{flushleft}
\caption{\label{apptab:abspar:wasp2} Derived physical properties of the WASP-2 system compared to literature determinations.}
\begin{tabular}{l r@{\,$\pm$\,}l r@{\,$\pm$\,}l r@{\,$\pm$\,}l r@{\,$\pm$\,}l r@{\,$\pm$\,}l r@{\,$\pm$\,}l r@{\,$\pm$\,}l}
\hline \hline
                    & \mc{This work} & \mc{This work} & \mc{This work} & \mc{This work} & \mc{This work} & \mc{This work} \\
                      & \mc{(mass--radius)} & \mc{(Padova models)} & \mc{(Siess models)} & \mc{(Y$^2$ models)}
                            & \mc{(Cambridge models)}  & \mc{(Claret models)} \\
\hline                                                                                                                            
$a$         (AU)      & 0.0301&0.0011 & 0.0309&0.0013 & 0.0321&0.0010 & 0.0310&0.0011 & 0.0312&0.0009 & 0.0312&0.0013 \\[2pt]     
$M_{\rm A}$ (\Msun)   & 0.784 & 0.085 & 0.846 & 0.110 & 0.951 & 0.089 & 0.858 & 0.090 & 0.877 & 0.078 & 0.875 & 0.109 \\          
$R_{\rm A}$ (\Rsun)   & 0.805 & 0.061 & 0.852 & 0.044 & 0.859 & 0.046 & 0.830 & 0.044 & 0.836 & 0.042 & 0.835 & 0.045 \\          
$\log g_{\rm A}$ (cgs)& 4.520 & 0.032 & 4.531 & 0.050 & 4.548 & 0.044 & 4.534 & 0.046 & 4.537 & 0.045 & 4.536 & 0.049 \\          
$\rho_{\rm A}$ (\psun)& 1.50  & 0.21  & 1.50  & 0.21  & 1.50  & 0.21  & 1.50  & 0.21  & 1.50  & 0.21  & 1.50  & 0.21  \\[2pt]     
$M_{\rm b}$ (\Mjup)   & 0.841 & 0.085 & 0.885 & 0.100 & 0.956 & 0.094 & 0.893 & 0.089 & 0.906 & 0.084 & 0.905 & 0.099 \\          
$R_{\rm b}$ (\Rjup)   & 1.030 & 0.069 & 1.056 & 0.076 & 1.098 & 0.073 & 1.061 & 0.071 & 1.069 & 0.069 & 1.068 & 0.075 \\          
$g_{\rm b}$ (\ms)     & 19.7  & 2.7   & 19.7  & 2.7   & 19.7  & 2.7   & 19.7  & 2.7   & 19.7  & 2.7   & 19.7  & 2.7   \\          
$\rho_{\rm b}$ (\pjup)& 0.77  & 0.15  & 0.75  & 0.14  & 0.72  & 0.14  & 0.75  & 0.14  & 0.74  & 0.14  & 0.74  & 0.14  \\          
\hline
\end{tabular}
\begin{tabular}{l r@{\,$\pm$\,}l r@{\,$\pm$\,}l r@{\,$\pm$\,}l r@{\,$\pm$\,}l r@{\,$\pm$\,}l r@{\,$\pm$\,}l}
                      & \mc{\citet{Cameron+07mn}} & \mc{\citet{Charbonneau+07apj}} & \mc{TWH08} \\
\hline
$a$         (AU)      & \mc{0.0296 to 0.0318}  & \mc{ }        & \erm{0.03138}{0.00130}{0.00154} \\[2pt]
$M_{\rm A}$ (\Msun)   & \erm{0.79}{0.15}{0.04} & 0.813 & 0.032 & \erm{0.89}{0.12}{0.12}          \\
$R_{\rm A}$ (\Rsun)   & \mc{0.78 $\pm$ 0.06}   & 0.81 & 0.04   & \erm{0.840}{0.062}{0.065}       \\
$\log g_{\rm A}$ (cgs)& \mc{ }                 & \mc{ }        & \erm{4.537}{0.035}{0.046}       \\
$\rho_{\rm A}$ (\psun)& \mc{ }                 & \mc{ }        & \erm{1.45}{0.19}{0.11}          \\[2pt]
$M_{\rm b}$ (\Mjup)   & \mc{0.81 to 0.95}      & \mc{ }        & \erm{0.915}{0.090}{0.093}       \\
$R_{\rm b}$ (\Rjup)   & \mc{0.65 to 1.26}      & 1.038 & 0.050 & \erm{1.071}{0.080}{0.083}       \\
$g_{\rm b}$ (\ms)     & \mc{ }                 & \mc{ }        & \erm{19.4}{1.8}{1.4}            \\
$\rho_{\rm b}$ (\pjup)& \mc{ }                 & \mc{ }        & \erm{0.74}{0.22}{0.16}          \\
\hline \hline \end{tabular} \end{flushleft} \end{table*}

\begin{table*} \begin{flushleft}
\caption{\label{apptab:abspar:hat1}Derived physical properties of the HAT-P-1 system compared to literature determinations.}
\begin{tabular}{l r@{\,$\pm$\,}l r@{\,$\pm$\,}l r@{\,$\pm$\,}l r@{\,$\pm$\,}l r@{\,$\pm$\,}l r@{\,$\pm$\,}l}
\hline \hline
                    & \mc{This work} & \mc{This work} & \mc{This work} & \mc{This work} & \mc{This work} & \mc{This work} \\
                      & \mc{(mass--radius)} & \mc{(Padova models)} & \mc{(Siess models)} & \mc{(Y$^2$ models)}
                            & \mc{(Cambridge models)}  & \mc{(Claret models)} \\
\hline                                                                                                                                                   
$a$         (AU)      & 0.05401&0.00149 & 0.05540&0.00038 & 0.05669&0.00033 & 0.05529&0.00043 & 0.05464&0.00050 & 0.05570&0.00049 \\[2pt]    
$M_{\rm A}$ (\Msun)   & 1.054  & 0.087  & 1.137  & 0.023  & 1.218  & 0.021  & 1.130  & 0.026  & 1.091  & 0.030  & 1.156  & 0.030  \\         
$R_{\rm A}$ (\Rsun)   & 1.080  & 0.055  & 1.107  & 0.034  & 1.133  & 0.040  & 1.105  & 0.034  & 1.092  & 0.032  & 1.113  & 0.032  \\         
$\log g_{\rm A}$ (cgs)& 4.394  & 0.022  & 4.405  & 0.026  & 4.415  & 0.024  & 4.404  & 0.026  & 4.399  & 0.027  & 4.408  & 0.027  \\         
$\rho_{\rm A}$ (\psun)& 0.837  & 0.076  & 0.837  & 0.076  & 0.837  & 0.076  & 0.837  & 0.076  & 0.837  & 0.076  & 0.837  & 0.076  \\[2pt]    
$M_{\rm b}$ (\Mjup)   & 0.507  & 0.033  & 0.533  & 0.020  & 0.558  & 0.021  & 0.531  & 0.020  & 0.519  & 0.020  & 0.539  & 0.021  \\
\hline
\end{tabular}
\begin{tabular}{l r@{\,$\pm$\,}l r@{\,$\pm$\,}l r@{\,$\pm$\,}l r@{\,$\pm$\,}l r@{\,$\pm$\,}l r@{\,$\pm$\,}l}
                      & \mc{\citet{Bakos+07apj}} & \mc{\citet{Winn+07aj2}} & \mc{TWH08} & \mc{\citet{Johnson+08apj}} \\
\hline
$a$         (AU)      & \mc{0.0551 $\pm$ 0.0015}         & 0.0551 & 0.0015            & \erm{0.0553}{0.0012}{0.0013} & \mc{ }                       \\[2pt]
$M_{\rm A}$ (\Msun)   & \mc{1.12 $\pm$ 0.09 fixed}       & \mc{1.12 $\pm$ 0.09 fixed} & \erm{1.133}{0.075}{0.079}    & \mc{$1.133 \pm 0.077$ fixed} \\
$R_{\rm A}$ (\Rsun)   & \mc{\er{1.15}{0.10}{0.07} fixed} & 1.115 & 0.043              & \erm{1.135}{0.048}{0.048}    & 1.115 & 0.050                \\
$\log g_{\rm A}$ (cgs)& \mc{ }                           & \mc{ }                     & \erm{4.382}{0.027}{0.030}    & \mc{ }                       \\
$\rho_{\rm A}$ (\psun)& \mc{ }                           & 0.809 & 0.071              & \erm{0.775}{0.053}{0.050}    & \mc{ }                       \\[2pt]
$M_{\rm b}$ (\Mjup)   & \mc{0.53 $\pm$ 0.04}             & \mc{0.53 $\pm$ 0.04 fixed} & \erm{0.532}{0.030}{0.03}     & 0.524 & 0.031                \\
$R_{\rm b}$ (\Rjup)   & \erm{1.36}{0.11}{0.09}           & 1.203 & 0.051              & \erm{1.242}{0.053}{0.053}    & 1.225 & 0.059                \\
$g_{\rm b}$ (\ms)     & \mc{ }                           & 9.045 & 0.0661             & \erm{8.53}{0.51}{0.48}       & \mc{ }                       \\
$\rho_{\rm b}$ (\pjup)& \mc{ }                           & 0.303 & 0.025              & \erm{0.278}{0.043}{0.024}    & \mc{ }                       \\
\hline \hline \end{tabular} \end{flushleft} \end{table*}

\begin{table*} \begin{flushleft}
\caption{\label{apptab:abspar:ogle10} Derived physical properties of the OGLE-TR-10 system compared to literature determinations.}
\begin{tabular}{l r@{\,$\pm$\,}l r@{\,$\pm$\,}l r@{\,$\pm$\,}l r@{\,$\pm$\,}l r@{\,$\pm$\,}l r@{\,$\pm$\,}l}
\hline \hline
                      & \mc{This work} & \mc{This work} & \mc{This work} & \mc{This work} & \mc{This work} \\
                      & \mc{(mass--radius)} & \mc{(Siess models)} & \mc{(Y$^2$ models)} & \mc{(Cambridge models)} & \mc{(Claret 2008)} \\
\hline
$a$         (AU)      & 0.04492&0.0017 & 0.04464&0.00051 & 0.04476&0.00064 & 0.04278&0.00071 & 0.04471&0.00059 \\[2pt]
$M_{\rm A}$ (\Msun)   & 1.256  & 0.141 & 1.233  & 0.042  & 1.243  & 0.053  & 1.085  & 0.054  & 1.239  & 0.049  \\
$R_{\rm A}$ (\Rsun)   & 1.286  & 0.121 & 1.278  & 0.079  & 1.281  & 0.086  & 1.225  & 0.082  & 1.280  & 0.082  \\
$\log g_{\rm A}$ (cgs)& 4.319  & 0.041 & 4.316  & 0.053  & 4.317  & 0.051  & 4.297  & 0.052  & 4.317  & 0.053  \\
$\rho_{\rm A}$ (\psun)& 0.59   & 0.11  & 0.59   & 0.11   & 0.59   & 0.11   & 0.59   & 0.11   & 0.59   & 0.11   \\[2pt]
$M_{\rm b}$ (\Mjup)   & 0.67   & 0.15  & 0.66   & 0.14   & 0.67   & 0.14   & 0.61   & 0.13   & 0.66   & 0.14   \\
$R_{\rm b}$ (\Rjup)   & 1.28   & 0.11  & 1.27   & 0.10   & 1.27   & 0.10   & 1.22   & 0.10   & 1.27   & 0.10   \\
$g_{\rm b}$ (\ms)     & 10.2   & 2.7   & 10.2   & 2.7    & 10.2   & 2.7    & 10.2   & 2.7    & 10.1   & 2.7    \\
$\rho_{\rm b}$ (\pjup)& 0.32   & 0.11  & 0.32   & 0.11   & 0.32   & 0.11   & 0.34   & 0.11   & 0.32   & 0.11   \\[2pt]
Age (Gyr)             & \mc{ } & \erm{2.0}{1.2}{1.5} & \ \ \ \ \ \ 2.0 & 1.1 & \erm{2.6}{1.2}{1.1} & \erm{2.0}{1.5}{1.4} \\
\end{tabular}
\begin{tabular}{l r@{\,$\pm$\,}l r@{\,$\pm$\,}l r@{\,$\pm$\,}l r@{\,$\pm$\,}l r@{\,$\pm$\,}l r@{\,$\pm$\,}l}
\hline
                       & \mc{\citet{Bouchy+05aa2}}& \mc{\citet{Konacki+05apj}} & \mc{\citet{Pont+07aa}}
                             & \mc{\citet{Holman+07apj2}} & \mc{TWH08} \\
\hline
$a$         (AU)       & \mc{ }       & 0.04162 & 0.00069          & \mc{ }                 & \mc{ }                     & \erm{0.0434}{0.0013}{0.0015} \\[2pt]
$M_{\rm A}$ (\Msun)    & 1.22 & 0.045 & \mc{1.00 $\pm$ 0.05 fixed} & \mc{1.18 $\pm$ 0.04}   & \erm{1.03}{0.13}{0.12}     & \erm{1.14}{0.10}{0.12}       \\
$R_{\rm A}$ (\Rsun)    & 1.21 & 0.066 & \mc{1.00 $\pm$ 0.10 fixed} & \mc{1.16 $\pm$ 0.06}   & \erm{1.06}{0.19}{0.16}     & \erm{1.17}{0.13}{0.11}       \\
$\log g_{\rm A}$ (cgs) & \mc{ }       & \mc{ }                     & \mc{ }                 & \mc{ }                     & \erm{4.358}{0.064}{0.082}    \\
$\rho_{\rm A}$ (\psun) & \mc{ }       & \mc{ }                     & \mc{ }                 & \mc{ }                     & \erm{0.73}{0.13}{0.17}       \\[2pt]
$M_{\rm b}$ (\Mjup)    & 0.66 & 0.21  & 0.57 & 0.12                & \mc{0.63 $\pm$ 0.14}   & \mc{0.57 $\pm$ 0.12 fixed} & \erm{0.62}{0.14}{0.14}       \\
$R_{\rm b}$ (\Rjup)    & 1.54 & 0.12  & 1.24 & 0.09                & \mc{1.26 $\pm$ 0.07}   & \mc{1.06 $\pm$ 0.08}       & \erm{1.25}{0.14}{0.12}       \\
$g_{\rm b}$ (\ms)      & \mc{ }       & \mc{ }                     & \mc{ }                 & \mc{ }                     & \erm{10.0}{2.9}{2.6}         \\
$\rho_{\rm b}$ (\pjup) & \mc{ }       & 0.31 & 0.08                & \erm{0.34}{0.15}{0.11} & \mc{ }                     & \erm{0.32}{0.15}{0.10}       \\
\hline \hline \end{tabular} \end{flushleft} \end{table*}

\begin{table*} \begin{flushleft}
\caption{\label{apptab:abspar:ogle56} Derived physical properties of the OGLE-TR-56 system compared to literature determinations.}
\begin{tabular}{l r@{\,$\pm$\,}l r@{\,$\pm$\,}l r@{\,$\pm$\,}l r@{\,$\pm$\,}l r@{\,$\pm$\,}l r@{\,$\pm$\,}l}
\hline \hline
       & \mc{This work} & \mc{This work} & \mc{This work} & \mc{This work} & \mc{This work} \\
       & \mc{(mass--radius)} & \mc{(Siess models)} & \mc{(Y$^2$ models)} & \mc{(Cambridge models)} & \mc{(Claret models)} \\
\hline
$a$         (AU)       & 0.02381&0.00137 & 0.02407&0.00030 & 0.02390&0.00034 & 0.02303&0.00031 & 0.02395&0.00029\\[2pt]
$M_{\rm A}$ (\Msun)    & 1.225  & 0.212  & 1.266  & 0.047  & 1.238  & 0.053  & 1.108  & 0.045  & 1.247  & 0.045 \\
$R_{\rm A}$ (\Rsun)    & 1.294  & 0.201  & 1.268  & 0.144  & 1.258  & 0.148  & 1.213  & 0.139  & 1.261  & 0.140 \\
$\log g_{\rm A}$ (cgs) & 4.330  & 0.070  & 4.334  & 0.089  & 4.331  & 0.088  & 4.315  & 0.089  & 4.332  & 0.089 \\
$\rho_{\rm A}$ (\psun) & 0.62   & 0.21   & 0.62   & 0.21   & 0.62   & 0.21   & 0.62   & 0.21   & 0.62   & 0.21  \\[2pt]
$M_{\rm b}$ (\Mjup)    & 1.29   & 0.18   & 1.32   & 0.14   & 1.30   & 0.14   & 1.21   & 0.13   & 1.31   & 0.14  \\
$R_{\rm b}$ (\Rjup)    & 1.20   & 0.18   & 1.21   & 0.17   & 1.21   & 0.17   & 1.16   & 0.17   & 1.21   & 0.17  \\
$g_{\rm b}$ (\ms)      & 22.3   & 7.0    & 22.3   & 7.0    & 22.3   & 7.0    & 22.3   & 7.0    & 22.3   & 7.0   \\
$\rho_{\rm b}$ (\pjup) & 0.75   & 0.35   & 0.74   & 0.34   & 0.74   & 0.35   & 0.77   & 0.36   & 0.74   & 0.35  \\[2pt]
Age (Gyr)              & \mc{ } & \erm{1.2}{1.4}{1.2} & \erm{1.8}{0.9}{1.4} & \erm{2.0}{0.8}{2.0} & \erm{1.6}{1.3}{1.5} \\
\end{tabular}
\begin{tabular}{l r@{\,$\pm$\,}l r@{\,$\pm$\,}l r@{\,$\pm$\,}l r@{\,$\pm$\,}l r@{\,$\pm$\,}l r@{\,$\pm$\,}l}
\hline
            & \mc{Konacki et al.} & \mc{Torres et al.} & \mc{Bouchy et al.}
                  & \mc{Santos et al.} & \mc{Pont et al.} & \mc{TWH08} \\
            & \mc{\citeyearpar{Konacki+03nat}} & \mc{\citeyearpar{Torres+04apj}} & \mc{\citeyearpar{Bouchy+05aa2}}
                  & \mc{\citeyearpar{Santos+06aa}} & \mc{\citeyearpar{Pont+07aa}} & \mc{ } \\
\hline
$a$         (AU)      & \mc{0.0225} & 0.0225 & 0.0004            &    \mc{ }    &    \mc{ }   &    \mc{ }      & \erm{0.02383}{0.00046}{0.00051} \\[2pt]
$M_{\rm A}$ (\Msun)   & 1.04 & 0.05 & \mc{1.04 $\pm$ 0.05 fixed} & 1.10 & 0.078 & 1.17 & 0.04 & 1.17 & 0.04     & \erm{1.228}{0.072}{0.078}      \\
$R_{\rm A}$ (\Rsun)   & 1.10 & 0.10 & \mc{1.10 $\pm$ 0.10 fixed} & 1.12 & 0.069 & 1.15 & 0.06 & 1.32 & 0.06      & \erm{1.363}{0.089}{0.086}     \\
$\log g_{\rm A}$ (cgs)&   \mc{ }    & \mc{ }                     &    \mc{ }    &    \mc{ }   &    \mc{ }         & \erm{4.258}{0.043}{0.043}    \\
$\rho_{\rm A}$ (\psun)&   \mc{ }    & \mc{ }                     &    \mc{ }    &    \mc{ }   &    \mc{ }          & \erm{0.479}{0.078}{0.055}   \\[2pt]
$M_{\rm b}$ (\Mjup)   & 0.9  & 0.3  & 1.45 & 0.23                & 1.18 & 0.13  & 1.24 & 0.13 & 1.29 & 0.12         & \erm{1.39}{0.18}{0.17}     \\
$R_{\rm b}$ (\Rjup)   & 1.30 & 0.15 & 1.23 & 0.16                & 1.25 & 0.09  & 1.25 & 0.08 & 1.30 & 0.05          & \erm{1.363}{0.092}{0.090} \\
$g_{\rm b}$ (\ms)     &   \mc{ }    & \mc{ }                     &    \mc{ }    &  \mc{ }     &    \mc{ }             & \erm{18.4}{3.0}{2.7}     \\
$\rho_{\rm b}$ (\pjup)& 0.7  & 0.4  & 1.4 & 0.4                  &    \mc{ }    &  \mc{ }     & \erm{0.59}{0.27}{0.20} & \erm{0.55}{0.15}{0.11}  \\
\hline \hline \end{tabular} \end{flushleft} \end{table*}

\begin{table*} \begin{flushleft}
\caption{\label{apptab:abspar:ogle111} Derived physical properties of the OGLE-TR-111 system compared to literature determinations.}
\begin{tabular}{l r@{\,$\pm$\,}l r@{\,$\pm$\,}l r@{\,$\pm$\,}l r@{\,$\pm$\,}l r@{\,$\pm$\,}l r@{\,$\pm$\,}l r@{\,$\pm$\,}l}
\hline \hline
                    & \mc{This work} & \mc{This work} & \mc{This work} & \mc{This work} & \mc{This work} & \mc{This work} \\
                    & \mc{(mass--radius)} & \mc{(Padova models)} & \mc{(Siess models)} & \mc{(Y$^2$ models)}
                          & \mc{(Cambridge models)} & \mc{(Claret models)} \\
\hline
$a$         (AU)      & 0.04616&0.00162 & 0.04650&0.00112 & 0.04745&0.00088 & 0.04676&0.00067 & 0.04511&0.00077 & 0.04702&0.00073 \\[2pt]
$M_{\rm A}$ (\Msun)   & 0.814  & 0.085  & 0.832  & 0.060  & 0.884  & 0.049  & 0.846  & 0.036  & 0.760  & 0.039  & 0.860  & 0.040  \\
$R_{\rm A}$ (\Rsun)   & 0.835  & 0.061  & 0.842  & 0.035  & 0.859  & 0.038  & 0.846  & 0.037  & 0.816  & 0.043  & 0.851  & 0.032  \\
$\log g_{\rm A}$ (cgs)& 4.505  & 0.031  & 4.508  & 0.045  & 4.517  & 0.042  & 4.510  & 0.041  & 4.495  & 0.037  & 4.513  & 0.044  \\
$\rho_{\rm A}$ (\psun)& 1.40   & 0.19   & 1.40   & 0.19   & 1.97   & 0.27   & 1.40   & 0.19   & 1.40   & 0.19   & 1.40   & 0.19   \\[2pt]
$M_{\rm b}$ (\Mjup)   & 0.532  & 0.103  & 0.540  & 0.100  & 0.562  & 0.103  & 0.546  & 0.099  & 0.508  & 0.093  & 0.552  & 0.101  \\
$R_{\rm b}$ (\Rjup)   & 1.069  & 0.075  & 1.077  & 0.070  & 1.099  & 0.070  & 1.083  & 0.067  & 1.045  & 0.066  & 1.089  & 0.068  \\
$g_{\rm b}$ (\ms)     & 11.5   & 2.5    & 11.5   & 2.5    & 11.5   & 2.5    & 11.5   & 2.5    & 11.5   & 2.5    & 11.5   & 2.5    \\
$\rho_{\rm b}$ (\pjup)& 0.44   & 0.11   & 0.44   & 0.11   & 0.52   & 0.14   & 0.43   & 0.11   & 0.45   & 0.11   & 0.43   & 0.11   \\[2pt]
Age (Gyr)             & \mc{ } & \erm{10.8}{9.9}{7.9} & \erm{10.0}{3.4}{8.8} & \erm{9.9}{4.2}{4.4} & \erm{14.3}{2.0}{6.2} & \erm{11.5}{5.4}{7.0} \\
\end{tabular}
\begin{tabular}{l r@{\,$\pm$\,}l r@{\,$\pm$\,}l r@{\,$\pm$\,}l r@{\,$\pm$\,}l r@{\,$\pm$\,}l r@{\,$\pm$\,}l r@{\,$\pm$\,}l r@{\,$\pm$\,}l}
\hline
            & \mc{Pont et al.} & \mc{Gallardo et} & \mc{Santos et al.} & \mc{Silva \& Cruz}
                  & \mc{Winn et al.} & \mc{Minniti et al.} & \mc{TWH08} & \mc{Diaz et al.} \\
            & \mc{\citeyearpar{Pont+04aa}} & \mc{al.\ \citeyearpar{Gallardo+05aa}} & \mc{\citeyearpar{Santos+06aa}}
                  & \mc{\citeyearpar{SilvaCruz06apj}} & \mc{\citeyearpar{Winn++07aj}} & \mc{\citeyearpar{Minniti+07apj}}
                        & \mc{ } & \mc{\citeyearpar{Diaz++08apj}} \\
\hline
$a$         (AU)       & 0.047 & 0.001          & \mc{ }      & \mc{ }      & 0.047 & 0.001 & \mc{ }                    & \erm{0.0467}{0.050}{0.065} & \erm{0.04689}{0.0010}{0.00097} & \mc{ }                    \\[2pt]
$M_{\rm A}$ (\Msun)    & \erm{0.82}{0.15}{0.02} & \mc{ }      & 0.81 & 0.02 & 0.96 & 0.21   & \mc{0.81 $\pm$ 0.02 fixed}& \mc{0.82 $\pm$ 0.02 fixed} & \erm{0.852}{0.058}{0.052}      & \mc{0.81 fixed}           \\
$R_{\rm A}$ (\Rsun)    & \erm{0.85}{0.10}{0.03} & 0.71 & 0.02 & 0.83 & 0.02 & \mc{ }        & \mc{0.831 $\pm$ 0.031}    & \mc{0.83 $\pm$ 0.03 fixed} & \erm{0.831}{0.045}{0.040}      & \erm{0.81}{0.041}{0.048}  \\
$\log g_{\rm A}$ (cgs) & \mc{ }                 & \mc{ }      & \mc{ }      & \mc{ }        & \mc{ }                    & \mc{ }                     & \erm{4.529}{0.038}{0.042}      & \mc{ }                    \\
$\rho_{\rm A}$ (\psun) & \mc{ }                 & \mc{ }      & \mc{ }      & \mc{ }        & \mc{ }                    & \mc{ }                     & \erm{1.47}{0.17}{0.16}         & \mc{ }                    \\[2pt]
$M_{\rm b}$ (\Mjup)    & 0.53 & 0.11            & \mc{ }      & 0.52 & 0.13 & \mc{ }        & \mc{0.52 fixed}           & \mc{0.53 fixed}            & \erm{0.55}{0.10}{0.10}         & \mc{0.52 fixed}           \\
\hline \hline \end{tabular} \end{flushleft} \end{table*}

\begin{table*} \begin{flushleft}
\caption{\label{apptab:abspar:ogle132} Derived physical properties of the OGLE-TR-132 system compared to literature determinations.}
\begin{tabular}{l r@{\,$\pm$\,}l r@{\,$\pm$\,}l r@{\,$\pm$\,}l r@{\,$\pm$\,}l r@{\,$\pm$\,}l r@{\,$\pm$\,}l r@{\,$\pm$\,}l}
\hline \hline
                      & \mc{This work (mass--radius)} & \mc{This work (Y$^2$ models)} & \mc{This work (Cambridge models)} & \mc{This work (Claret models)} \\
\hline
$a$         (AU)      & 0.03081&0.00160 & 0.03050&0.00034 & 0.02879&0.00044 & 0.03040&0.00033 \\[2pt]
$M_{\rm A}$ (\Msun)   & 1.365  & 0.213  & 1.325  & 0.044  & 1.114  & 0.051  & 1.311  & 0.043  \\
$R_{\rm A}$ (\Rsun)   & 1.397  & 0.201  & 1.38   & 0.14   & 1.31   & 0.14   & 1.38   & 0.14   \\
$\log g_{\rm A}$ (cgs)& 4.283  & 0.062  & 4.278  & 0.079  & 4.253  & 0.079  & 4.277  & 0.079  \\
$\rho_{\rm A}$ (\psun)& 0.50   & 0.15   & 0.50   & 0.15   & 0.50   & 0.15   & 0.50   & 0.15   \\[2pt]
$M_{\rm b}$ (\Mjup)   & 1.02   & 0.32   & 1.00   & 0.30   & 0.89   & 0.27   & 1.00   & 0.30   \\
$R_{\rm b}$ (\Rjup)   & 1.28   & 0.17   & 1.26   & 0.15   & 1.19   & 0.15   & 1.26   & 0.15   \\
$g_{\rm b}$ (\ms)     & 15.6   & 6.1    & 15.6   & 6.1    & 15.6   & 6.1    & 15.6   & 6.1    \\
$\rho_{\rm b}$ (\pjup)& 0.49   & 0.24   & 0.50   & 0.24   & 0.53   & 0.26   & 0.50   & 0.24   \\[2pt]
Age (Gyr)             & \mc{ } & \erm{1.4}{0.6}{1.1} & \erm{1.9}{0.5}{1.3} & \erm{1.5}{0.6}{1.4} \\
\hline
\end{tabular}
\begin{tabular}{l r@{\,$\pm$\,}l r@{\,$\pm$\,}l r@{\,$\pm$\,}l r@{\,$\pm$\,}l r@{\,$\pm$\,}l r@{\,$\pm$\,}l r@{\,$\pm$\,}l}
                      & \mc{\citet{Bouchy+04aa}} & \mc{\citet{Moutou++04aa}}
                            & \mc{\citet{Gillon+07aa3}} & \mc{TWH08} \\
\hline
$a$         (AU)       & 0.0306 & 0.0008        & \mc{ }      & \mc{ }         & \erm{0.03035}{0.00057}{0.00053}\\[2pt]
$M_{\rm A}$ (\Msun)    & 1.34 & 0.10            & 1.35 & 0.06 & 1.26 & 0.03     & \erm{1.305}{0.075}{0.067}     \\
$R_{\rm A}$ (\Rsun)    & \erm{1.41}{0.49}{0.10} & 1.43 & 0.10 & 1.34 & 0.08      & \erm{1.32}{0.17}{0.12}       \\
$\log g_{\rm A}$ (cgs) & \mc{ }                 & \mc{ }      & \mc{ }            & \erm{4.313}{0.063}{0.090}   \\
$\rho_{\rm A}$ (\psun) & \mc{ }                 & \mc{ }      & \mc{ }             & \erm{0.51}{0.21}{0.09}     \\[2pt]
$M_{\rm b}$ (\Mjup)    & 1.01 & 0.31            & 1.19 & 0.13 & 1.14 & 0.12         & \erm{1.18}{0.14}{0.13}    \\
$R_{\rm b}$ (\Rjup)    & \erm{1.15}{0.80}{0.13} & 1.13 & 0.08 & 1.18 & 0.07          & \erm{1.20}{0.15}{0.11}   \\
$g_{\rm b}$ (\ms)      & \mc{ }                 & \mc{ }      & \mc{ }                & \erm{18.9}{4.9}{3.3}    \\
$\rho_{\rm b}$ (\pjup) & \erm{0.6}{0.3}{0.5}    & 0.82 & 0.27 & \erm{0.69}{0.22}{0.17} & \erm{0.69}{0.26}{0.21} \\
\hline \hline \end{tabular} \end{flushleft} \end{table*}

\begin{table*} \begin{flushleft}
\caption{\label{apptab:abspar:gj436} Derived physical properties of the GJ\,436 system compared to literature determinations.}
\setlength{\tabcolsep}{2pt}
\begin{tabular}{l r@{\,$\pm$\,}l r@{\,$\pm$\,}l r@{\,$\pm$\,}l r@{\,$\pm$\,}l r@{\,$\pm$\,}l r@{\,$\pm$\,}l}
\hline \hline
          & \mc{This work} & \mc{This work} & \mc{This work} & \mc{This work} & \mc{This work} \\
          & \mc{(mass--radius)} & \mc{(Padova models)} & \mc{(Siess models)} & \mc{(Y$^2$ models)} & \mc{(Claret models)} \\
\hline
$a$         (AU)      & 0.0282 & 0.0013 & 0.02928&0.00070 & 0.02995&0.00042 & 0.02882&0.00066 & 0.02972& 0.0010 \\[2pt]
$M_{\rm A}$ (\Msun)   & 0.429  & 0.060  & 0.479  & 0.035  & 0.513  & 0.022  & 0.457  & 0.031  & 0.501  & 0.053  \\
$R_{\rm A}$ (\Rsun)   & 0.443  & 0.031  & 0.460  & 0.026  & 0.471  & 0.023  & 0.453  & 0.021  & 0.467  & 0.025  \\
$\log g_{\rm A}$ (cgs)& 4.777  & 0.031  & 4.793  & 0.025  & 4.803  & 0.027  & 4.786  & 0.032  & 4.799  & 0.034  \\
$\rho_{\rm A}$ (\psun)& 4.92   & 0.55   & 4.92   & 0.55   & 4.92   & 0.55   & 4.92   & 0.55   & 4.92   & 0.55   \\[2pt]
$M_{\rm b}$ (\Mjup)   & 0.0704 & 0.0069 & 0.0757 & 0.0042 & 0.0793 & 0.0032 & 0.0734 & 0.0039 & 0.0781 & 0.0059 \\
$R_{\rm b}$ (\Rjup)   & 0.357  & 0.022  & 0.371  & 0.017  & 0.379  & 0.015  & 0.365  & 0.016  & 0.376  & 0.019  \\
$g_{\rm b}$ (\ms)     & 13.7   & 1.1    & 13.7   & 1.1    & 13.7   & 1.1    & 13.7   & 1.1    & 13.7   & 1.1    \\
$\rho_{\rm b}$ (\pjup)& 1.54   & 0.20   & 1.49   & 0.18   & 1.45   & 0.17   & 1.51   & 0.18   & 1.47   & 0.18   \\
\end{tabular}
\begin{tabular}{l r@{\,$\pm$\,}l r@{\,$\pm$\,}l r@{\,$\pm$\,}l r@{\,$\pm$\,}l r@{\,$\pm$\,}l r@{\,$\pm$\,}l r@{\,$\pm$\,}l r@{\,$\pm$\,}l r@{\,$\pm$\,}l r@{\,$\pm$\,}l}
\hline
            & \mc{\citet{Gillon+07aa1}} & \mc{\citet{Gillon+07aa2}} & \mc{Deming et al.}
                  & \mc{\citet{Torres07apj}} & \mc{TWH08} & \mc{Alonso et al.}
                        & \mc{Shporer et al.} & \mc{\citet{Bean+08aa}} \\
            & \mc{ } & \mc{ } & \mc{\citeyearpar{Deming+07apj}} & \mc{ } & \mc{ }
                  & \mc{\citeyearpar{Alonso+08aa}} & \mc{\citeyearpar{Shporer+08xxx}} & \mc{ } \\
\hline
$a$         (AU)      & \mc{ }                     & \mc{ }                    & 0.0291&0.0004 & 0.02872 & 0.00027         & \erm{0.02872}{0.00029}{0.00026} & \mc{0.02872} & \erm{0.02872}{0.00030}{0.00025}     & \mc{ }                     \\[2pt]
$M_{\rm A}$ (\Msun)   & \mc{0.44 $\pm$ 0.04 fixed} & \mc{0.44 $\pm$ 0.04 fixed}& 0.47 & 0.02   & \erm{0.452}{0.014}{0.012} & \erm{0.452}{0.014}{0.012}       & \mc{ }       & \mc{\er{0.452}{0.014}{0.012} fixed} & \mc{$0.44 \pm 0.04$ fixed} \\
$R_{\rm A}$ (\Rsun)   & \mc{0.44 $\pm$ 0.04 fixed} & \erm{0.463}{0.22}{0.17}   & 0.47 & 0.02   & \erm{0.464}{0.009}{0.011} & \erm{0.464}{0.009}{0.011}       & \mc{ }       & 0.45 & 0.02                         & \erm{0.505}{0.029}{0.020}  \\
$\log g_{\rm A}$ (cgs)& \mc{ }                     & \mc{ }                    & \mc{ }        & \mc{ }                    & \erm{4.843}{0.018}{0.011}       & \mc{ }       & \mc{ }                              & \mc{ }                     \\
$\rho_{\rm A}$ (\psun)& \mc{ }                     & \mc{ }                    & \mc{ }        & \mc{ }                    & \erm{4.565}{0.618}{0.568}       & \mc{ }       & \mc{ }                              & \mc{ }                     \\
\hline \hline \end{tabular} \end{flushleft} \end{table*}

\begin{table*} \begin{flushleft}
\caption{\label{apptab:abspar:149026} Derived physical properties of the HD\,149026 system compared to literature determinations.}
\begin{tabular}{l r@{\,$\pm$\,}l r@{\,$\pm$\,}l r@{\,$\pm$\,}l r@{\,$\pm$\,}l r@{\,$\pm$\,}l}
\hline \hline
                      & \mc{This work (mass--radius)} & \mc{This work (Y$^2$ models)} & \mc{This work (Cambridge models)} & \mc{This work (Claret models)} \\
\hline
$a$         (AU)      & \erm{0.0427}{0.0020}{0.0013} & \erm{0.04303}{0.00034}{0.00026} & \erm{0.04060}{0.0050}{0.0032} & \erm{0.04294}{0.00037}{0.00021} \\[2pt]
$M_{\rm A}$ (\Msun)   & \erm{1.26}{0.19}{0.11}       & \erm{1.285}{0.031}{0.023}       & \erm{1.079}{0.040}{0.026}     & \erm{1.277}{0.033}{0.019}       \\
$R_{\rm A}$ (\Rsun)   & \erm{1.28}{0.17}{0.09}       & \erm{1.295}{0.121}{0.058}       & \erm{1.221}{0.116}{0.057}     & \erm{1.292}{0.12}{0.054}        \\
$\log g_{\rm A}$ (cgs)& \erm{4.319}{0.030}{0.054}    & \erm{4.322}{0.037}{0.068}       & \erm{4.297}{0.037}{0.068}     & \erm{4.321}{0.039}{0.069}       \\
$\rho_{\rm A}$ (\psun)& \erm{0.592}{0.083}{0.129}    & \erm{0.592}{0.083}{0.129}       & \erm{0.592}{0.083}{0.129}     & \erm{0.592}{0.083}{0.129}       \\[2pt]
$M_{\rm b}$ (\Mjup)   & \erm{0.353}{0.035}{0.024}    & \erm{0.358}{0.011}{0.011}       & \erm{0.319}{0.012}{0.010}     & \erm{0.357}{0.012}{0.011}       \\
$R_{\rm b}$ (\Rjup)   & \erm{0.608}{0.110}{0.047}    & \erm{0.612}{0.099}{0.072}       & \erm{0.578}{0.094}{0.068}     & \erm{0.611}{0.099}{0.072}       \\
$g_{\rm b}$ (\ms)     & \erm{23.7}{6.8}{6.2}         & \erm{23.7}{6.8}{6.2}            & \erm{23.7}{6.8}{6.2}          & \erm{23.7}{6.8}{6.2}            \\
$\rho_{\rm b}$ (\pjup)& \erm{1.57}{0.71}{0.58}       & \erm{1.56}{0.71}{0.57}          & \erm{1.65}{0.76}{0.60}        & \erm{1.56}{0.71}{0.57}          \\[2pt]
Age (Gyr)             & \mc{ } & \erm{1.2}{0.8}{0.6} & \ \ \ \ \ \ \ \ \ \ \ \ 1.2 & 1.0 & \erm{1.9}{0.7}{0.4} \\
\end{tabular}
\begin{tabular}{l r@{\,$\pm$\,}l r@{\,$\pm$\,}l r@{\,$\pm$\,}l r@{\,$\pm$\,}l r@{\,$\pm$\,}l r@{\,$\pm$\,}l r@{\,$\pm$\,}l}
\hline
                      & \mc{\citet{Sato+05apj}} & \mc{Charbonneau} & \mc{Masana et al.}
                            & \mc{\citet{Wolf+07apj}} & \mc{\citet{Winn+08apj}} & \mc{TWH08} & \mc{Nutzman et al.} \\
                      & \mc{ } & \mc{et al.\ \citeyearpar{Charbonneau+06apj}} & \mc{\citeyearpar{Masana++06aa}}
                            & \mc{ } & \mc{ } & \mc{ } & \mc{\citeyearpar{Nutzman+08xxx}} \\
\hline
$a$         (AU)      & \mc{0.042}   & \mc{ }                     & \mc{ }        & \mc{ }                     & \mc{ }                     & \erm{0.04313}{0.0028}{0.0037} & \mc{ }                     \\[2pt]
$M_{\rm A}$ (\Msun)   & 1.3 & 0.1    & \mc{1.30 $\pm$ 0.10 fixed} & \mc{ }        & \mc{1.30 $\pm$ 0.10 fixed} & \mc{1.30 $\pm$ 0.06 fixed} & \erm{1.294}{0.060}{0.050}     & \mc{$1.30 \pm 0.10$ fixed} \\
$R_{\rm A}$ (\Rsun)   & 1.45 & 0.1   & \mc{1.45 $\pm$ 0.10 fixed} & 1.515 & 0.096 & \mc{1.45 $\pm$ 0.10 fixed} & \mc{1.45 $\pm$ 0.10 fixed} & \erm{1.368}{0.120}{0.083}     & 1.497 & 0.069              \\
$\log g_{\rm A}$ (cgs)& \mc{ }       & \mc{ }                     & \mc{ }        & \mc{ }                     & \mc{ }                     & \erm{4.278}{0.045}{0.063}     & \mc{ }                     \\
$\rho_{\rm A}$ (\psun)& \mc{ }       & \mc{ }                     & \mc{ }        & \mc{ }                     & \mc{ }                     & \erm{0.584}{0.008}{0.177}     & \mc{ }                     \\[2pt]
$M_{\rm b}$ (\Mjup)   & 0.36 & 0.04  & \mc{ }                     & \mc{ }        & \mc{0.352 $\pm$ 0.025}     & \mc{0.36 $\pm$ 0.03}       & \erm{0.359}{0.022}{0.021}     & 0.359 & 0.006              \\
\hline \hline \end{tabular} \end{flushleft} \end{table*}

\begin{table*} \begin{flushleft}
\caption{\label{apptab:abspar:189733} Derived physical properties of the HD\,189733 system compared to literature determinations.}
\begin{tabular}{l r@{\,$\pm$\,}l r@{\,$\pm$\,}l r@{\,$\pm$\,}l r@{\,$\pm$\,}l r@{\,$\pm$\,}l r@{\,$\pm$\,}l}
\hline \hline
                    & \mc{This work} & \mc{This work} & \mc{This work} & \mc{This work} & \mc{This work} & \mc{This work} \\
                    & \mc{(mass--radius)} & \mc{(Padova models)} & \mc{(Siess models)} & \mc{(Y$^2$ models)}
                          & \mc{(Cambridge models)} & \mc{(Claret models)} \\
\hline
$a$         (AU)      & 0.02932&0.00097 & 0.03129&0.00019 & 0.03189$^*$&0.00024 & 0.03122&0.00036 & 0.03144$^*$&0.00021 & 0.03175&0.00019 \\[2pt]     
$M_{\rm A}$ (\Msun)   & 0.682 & 0.067   & 0.830 & 0.016   & 0.878$^*$  & 0.020  & 0.824  & 0.029  & 0.841$^*$  & 0.017  & 0.866  & 0.015  \\          
$R_{\rm A}$ (\Rsun)   & 0.701 & 0.036   & 0.749 & 0.024   & 0.763$^*$  & 0.027  & 0.747  & 0.022  & 0.752$^*$  & 0.025  & 0.760  & 0.022  \\          
$\log g_{\rm A}$ (cgs)& 4.580 & 0.022   & 4.608 & 0.023   & 4.616$^*$  & 0.021  & 4.607  & 0.025  & 4.610$^*$  & 0.022  & 4.615  & 0.024  \\          
$\rho_{\rm A}$ (\psun)& 1.98 & 0.16     & 1.98  & 0.16    & 1.98$^*$   & 0.16   & 1.98   & 0.17   & 1.98$^*$   & 0.17   & 1.98   & 0.17   \\[2pt]     
$M_{\rm b}$ (\Mjup)   & 1.023 & 0.074   & 1.165 & 0.037   & 1.210$^*$  & 0.040  & 1.159  & 0.043  & 1.176$^*$  & 0.034  & 1.199  & 0.038  \\
\hline
\end{tabular}
\setlength{\tabcolsep}{4pt}
\begin{tabular}{l r@{\,$\pm$\,}l r@{\,$\pm$\,}l r@{\,$\pm$\,}l r@{\,$\pm$\,}l r@{\,$\pm$\,}l r@{\,$\pm$\,}l r@{\,$\pm$\,}l r@{\,$\pm$\,}l}
                  & \mc{Bouchy et al.} & \mc{Bakos et al.} & \mc{Masana et al.} & \mc{Winn et al.} & \mc{Winn et al.}
                        & \mc{Baines et al.} & \mc{TWH08} & \mc{Miller-Ricci}\\
                  & \mc{\citeyearpar{Bouchy+05aa}} & \mc{\citeyearpar{Bakos+06apj2}} & \mc{\citeyearpar{Masana++06aa}}
                        & \mc{\citeyearpar{Winn+06apj}} & \mc{\citeyearpar{Winn++07apj}}
                              & \mc{\citeyearpar{Baines+07apj}} & \mc{ } & \mc{et al.\ \citeyearpar{Millerricci+08apj}} \\
\hline
$a$         (AU)      & 0.0313 & 0.0004            & \mc{ }                      & \mc{ }        & \mc{ }      & \mc{ }        & \mc{ }        & \erm{0.03099}{0.00060}{0.00063} & \mc{ }                     \\[2pt]
$M_{\rm A}$ (\Msun)   & \mc{0.83 $\pm$ 0.03 fixed} & \mc{0.83 $\pm$ 0.03 fixed}  & \mc{ }        & 0.82 & 0.03 & \mc{ }        & \mc{ }        & \erm{0.806}{0.0048}{0.0048}     & \mc{$0.82 \pm 0.03$ fixed} \\
$R_{\rm A}$ (\Rsun)   & \mc{0.71 $\pm$ 0.02 fixed} & \mc{0.758 $\pm$ 0.016 fixed}& 0.758 & 0.016 & 0.73 & 0.02 & 0.753 & 0.025 & 0.779 & 0.052 & \erm{0.756}{0.018}{0.018}       & 0.749 & 0.009              \\
$\log g_{\rm A}$ (cgs)& \mc{ }                     & \mc{ }                      & \mc{ }        & \mc{ }      & \mc{ }        & \mc{ }        & \erm{4.587}{0.014}{0.015}       & \mc{ }                     \\
$\rho_{\rm A}$ (\psun)& \mc{ }                     & \mc{ }                      & \mc{ }        & \mc{ }      & \mc{ }        & \mc{ }        & \erm{1.867}{0.038}{0.038}       & \mc{ }                     \\[2pt]
$M_{\rm b}$ (\Mjup)   & 1.15 & 0.04                & \mc{ }                      & \mc{ }        & 1.13 & 0.03 & \mc{ }        & \mc{ }        & \erm{1.144}{0.057}{0.056}       & \mc{ }                     \\
\hline \hline \end{tabular} \\
$^*$\,These results are for the youngest ages available so cannot be regarded as robustly determined (see text for discussion). This is particularly important for the quoted uncertainties which will be rendered artifically low by this circumstance.
\end{flushleft} \end{table*}

\begin{table*} \begin{flushleft}
\caption{\label{apptab:abspar:209458} Derived physical properties of the HD\,209458 system compared to literature determinations.}
\begin{tabular}{l r@{\,$\pm$\,}l r@{\,$\pm$\,}l r@{\,$\pm$\,}l r@{\,$\pm$\,}l r@{\,$\pm$\,}l r@{\,$\pm$\,}l r@{\,$\pm$\,}l r@{\,$\pm$\,}l r@{\,$\pm$\,}l r@{\,$\pm$\,}l r@{\,$\pm$\,}l r@{\,$\pm$\,}l r@{\,$\pm$\,}l}
\hline \hline
                    & \mc{This work} & \mc{This work} & \mc{This work} & \mc{This work} & \mc{This work} & \mc{This work} \\
                    & \mc{(mass--radius)} & \mc{(Padova models)} & \mc{(Siess models)} & \mc{(Y$^2$ models)}
                          & \mc{(Cambridge models)} & \mc{(Claret models)} \\
\hline                                                                                                                                       
$a$         (AU)       & 0.04724&0.00105 & 0.04742&0.00045 & 0.04888&0.00032 & 0.04757&0.00034 & 0.04799&0.00029 & 0.04770&0.00041 \\[2pt]   
$M_{\rm A}$ (\Msun)    & 1.132  & 0.076  & 1.144  & 0.033  & 1.253  & 0.025  & 1.155  & 0.025  & 1.186  & 0.021  & 1.165  & 0.030  \\        
$R_{\rm A}$ (\Rsun)    & 1.159  & 0.026  & 1.163  & 0.011  & 1.199  & 0.008  & 1.167  & 0.009  & 1.177  & 0.008  & 1.170  & 0.010  \\        
$\log g_{\rm A}$ (cgs) & 4.364  & 0.010  & 4.365  & 0.005  & 4.378  & 0.003  & 4.367  & 0.004  & 4.370  & 0.003  & 4.368  & 0.004  \\        
$\rho_{\rm A}$ (\psun) & 0.727  & 0.005  & 0.727  & 0.005  & 0.727  & 0.005  & 0.727  & 0.005  & 0.727  & 0.005  & 0.727  & 0.005  \\[2pt]   
$M_{\rm b}$ (\Mjup)    & 0.693  & 0.032  & 0.698  & 0.016  & 0.742  & 0.013  & 0.703  & 0.013  & 0.715  & 0.012  & 0.707  & 0.015  \\
\hline
\end{tabular}
\begin{tabular}{l r@{\,$\pm$\,}l r@{\,$\pm$\,}l r@{\,$\pm$\,}l r@{\,$\pm$\,}l r@{\,$\pm$\,}l r@{\,$\pm$\,}l r@{\,$\pm$\,}l r@{\,$\pm$\,}l r@{\,$\pm$\,}l r@{\,$\pm$\,}l r@{\,$\pm$\,}l r@{\,$\pm$\,}l r@{\,$\pm$\,}l}
                  & \mc{Charbonneau} & \mc{Henry et al.} & \mc{Mazeh et al.} & \mc{Queloz et al.} & \mc{Brown et al.}
                        & \mc{Cody \& Sasselov} & \mc{Mandel \& Agol} \\
                  & \mc{et al.\ \citeyearpar{Charbonneau+00apj}} & \mc{\citeyearpar{Henry+00apj}}
                        & \mc{\citeyearpar{Mazeh+00apj}} & \mc{\citeyearpar{Queloz+00aa}} & \mc{\citeyearpar{Brown+01apj}}
                              & \mc{\citeyearpar{CodySasselov02apj}} & \mc{\citeyearpar{MandelAgol02apj}} \\
\hline
$a$         (AU)       & \mc{ }         & \mc{ }          & \mc{ }       & 0.048&0.001 & \mc{ }                   & \mc{ }                & \mc{ }                   \\[2pt]
$M_{\rm A}$ (\Msun)    & \mc{1.1 fixed} & \mc{1.03 fixed} & 1.1 & 0.1    & \mc{ }       &\mc{1.1 $\pm$ 0.1 fixed} & 1.06 & 0.13           & \mc{1.1 $\pm$ 0.1 fixed} \\
$R_{\rm A}$ (\Rsun)    & \mc{1.1 fixed} & \mc{1.15 fixed} & 1.2 & 0.1    & 1.2 & 0.1   & 1.146 & 0.050            & 1.18 & 0.10           & 1.145 & 0.035            \\
$\log g_{\rm A}$ (cgs) & \mc{ }         & \mc{ }          & \mc{ }       & \mc{ }      & \mc{ }                   & \mc{ }                & \mc{ }                   \\
$\rho_{\rm A}$ (\psun) & \mc{ }         & \mc{ }          & \mc{ }       & \mc{ }      & \mc{ }                   & \mc{ }                & \mc{ }                   \\[2pt]
$M_{\rm b}$ (\Mjup)    & \mc{0.63}      & \mc{ }          & 0.69 & 0.05  & \mc{ }      & \mc{ }                   & 0.685 & 0.02          & \mc{ }                   \\
\end{tabular}
\setlength{\tabcolsep}{4pt}
\begin{tabular}{l r@{\,$\pm$\,}l r@{\,$\pm$\,}l r@{\,$\pm$\,}l r@{\,$\pm$\,}l r@{\,$\pm$\,}l r@{\,$\pm$\,}l r@{\,$\pm$\,}l r@{\,$\pm$\,}l r@{\,$\pm$\,}l r@{\,$\pm$\,}l r@{\,$\pm$\,}l r@{\,$\pm$\,}l r@{\,$\pm$\,}l}
\hline
                       & \mc{Winn et al.} & \mc{Wittenmyer et al.} & \mc{Richardson et al.} & \mc{Masana et al.}
                             & \mc{Knutson et al.} & \mc{TWH08} & \mc{Miller-Ricci} & \mc{Kipping} \\
                       & \mc{\citeyearpar{Winn+05apj}} & \mc{\citeyearpar{Wittenmyer+05apj}}
                             & \mc{\citeyearpar{Richardson+06apj}} & \mc{\citeyearpar{Masana++06aa}}
                                   & \mc{\citeyearpar{Knutson+07apj}} & \mc{ } & \mc{et al.\ \citeyearpar{Millerricci+08apj}}
                                          & \mc{\citeyearpar{Kipping08mn}} \\
\hline
$a$         (AU)       & \mc{ }          & \mc{ }        & \mc{ }        & \mc{ }        & \mc{ }        & \erm{0.04707}{0.00046}{0.00047} & \mc{ }        & \mc{ }                    \\[2pt]
$M_{\rm A}$ (\Msun)    & \mc{1.06 fixed} & 1.093 & 0.092 & \mc{1.171}    & \mc{ }        & 1.101 & 0.066 & \erm{1.119}{0.033}{0.033}       & 1.101 & 0.064 & \erm{1.163}{0.096}{0.079} \\
$R_{\rm A}$ (\Rsun)    & 1.15 & 0.05     & 1.145 & 0.056 & 1.064 & 0.069 & 1.164 & 0.057 & 1.125 & 0.023 & \erm{1.155}{0.014}{0.016}       & 1.118 & 0.03  & \erm{1.072}{0.055}{0.052} \\
$\log g_{\rm A}$ (cgs) & \mc{ }          & \mc{ }        & \mc{ }        & \mc{ }        & \mc{ }        & \erm{4.361}{0.007}{0.008}       & \mc{ }        & \mc{ }                    \\
$\rho_{\rm A}$ (\psun) & \mc{ }          & \mc{ }        & \mc{ }        & \mc{ }        & \mc{ }        & \erm{0.727}{0.010}{0.010}       & \mc{ }        & 0.94 & 0.16               \\[2pt]
$M_{\rm b}$ (\Mjup)    & 0.657 & 0.064   & 0.657 & 0.052 & \mc{ }        & \mc{ }        & 0.64 & 0.06   & \erm{0.685}{0.015}{0.014}       & \mc{ }        & 0.681 & 0.039             \\
\hline \hline \end{tabular} \end{flushleft} \end{table*}

\clearpage

\begin{table*}
\caption{\label{tab:abspar:MRstars} The physical properties of the stellar components
of the TEPs studied in this work, derived using the empirical mass--radius relation.}
%
 \end{table*}

\begin{table*}
\caption{\label{tab:abspar:MRplanets} The physical properties of the planetary components
of the TEPs studied in this work, derived using the empirical mass--radius relation.}
%
 \end{table*}

\clearpage

\begin{table*}
\caption{\label{tab:abspar:pub1} \reff{Properties of the stellar components of the
TEPs not studied in this work. References are given in Table\,\ref{tab:abspar:pub3}.}}
%
 \end{table*}

\begin{table*}
\caption{\label{tab:abspar:pub2} \reff{Properties of the planetary components of the
TEPs not studied in this work. References are given in Table\,\ref{tab:abspar:pub3}.}}
%
 \end{table*}

\begin{table*}
\caption{\label{tab:abspar:pub3} \reff{References to Tables
\ref{tab:abspar:pub1} and \ref{tab:abspar:pub2}.} }
%
 \end{table*}

\end{document}